\documentclass{aa}  

\usepackage{graphicx}
\usepackage{graphbox}
\usepackage{txfonts}
\usepackage{amssymb}
\usepackage{mathtools}
\usepackage{color}
\usepackage{siunitx}

\usepackage{footnote}
\makesavenoteenv{tabular}
\makesavenoteenv{table}
\usepackage[bottom]{footmisc}
\usepackage{tablefootnote}
\usepackage{longtable}
\usepackage[flushleft]{threeparttable}
\usepackage{threeparttablex}
\usepackage{adjustbox}
\usepackage{subcaption}
\usepackage{float}
\usepackage{soul}
\usepackage{booktabs}
\usepackage{tabularx}
\usepackage{tabularray}
\usepackage{afterpage}
\usepackage{placeins}
\usepackage{hyperref}
\hypersetup{
colorlinks = true,
citecolor = blue
}
\begin{document}

\title{Radial velocities and stellar population properties of  56 MATLAS dwarf galaxies observed with MUSE} 
    \author{Nick Heesters\inst{1}, Oliver M\"uller\inst{1}, Francine R. Marleau\inst{2}, Pierre-Alain Duc\inst{3}, Rubén Sánchez-Janssen\inst{4}, M\'elina Poulain\inst{5}, Rebecca Habas\inst{3}, Sungsoon Lim\inst{6}, Patrick R. Durrell\inst{7}}

    \authorrunning{Heesters et al.}
    \titlerunning{MATLAS dwarfs with MUSE}

    \institute{Institute of Physics, Laboratory of Astrophysics, École Polytechnique Fédérale de Lausanne (EPFL), 1290 Sauverny, Switzerland
    \and
    Institut f{\"u}r Astro- und Teilchenphysik, Universit{\"a}t Innsbruck, Technikerstra{\ss}e 25/8, Innsbruck, A-6020, Austria
    \and
    Observatoire Astronomique de Strasbourg  (ObAS), Universite de Strasbourg - CNRS, UMR 7550 Strasbourg, France
    \and
    UK Astronomy Technology Centre, Royal Observatory, Blackford Hill, Edinburgh, EH9 3HJ, UK
    \and
    Space Physics and Astronomy Research Unit, University of Oulu, P.O. Box 3000, FI-90014, Oulu, Finland
    \and
    Department of Astronomy and Center for Galaxy Evolution Research, Yonsei University, Seoul 03722
    \and
    Youngstown State University, One University Plaza, Youngstown, OH 44555 USA
     }

   \date{Received ; accepted }

 
  \abstract 
   {Dwarf galaxies have been extensively studied in the Local Group, in nearby groups, and selected clusters, giving us a robust picture of their global stellar and dynamical properties, such as their circular velocity, stellar mass, surface brightness, age and metallicity in particular locations in the Universe. Intense study of these properties has revealed correlations between them, the so-called scaling relations, including the well known universal stellar mass-metallicity relation. However, since dwarfs play a role in a vast range of different environments, much can be learned about galaxy formation and evolution through extending the study of these objects to various locations. We present MUSE spectroscopy of a sample of 56 dwarf galaxies as a follow-up to the MATLAS survey in low-to-moderate density environments beyond the Local Volume. The dwarfs have stellar masses in the range of $M_{*}/M_{\odot}$ = 10$^{6.1}$-10$^{9.4}$ and show a distance range of D = 14-148\,Mpc, the majority (75\%) of which are located in the range targeted by the MATLAS survey (10-45\,Mpc). We thus report a 75\% (79\% for dwarf ellipticals) success rate for the semi-automatic identification of dwarf galaxies in the MATLAS survey on the here presented subsample. Using pPXF full spectrum fitting, we determine their line-of-sight velocity and can match the majority of them with their massive host galaxy. Due to the observational setup of the MATLAS survey the dwarfs are located in the vicinity of massive galaxies. Therefore, we are able to confirm their association through recessional velocity measurements. Close inspection of their spectra reveals that $\sim$ 30\% show clear emission lines and thus star formation activity. We estimate their stellar population properties (age and metallicity) and compare our results with other works investigating Local Volume and cluster dwarf galaxies. We find that the dwarf galaxies presented in this work show a systematic offset from the universal stellar mass-metallicity relation towards lower metallicities at the same stellar mass. A similar deviation is present in other works in the stellar mass range probed in this work and might be attributed to the use of different methodologies for deriving the metallicity.}


   \maketitle
%

\section{Introduction}
\label{sec:intro}

Dwarf galaxies are regarded as the oldest and most numerous galaxy type in the Universe \citep{1990A&A...228...42B,1994A&ARv...6...67F}, responsible for the formation of the more luminous and higher mass galaxies we see today \citep{2012AnP...524..507F}. They are typically defined as galaxies with stellar masses $\leq$ 10$^{9}$\,M$_{\odot}$ \citep{bullock2017small}, small physical sizes and magnitudes fainter than -17\,mag in the \emph{V}-band \citep{tammann1994dwarf,tolstoy2009star}. 

Since most dwarf galaxies have low surface brightness they are elusive when compared to massive host galaxies. Thus their study has been limited by instrumental constraints for a long time, leading to well studied populations only in the Local Group \citep[LG; e.g.,][]{1998ARA&A..36..435M,2006MNRAS.371.1983M,2007ApJ...671.1591I,2008ApJ...686..279K,2009Natur.461...66M,2011ApJ...742L..15B,2012AJ....144....4M,2014ApJ...780..128I,2016ApJ...833..167M,2018ApJ...868...55M,2019ARA&A..57..375S,2020ApJ...893...47D} and a handful of nearby groups in the Local Volume (LV; D $\lesssim$ 10\,Mpc) \citep[e.g.,][]{2013AJ....146..126C,2017ApJ...837..136D,2019ApJ...872...80C,2019ApJ...878L..16C,2020ApJ...893L...9B,muller2021properties}, and some galaxy clusters \citep[e.g.,][]{ferrarese2012next,eigenthaler2018next,venhola2019fornax}. In order to model and understand galaxy formation and evolution across cosmic time, it is essential to answer the question of whether the dwarfs studied in the Local Volume are representative of dwarfs in the nearby Universe at large. 

Even though galaxy formation and evolution is thought to depend on a number of different factors and processes, galaxies show remarkably tight correlations between some of their basic stellar and dynamical properties \citep[e.g.,][]{binggeli1997photometric,tassis2008scaling}. These so-called scaling relations have been extensively studied for different galaxy types, in a range of environments, and in particular in light of a possible evolution with time \citep[see e.g.,][for a recent review]{d2021past}. A few examples of well established galaxy scaling relations are the velocity-luminosity or Tully-Fisher relation \citep{tully1977new,courteau2007scaling}, the Faber-Jackson relation \citep{faber1976velocity}, the Kormendy relation \citep{1977ApJ...218..333K}, the fundamental plane of galaxies \citep[e.g.,][]{1987ApJ...313...59D,1987ApJ...313...42D,cappellari2006sauron,la2008sdss}, the bulge to black hole mass relation \citep{1998AJ....115.2285M} and the mass-radius relation \citep{chiosi2020parallelism}. 

The mass-metallicity relation (MZR) is another long known and studied scaling relation which exists for both gas- and stellar metallicities \citep[see e.g.,][for a recent review]{2019A&ARv..27....3M}. Stellar spectroscopy and the analysis of color-magnitude diagrams first revealed this connection in nearby elliptical galaxies \citep{1968AJ.....73.1008M,1972ApJ...176...21S,1983ApJ...270..471M,buonanno1985color}. Analysis of the Sloan Digital Sky Survey (SDSS) optical spectra has shown that this relation persists in galaxies with stellar masses in the range $M_{*}/M_{\odot}$ = 10$^{9}$-10$^{12}$ for stellar and gas metallicities \citep{tremonti2004origin,2005MNRAS.362...41G,2006MNRAS.370.1106G,2006ApJ...647..970L,2008MNRAS.391.1117P,2010MNRAS.408.2115M,2014ApJ...791L..16G}. The study of this correlation was extended to dwarf galaxies in the LG, where \citet{kirby2013universal} found that dwarf galaxies follow the same stellar MZR as more massive galaxies. This result was also found in semi-analytical galaxy formation and evolution models (SAMs) by several authors \citep[e.g.,][]{2010MNRAS.401.2036L,2011MNRAS.417.1260F,2014ApJ...791....8H,2014ApJ...795..123L,2017ApJ...846...66L}.  \cite{xia2019stellar,xia2019origin} have demonstrated that the relation is universal in SAMs for different types of galaxies and over a large range of stellar ($M_{*}/M_{\odot}$ $\sim$ 10$^3$-10$^{11}$) and dark matter halo masses ($M_{halo}/M_{\odot}$ $\sim$ 10$^{9}$-10$^{15}$\,h$^{-1}$).

There are various mechanisms which are potentially driving the MZR and have been discussed in the literature \citep[see][and references therein]{2019A&ARv..27....3M}. The most important ones are: outflows due to stellar feedback \citep[e.g.,][]{garnett2002luminosity,brooks2007origin}, downsizing \citep[e.g.,][]{cowie1996new}, low mass galaxies being in an earlier evolutionary stage and therefore showing larger gas fractions \citep[e.g.,][]{erb2006stellar}, a potentially mass-dependent initial mass function for higher mass galaxies \citep[e.g.,][]{koppen2007possible,trager2000stellar,molla2015galactic,vincenzo2016modern,lian2018modelling} and finally metal-rich accreted gas from previous burst of star formation in larger mass systems compared to lower mass ones \citep{brook2014magicc,ma2016origin}.

In addition to the mass of a galaxy, its environment is one of the main independent factors when it comes to galaxy evolution and as such has been studied extensively in the context of the MZR \citep{trager2000stellar,kuntschner2001dependence,thomas2005epochs,thomas2010environment,sheth2006environment,sanchez2006stellar,pasquali2010ages,zhang2018stellar}. It has been found that while the environment has a significant influence on the morphology, age and star formation activity of a galaxy, its direct contribution to the shape of the MZR is small \citep{thomas2010environment,mouhcine2011galaxy,fitzpatrick2015early,sybilska2017helena}. \cite{peng2015strangulation} and \cite{trussler2020both}, however, find that dwarf satellite galaxies in high density environments are more metal rich compared to dwarfs residing in lower density environments. This observation is attributed to the process of starvation, i.e., the lack of cold gas accretion in these high density regions.

One caveat when calculating the MZR and comparing with results from other works is the variety of methods which can be used to determine the metallicity \citep[e.g.,][]{kewley2008metallicity}. Another important factor in the context of this work is that there have been few studies investigating the MZR in the dwarf galaxy mass regime \citep{lequeux1979chemical,2006ApJ...647..970L,vaduvescu2007chemical,zahid2012metallicities,kirby2013universal,andrews2013mass} and these have mostly been limited to the LG. Therefore, increasing the sample size of dwarf galaxies in different environments will greatly benefit these discussions. In recent years there has been a great effort to advance our knowledge of dwarf galaxies beyond the boundaries of the LG \citep[e.g.,][]{2009ApJ...692.1447I,2009AJ....138..338S,2011MNRAS.412.1881K,2013AJ....146..126C,2017ApJ...850..109B,2017ApJ...837..136D,2017ApJ...848...19P,2018ApJ...868...96C,2019ApJ...872...80C,2020A&A...644A..91M,2021MNRAS.500.3854D,2021ApJS..256....2D,muller2021properties,2022ApJ...926...77M,2022ApJ...933...47C} and further beyond the LV \citep[e.g.,][]{2017ApJ...847....4G,2018ApJ...857..104G,2019ApJS..240....1Z,habas2020newly,2021A&A...647A.100S,2021MNRAS.500.2049P,2021ApJS..252...18T,2021ApJ...907...85M,2022A&A...659A..92L}. The MATLAS survey \citep{duc2015atlas3d,2020MNRAS.498.2138B}, the basis for this study, is among the latter and delivers a large number (2210) of newly discovered dwarf galaxies beyond the LV. The vast majority of the dwarf galaxies identified in the MATLAS fields around $\sim$ 140 targeted ETGs, however, only have photometric data and thus their distance, satellite nature and environment is uncertain. In the case of the M101 group, in the Local Volume, it has been shown that $\sim$ 80\,\% of the candidates in a dwarf catalog have been contaminants \citep{bennet2017discovery,bennet2019m101}. However, due to a careful detection and selection procedure \citep[see][]{habas2020newly}, the degree of contamination in the MATLAS dwarf catalog is likely significantly lower. In order to advance our understanding of structure formation as a function of the environment, it is therefore of great importance to obtain distance or recessional velocity estimates for dwarf galaxies in order to confirm their dwarf and satellite nature. 

In this study we aim to add information about line-of-sight velocities of dwarf galaxies identified in the MATLAS fields beyond the LV and to compare their extracted stellar population properties with results from other studies. This information contributes to the connection between host halo and number of subhalos \citep[e.g.,][]{2018PhRvL.121u1302K,2019ApJ...873...34N,2021ApJ...923...35M}, the morphology-density relation, i.e. the role the environment plays in the formation and evolution of dwarf galaxies \citep[e.g.,][]{ferguson1990nasa,2012AJ....144....4M,ferguson1989spatial,sawala2012local,steyrleithner2020effect}, the discussion on phase-space correlations in dwarf satellites \citep[e.g.,][]{kunkel1976magellanic,lynden1976dwarf,pawlowski2012vpos,ibata2013vast,muller2016testing,muller2018whirling,muller2019dwarf,muller2021coherent,2021A&A...654A.161H,2022NatAs.tmp..273S} and the study of scaling relations for low-mass galaxies beyond the LV \citep[see][on scaling relations in the MATLAS dwarfs based on photometry]{habas2020newly,poulain2021structure,2021A&A...654A.105M,2022A&A...659A..14P}.

This paper is structured as follows: in Section \ref{sec:obs_data} we describe the MATLAS survey as a precursor and base for this study, as well as the observational details utilizing the MUSE instrument. We then map out the individual steps we use to reduce the data, to fit the spectra and to estimate our errors. In Section \ref{sec:results} we present our findings regarding the dwarf line-of-sight velocities and resulting satellite nature. We then discuss background contamination as mentioned above and comment on the assumptions regarding satellite membership made in the MATLAS survey. We extract the stellar populations and discuss our results in light of the universal stellar mass-metallicity relation. In Section \ref{sec:conc} we summarize our results and give an outlook.

\section{Observations and data reduction}
\label{sec:obs_data}

\subsection{The MATLAS dwarf galaxy candidate sample}

The dwarf galaxies were identified \citep{habas2020newly} in the MegaCam images of the MATLAS survey \citep{duc2015atlas3d}, an extension of the ATLAS$^{3D}$ project \citep{cappellari2011atlas3d}, which aims to characterize the morphology and the kinematics of 260 early-type galaxies (ETG) in the context of galaxy formation and evolution. The ETGs are between 10 and 45 Mpc, have declinations within $|\delta - 29| < 35$ degrees, galactic latitudes $> 15$ degrees, and \emph{K}-band absolute magnitudes below $-$21.5. The hosts mostly reside in group environments and a few are isolated. 

The MATLAS fields were observed between 2012 and 2015. Each pointing has an ETG in its center and may contain additional ETGs and late-type galaxies. The MATLAS data was observed in the \emph{g}, \emph{r}, and \emph{i}-bands for 150, 148, and 78 fields, respectively, as well as in the \emph{u}-band for 12 fields. A surface brightness limit of $28.5\,-\,29.0$ mag arcsec$^{-2}$ was reached in the \emph{g}-band.

In total 2210 dwarf galaxies were identified in these fields using a visual and semi-automatic approach \citep{habas2020newly}. Their structural parameters were presented in \citet{poulain2021structure}. About 75\,\% of the dwarf galaxy candidates are dwarf ellipticals (dEs). The dwarfs are located in 1\,deg$^{2}$ fields around the targeted ETG with a median value of 17 dwarf galaxies per field \citep[see][for details on the MATLAS dwarf catalog]{habas2020newly}. Since there are no distance estimates for the majority ($\sim 85\,\%$) of the dwarfs and the fields often contain massive ETGs and/or LTGs in addition to the targeted ETG, the satellite nature and association of the dwarfs to the massive galaxies are uncertain. In \citet{habas2020newly}, distances were estimated for 14\,\% based on literature spectroscopic and HI measurements \citep{2022A&A...659A..14P}, of which 90\,\% were confirmed to be members of the host system based on their relative velocities being consistent with the hosts velocity. Out of the 2210 dwarf galaxies, 3\,\% fall into the regime of the ultra-diffuse galaxies \citep{2021A&A...654A.105M}. Based on their globular cluster count, one of the most extreme cases of these ultra-diffuse galaxies is MATLAS-2019, which has been observed with HST \citep{2021ApJ...923....9M,2022ApJ...927L..28D} and MUSE \citep{2020A&A...640A.106M}. Based on the MUSE observations, it has a metal-poor and old stellar population \citep{2020A&A...640A.106M}.

\subsection{MUSE observations}

In this work, we followed up 56 dwarf galaxies from the MATLAS survey. We obtained the data from the Multi Unit Spectroscopic Explorer \citep[MUSE;][]{bacon2010muse,bacon2012news} at the Very Large Telescope (VLT) of the European Southern Observatory (ESO) from four different proposals and observational periods: P103 (PI: Marleau, proposal ID: 0103.B-0635), P106 (PI: Marleau, proposal ID: 106.21A1), P108 (PI: Marleau, proposal ID: 108.2214) and P109 (PI: Marleau, proposal ID: 109.22ZV). The goal of these proposals was to obtain a reference sample of dwarf galaxies identified in the low-to-moderate density MATLAS fields and the galaxies were observed under relaxed seeing conditions (filler conditions), with an average seeing of 1.0\,arcsec. We selected our targets to have an average surface brightness $\langle\mu_{e,g}\rangle$ $< 25.5$\,mag/arcsec$^2$ in the \emph{g}-band and an effective radius $r_{eff} > 3$\,arcsec. All targets satisfy 2$r_{eff} < 1$\,arcmin and are thus well matched with the field of view (FOV) of the MUSE Wide Field Mode (WFM) of 1\,$\times$\,1\,arcmin$^2$ with a spatial sampling of 0.2\,$\times$\,0.2 arcsec$^2$. The instrument covers a spectral range of 4750-9350\,\AA{} with a sampling of 1.25\,\AA{} and a resolving power of 1770 (480\,nm)\,-\,3590 (930\,nm). Each galaxy was observed for a single Observational Block (OB) with four science exposures (O) amounting to an on-target integration time of 2700\,s. We chose an OOOO observing sequence with 90 degree rotations and small dithering. The size of the galaxies in relation to the MUSE FOV allows us to obtain the sky spectra directly from the science exposure by implementing an offset from the target of $\pm$ $\sim$10\,arcsec in right ascension (RA) and declination (Dec). All of our targets are thus situated in a corner of the MUSE FOV, so as to optimize sky exposure and minimize contamination from stars and background galaxies. This strategy was used in other works in the literature \citep[e.g.,][]{2019A&A...625A..76E,2019A&A...625A..77F,muller2020spectroscopic}.

\subsection{Sky subtraction}

In order to enhance the sky subtraction performed in the reduced MUSE data cubes by the ESO pipeline, we first use the MUSE Python Data Analysis Framework (MPDAF) to collapse the data cube along the wavelength axis. We then detect all sources in the produced 2D image and create a binary mask, with 1 corresponding to a detection and 0 to the background. 
For this task we use a combination of Source Extractor \citep{bertin1996sextractor} and MTObjects \citep{teeninga2015improved}, employing their respective strengths of detecting faint point sources (Source Extractor) and the low surface brightness galaxy itself (MTObjects). This mask is used as an input for the Zurich Atmosphere Purge \citep[ZAP;][]{2016MNRAS.458.3210S} developed for MUSE, which uses principal component analysis in order to isolate and subtract sky features from the data cube. 

\subsection{Spectral extraction}

\begin{figure*}[!htb]
\centering
\includegraphics[width=\linewidth]{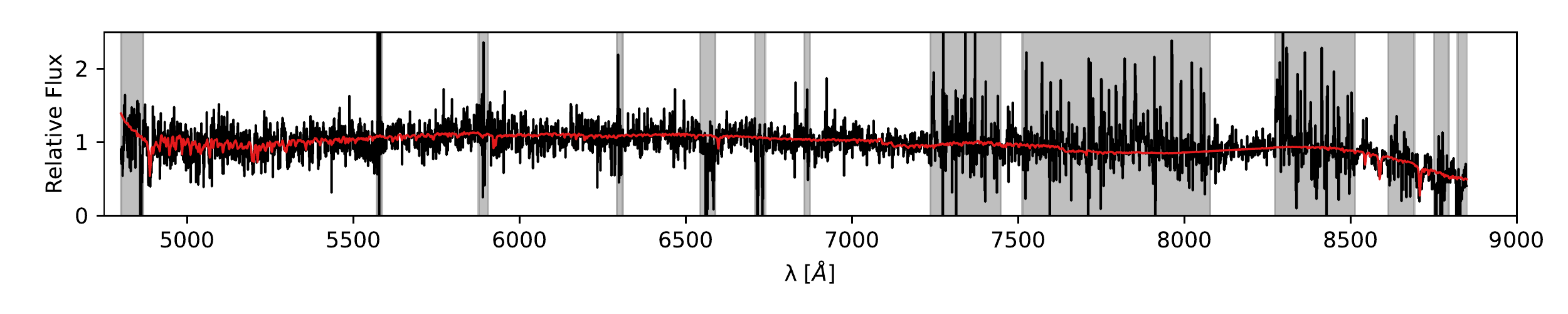}
\includegraphics[width=\linewidth]{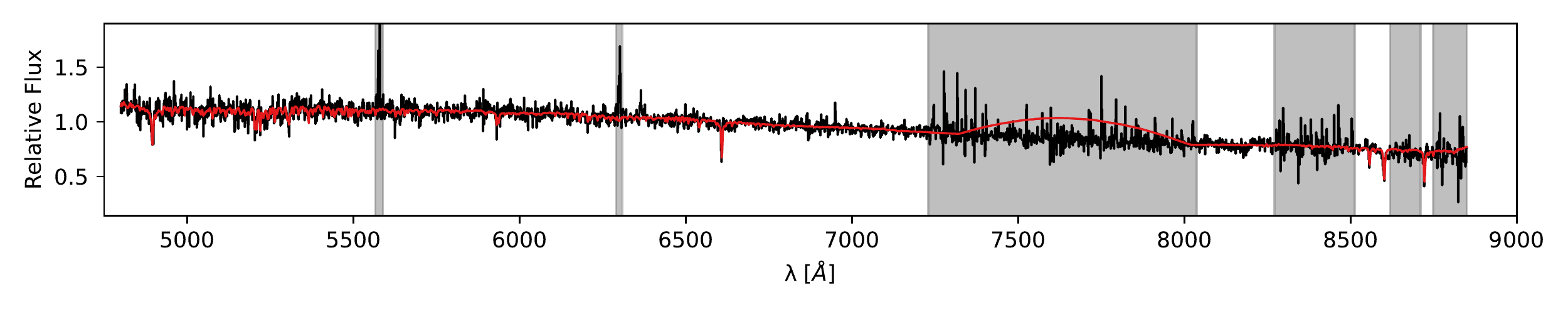}
\includegraphics[width=\linewidth]{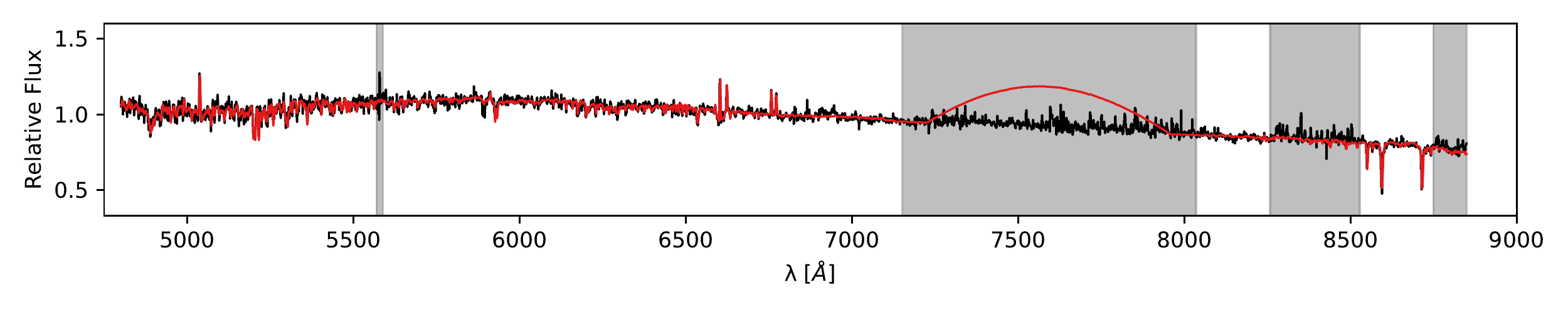}
\caption{Examples of different quality, i.e., different SNR, spectra. We show the relative flux on the y-axis and the wavelength on the x-axis in angstroms. We note that these spectra have not been shifted to the rest frame. The black line is the galaxy spectrum and the red line the fit produced by pPXF. Gray regions were manually masked out to improve the fit. From top to bottom we show the spectra for the galaxies MATLAS-269, MATLAS-1232 and MATLAS-10 with SNRs of 8.7, 26.0 and 61.9, respectively.}
\label{figure:snr_examples}
\end{figure*}

To extract the galaxies' spectra, we create masks for all of the dwarf galaxies on an individual basis, in order to isolate the dwarf spectra from the ones coming from other sources in the field of view. We use a combination of Source Extractor and manual masking in order to eliminate bright sources. Depending on the shape of the dwarf we draw circular or elliptical apertures centered on the galaxy. The resulting mask contains values 1 for the dwarf flux and 0 in all other regions, i.e., for radii beyond the defined aperture, where the collapsed image contains bright sources and where the median flux has values $\leq$ 0.2\,*10$^{-20}$\,erg/(\AA{}\,cm$^{2}$\,s). The latter constraint aids the optimization of the signal-to-noise ratio (SNR) per spaxel. We extract the noise from the second extension of the MUSE data cube using the same mask. In addition to a spatial mask, we also create a spectral one, manually eliminating residual sky lines which are not considered in the following full spectrum fitting (see the gray bands in Figure \ref{figure:snr_examples}).

\subsection{Full spectrum fitting}
\label{sec:fitting}

We fit the dwarf galaxy spectra using the Penalized Pixel-Fitting algorithm \citep[pPXF;][]{2004PASP..116..138C,2017MNRAS.466..798C}, a standard full spectrum fitting method to extract stellar and gas kinematics as well as stellar populations. Employing a similar strategy as in the literature \citep[e.g.,][]{2019A&A...625A..76E,2019A&A...625A..77F,2021A&A...645A..92M,2022A&A...667A.101F}, we use single stellar population (SSP) models form the E-MILES library \citep{vazdekis2016uv} with ages ranging from 70\,Myr to 14\,Gyr and metallicites from solar down to -2.27\,dex. We use a Kroupa initial mass function \citep[IMF;][]{2001MNRAS.322..231K} and account for the slightly varying MUSE resolution across the spectral range by convolving with a MUSE line spread function as described in \citet{2017A&A...608A...5G}. In order to extract the line-of-sight velocities we use additive and multiplicative polynomials of degrees 8 and 12, respectively \citep{2019A&A...625A..76E}. For the stellar population properties age and metallicity we fix the determined velocity and rerun pPXF, using only multiplicative polynomials of degree 12 \citep{2019A&A...625A..77F}. In dwarf galaxies featuring emission lines, we determine the line-of-sight velocity by fitting absorption and emission lines simultaneously. For the stellar populations we first mask all emission lines, only leaving the absorption spectrum. We utilize the weights returned by pPXF to calculate the mean age and metallicity as well as the mass-to-light ratio (ML) in the \emph{V}-band from the E-MILES SSPs for each galaxy. Here the ML is not a fitted parameter such as age and metallicity but is rather inferred from the age and metallicity grid returned by pPXF.



\begin{figure*}[htb]
\centering
\includegraphics[align = t,width=0.85\linewidth]{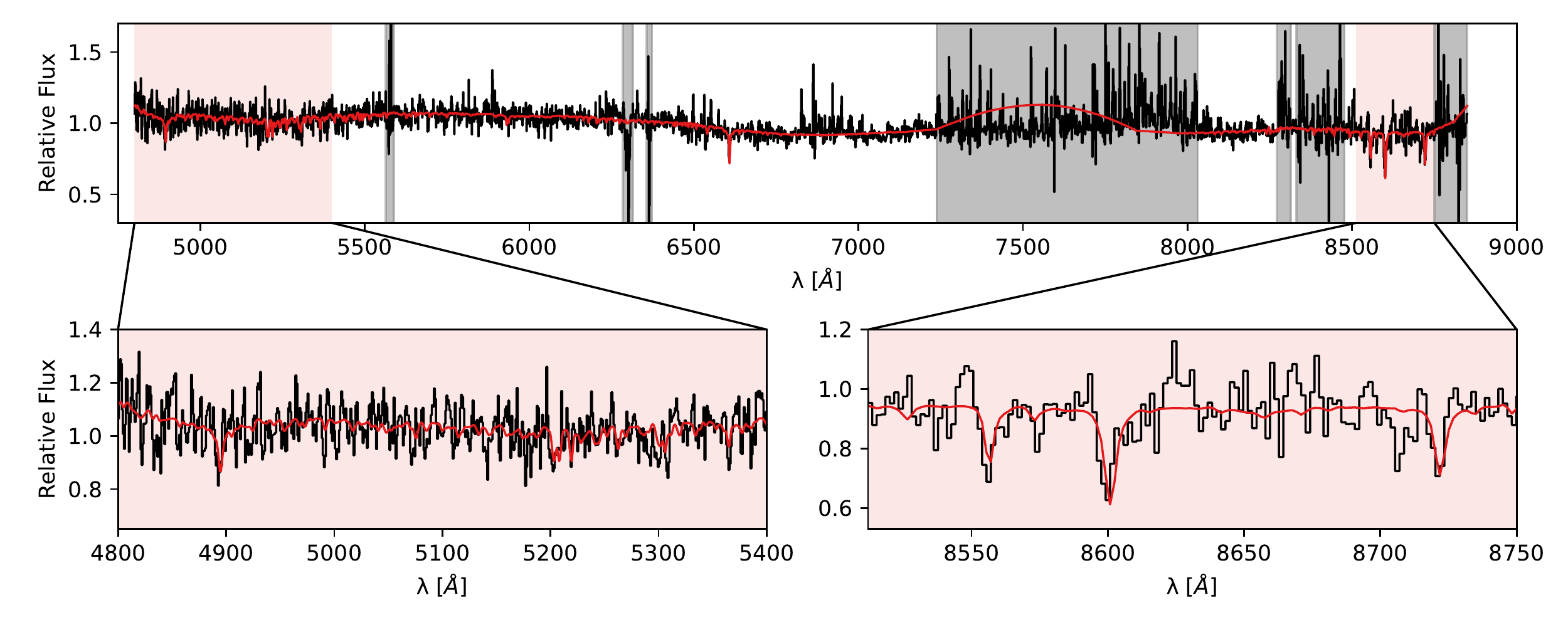}
\includegraphics[align = t,width=0.14\linewidth]{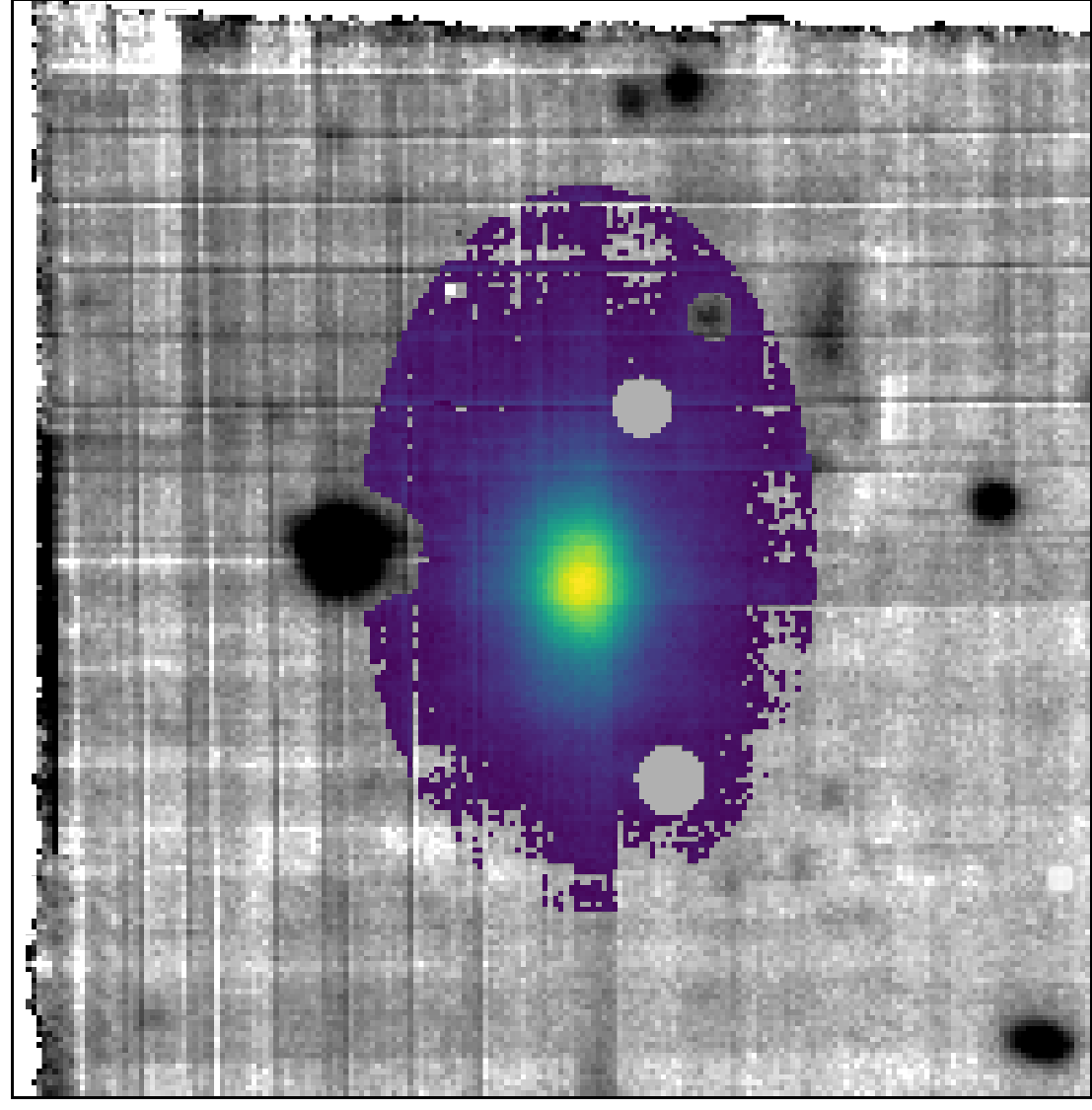}
\includegraphics[align = t,width=0.85\linewidth]{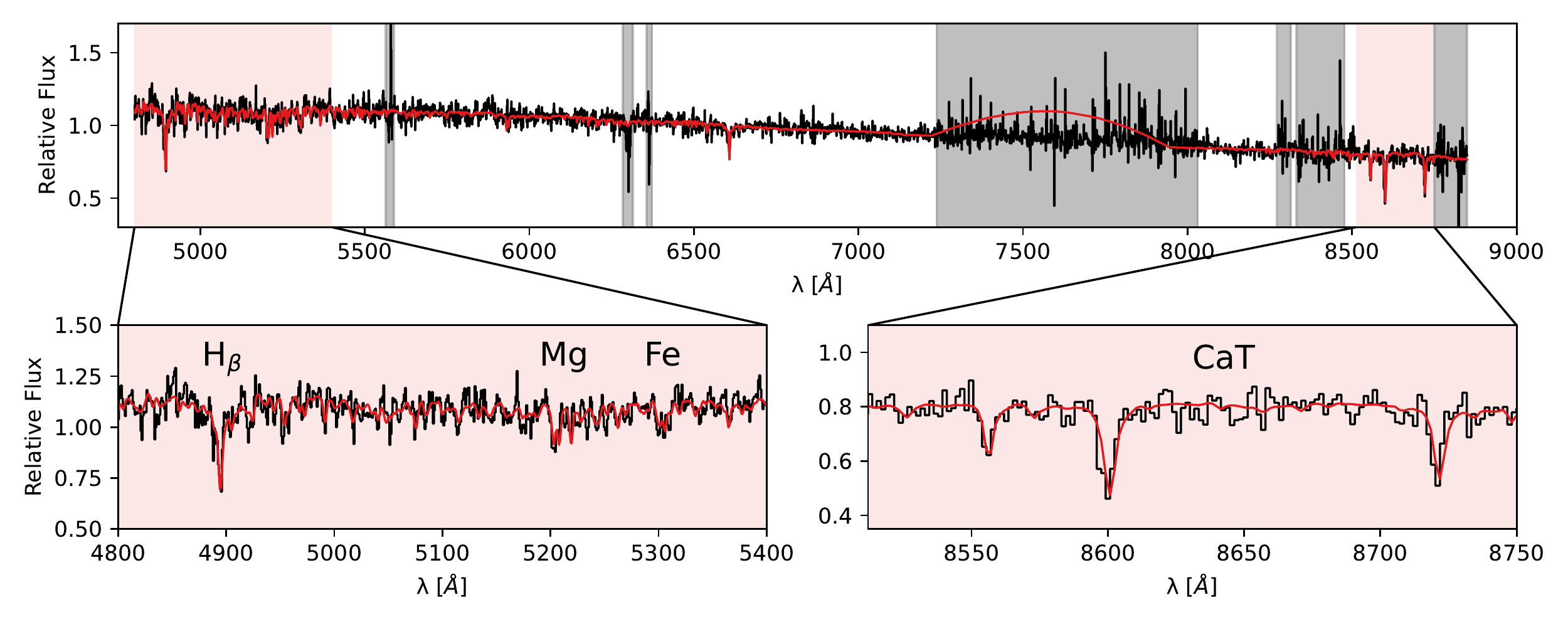}
\includegraphics[align = t,width=0.14\linewidth]{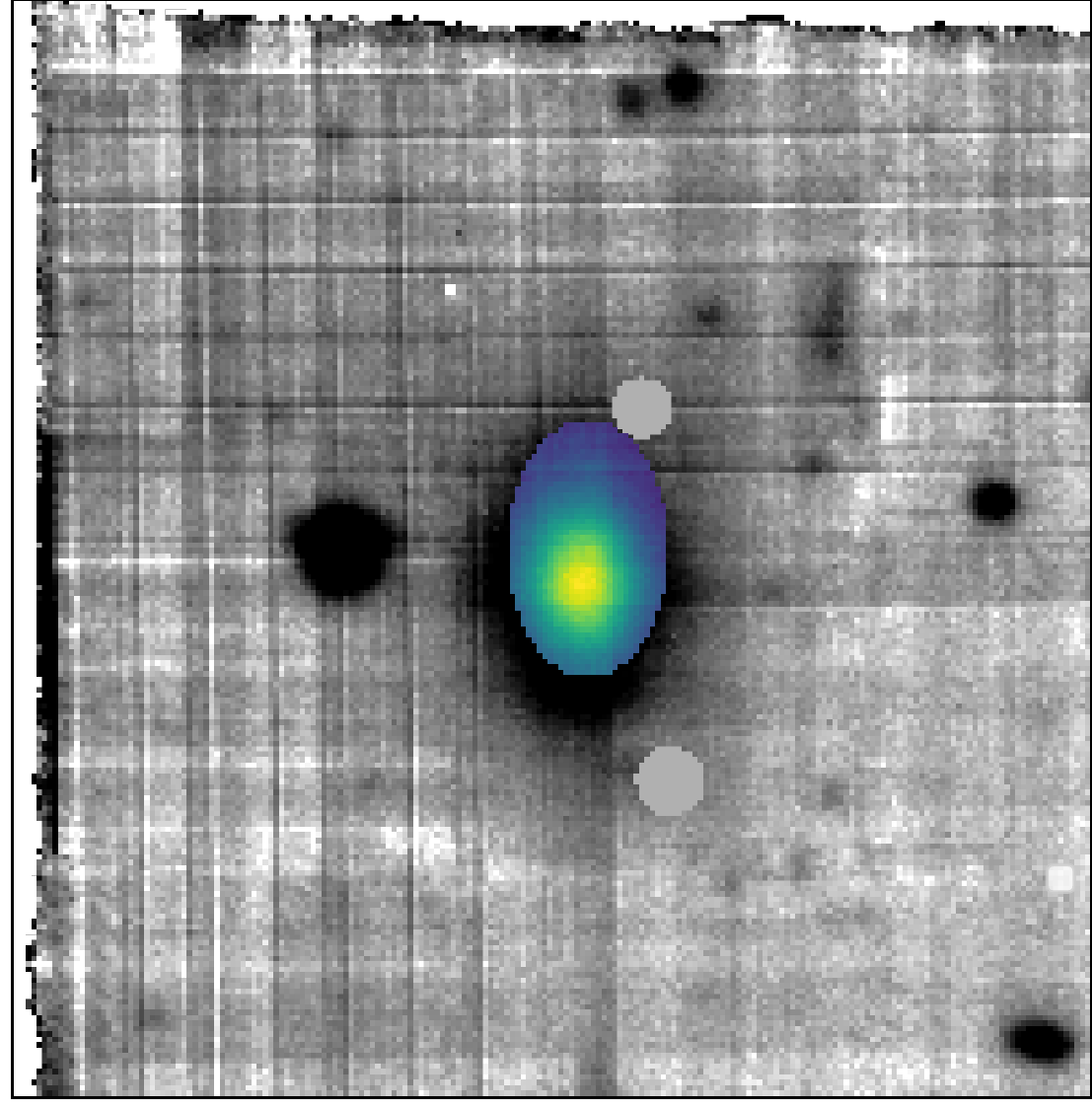}
\caption{Comparison of dwarf spectra from visually intuited aperture (top) vs. SNR optimized aperture (bottom). Right: stacked MUSE cube of the dwarf MATLAS-445. The spectra were extracted from the colored regions. The SNR was improved from 29.3 (visually intuited aperture) to 32.4 (optimized aperture) for this galaxy.}
\label{figure:snr_comp}
\end{figure*}

\subsection{Signal-to-noise ratio optimization}

The dwarf galaxies in this sample show a wide variety of SNRs. In Figure \ref{figure:snr_examples} we show examples of different quality dwarf spectra (i.e., different SNRs). The black lines show the galaxy flux while the red ones are the best fit lines returned by pPXF. Gray regions were manually masked and are not considered in the fit. This can lead to curves in the fit line in these unconstrained regions. An example of this can be seen in the second and third spectra in Figure \ref{figure:snr_examples} in the grayed out region ($\sim$\,7200-8000\,\AA). This does, however, not affect the quality of the fit.

In order to gain the maximal value from our MUSE data, we select the aperture from which we extract the spectrum so that the signal-to-noise ratio is optimized. To do this, we use the center of the dwarf galaxy and perform pPXF fitting for increasing aperture radii in a range $ r \in [10,100]$\,pix in steps of 10\,pix. We choose this range based on the apparent sizes these dwarfs have in the MUSE image ($\sim$ 300\,$\times$\,300\,pix). We use the aperture at the peak of the SNR to extract the spectrum and proceed with the analysis. To illustrate the difference between the spectrum extracted using a visually intuited aperture and the SNR optimized one, we show the corresponding spectra for one of the dwarfs (MATLAS-445) in Figure \ref{figure:snr_comp}. As is done in other studies \citep[e.g.,][]{2019A&A...625A..50F,muller2020spectroscopic,fahrion2020metal,muller2021properties}, we calculate the SNR in a region of the spectrum featuring no strong absorption or emission lines. Other works \citep[e.g.,][]{2019A&A...625A..50F,muller2021properties} have used the wavelength interval $[6600,6800]\,\AA$ for this purpose. Since the dwarf galaxies analyzed in this work show a range of different redshifts, we use this wavelength interval and shift it using the estimated redshift for each galaxy via the best fit recessional velocity. The SNR is calculated as the mean fraction between the flux and the square root of the variance. For this, the variance is multiplied by the $\chi^2$ value returned by pPXF, thus using a better estimate for the local noise \citep{2019A&A...625A..77F,muller2020spectroscopic}. 

For galaxies which are best described by an elliptical aperture, we increase its size by varying the semi-major axis $a \in [10,100]$ and keeping the axis ratio between minor and major axis of the ellipse constant. We use the radius or semi-major axis $a$ for which the SNR peaks for all further analysis. In some instances the SNR does not show a peak in the tested interval but diverges. Individual inspection of these cases reveals that this only occurs in low SNR ($\lesssim$10) and very faint galaxies. In these cases we manually set the aperture to best match visual intuition from the stacked data cube. It should be noted that for nucleated dwarfs we masked the nucleus for this SNR optimization in order to avoid a bias towards smaller apertures due to the nucleus dominating the SNR. We subsequently unmask the nucleus when extracting the dwarf properties.

\begin{figure}[!htb]
\centering
\includegraphics[width=\linewidth]{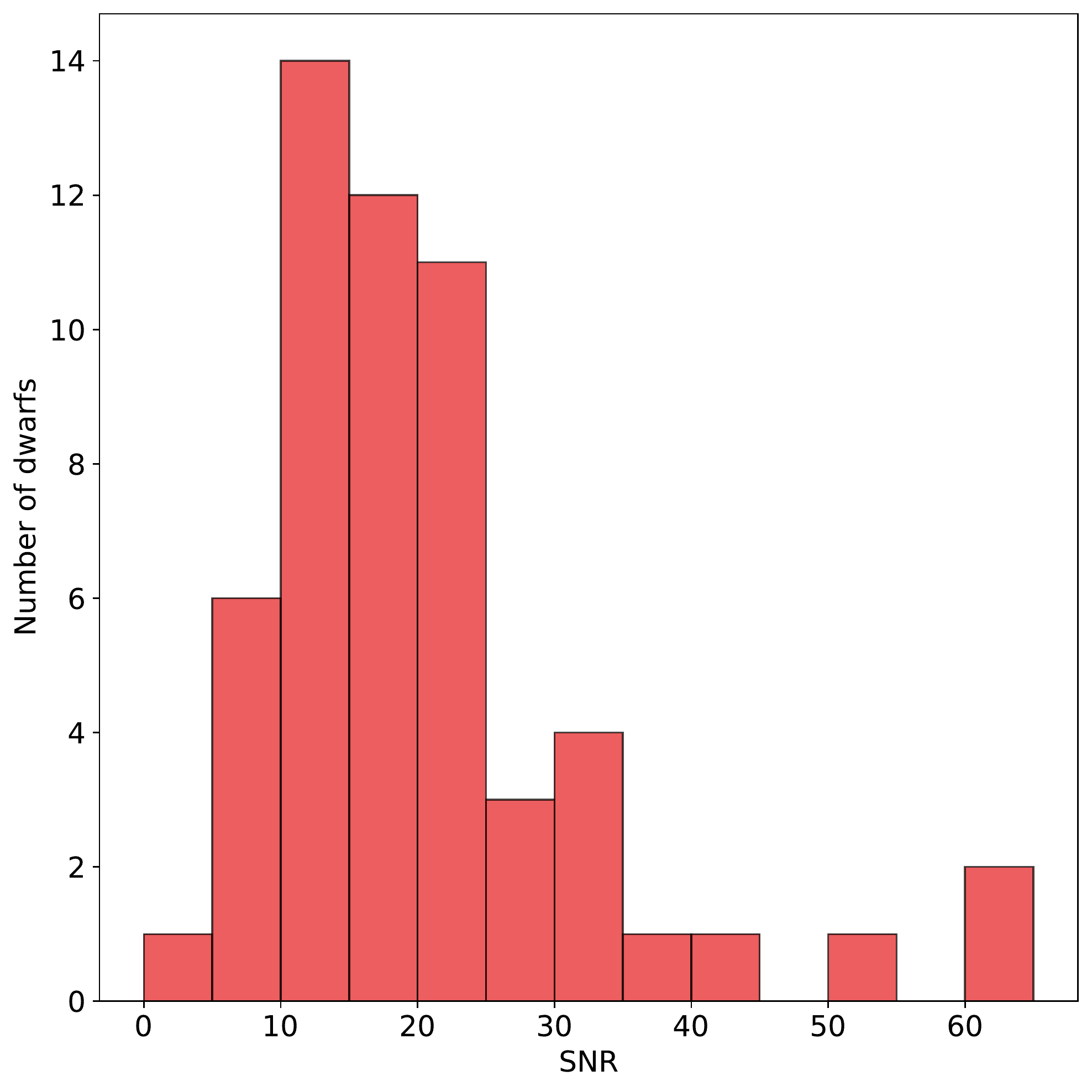}
\includegraphics[width=\linewidth]{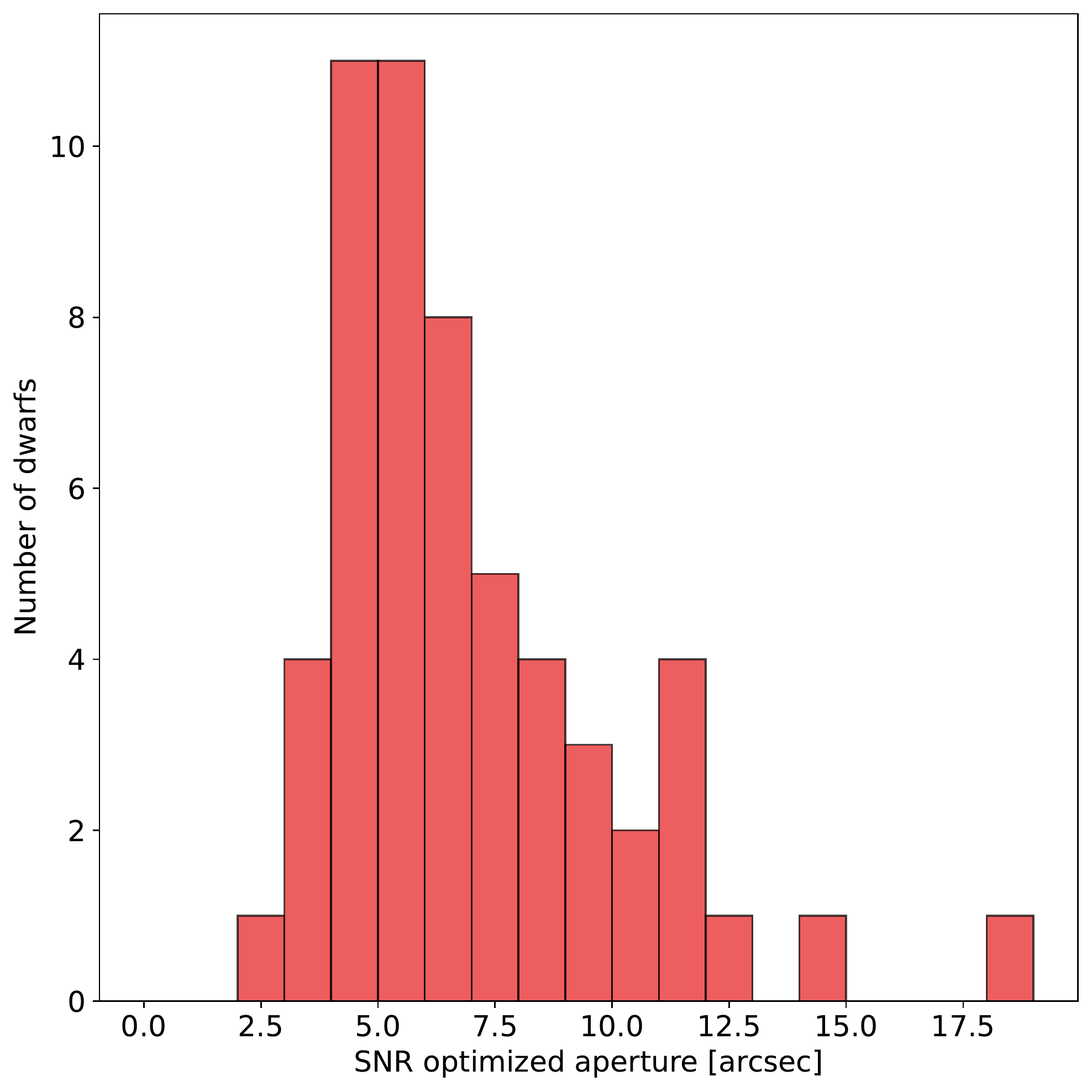}
\caption{Distributions of signal-to-noise ratios and corresponding optimized apertures from which the spectra were extracted. Top: distribution of SNRs for the sample of dwarf galaxies analyzed in this work. Bottom: distribution of SNR optimized apertures for the dwarf galaxies studied in this work.}
\label{figure:snr_apertures}
\end{figure}

\begin{figure}
\centering
\includegraphics[width=\linewidth]{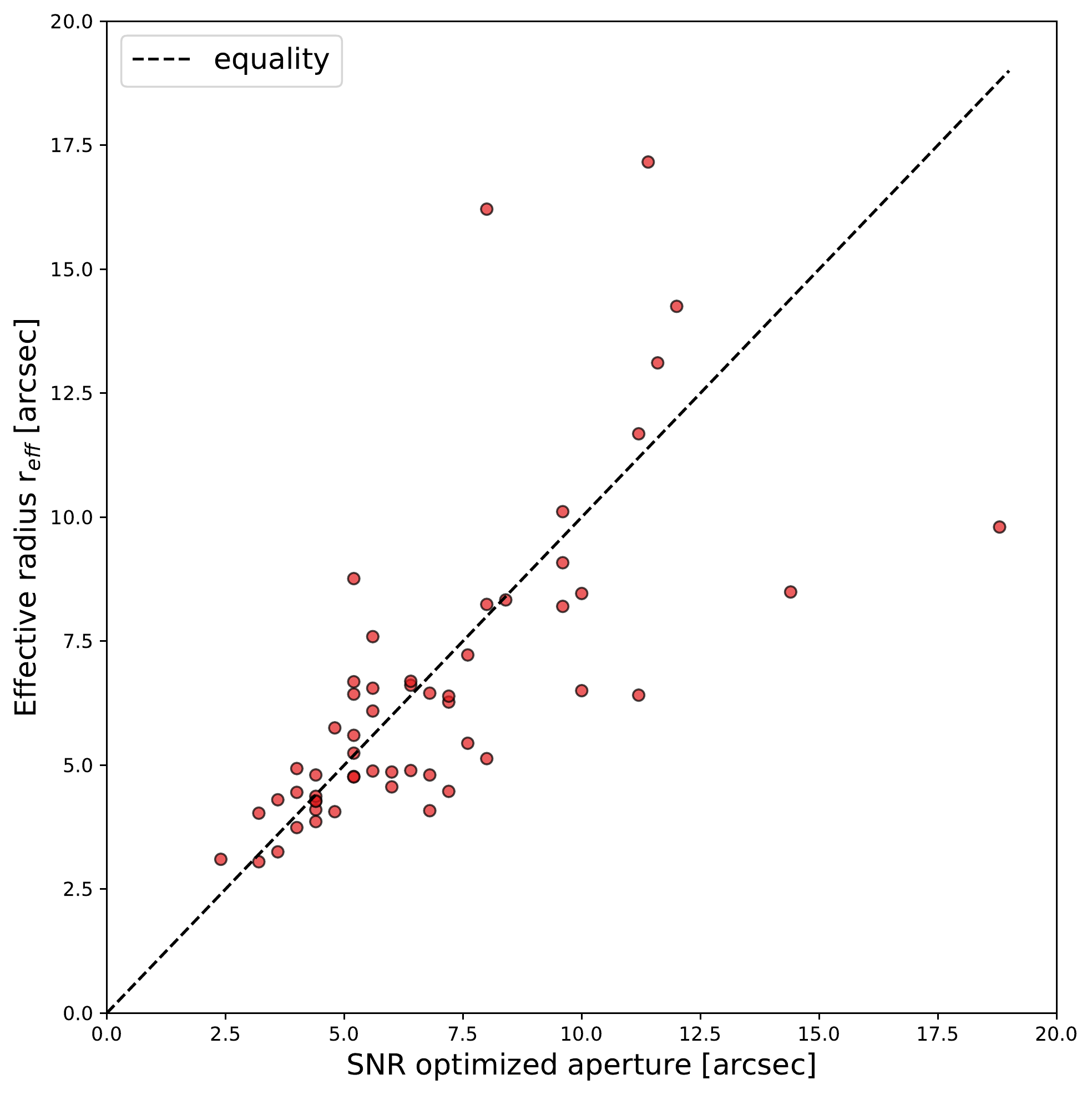}
\includegraphics[width=\linewidth]{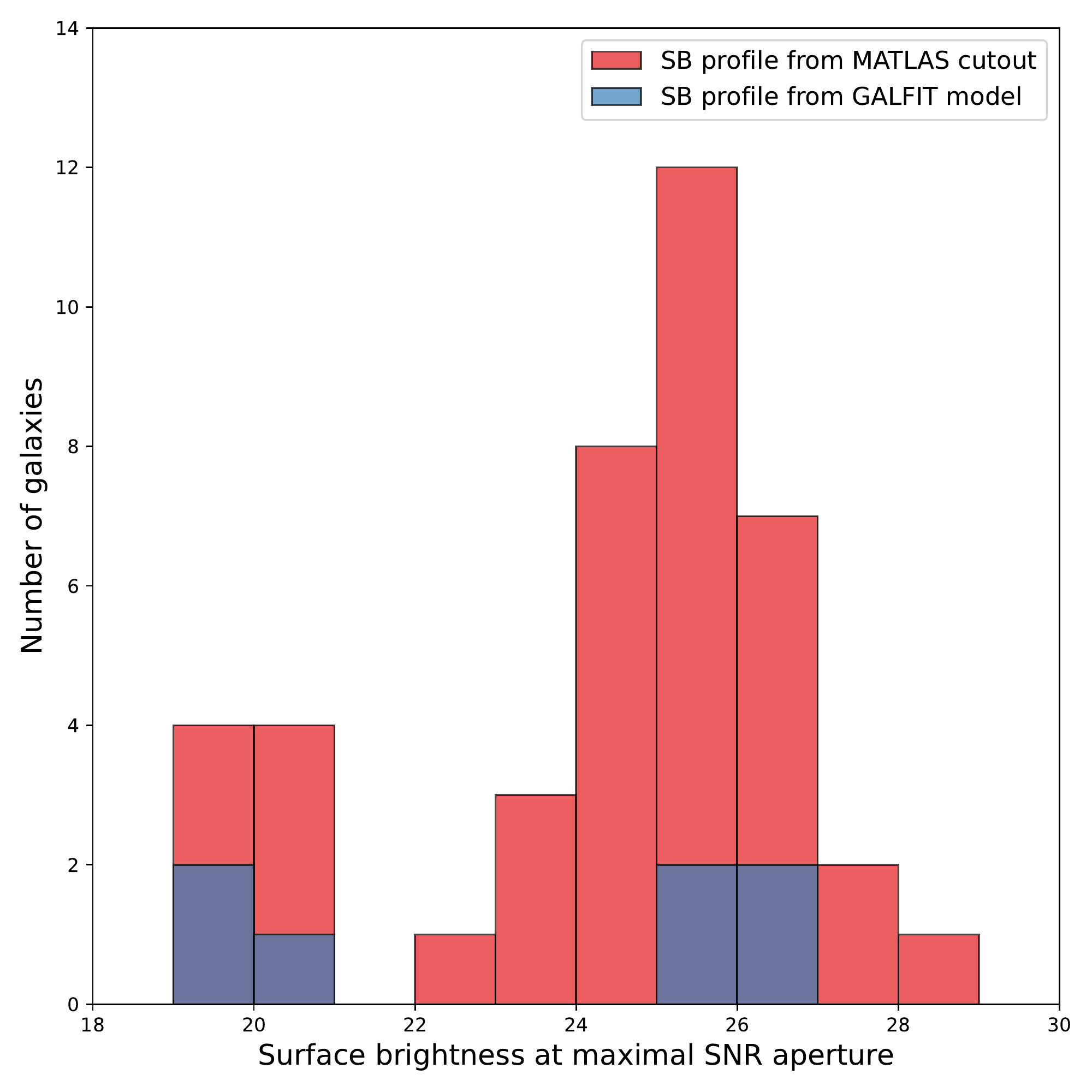}
\caption{Comparison of photometric properties with the SNR optimized apertures. Top: comparison between the SNR optimized radius obtained in this work and the effective radius from \citet{poulain2021structure}. The black dashed line indicates the equality of the two measurements. Bottom: distribution of the dwarf \emph{g}-band surface brightness from \citet{poulain2021structure} at the radius which optimizes the SNR of the extracted spectrum.}
\label{figure:reff_SB}
\end{figure}


In the top panel of Figure \ref{figure:snr_apertures} we show the SNR distribution of all dwarfs in our sample, which takes values in the range $\sim$[5,63]. In the bottom panel of the same figure we show the distribution of apertures (radius or semi-major axis) which lead to the maximum SNR for our dwarf sample and relate these apertures to the effective radius from their \textsc{galfit} model \citep{habas2020newly,poulain2021structure}. We can see a prominent peak around $\sim$\,5\,arcsec. In the top panel of Figure \ref{figure:reff_SB} we see that the effective radius $r_{eff}$ roughly traces the SNR optimized aperture. However, with a rather large scatter, which increases significantly as the dwarf size increases. Since we have obtained profiles describing the SNR as a function of the radius and have similar profiles of the surface brightness (measured in the \emph{g}-band) as a function of the radius \citep[see][]{poulain2021structure}, we can relate the SNR with the surface brightness. In the bottom panel of Figure \ref{figure:reff_SB} we show the distribution of the surface brightness at the SNR optimized dwarf radius. For eight dwarfs the SNR curve diverged and we set the optimal aperture manually at the first peak (if applicable) or based on visual intuition. We omit these dwarfs from the plot. In a few cases we are not able to obtain the surface brightness profile from the MATLAS images directly and measure it on the dwarf's \textsc{galfit} model. These cases are shown in blue. We note a clear peak at $\sim$\,25.5\,mag arcsec$^{-2}$ and a second smaller one at $\sim$\,20 mag arcsec$^{-2}$. Out of these 11 dwarfs in the lower peak, 8 show strong emission lines, which may explain the optimal radius at such high surface brightnesses. This distribution illustrates the gains expected in terms of SNR by probing deeper surface brightnesses.

\subsection{Error estimation}
We estimate the errors for all properties by running a Monte Carlo (MC) simulation. We determine the best fit to the galaxy spectrum using pPXF and calculate the residuals between best fit and input spectrum at each wavelength. We then create new realizations of the spectrum by using the best fit as a base line. At each wavelength we randomly add the residual or subtract it from the best fit. We re-run pPXF for the newly generated spectrum and compare the original values for the radial velocity, age and metallicity with the returned values for the new realization. We repeat this 400 times for each galaxy. This number is motivated by the error estimation for the lowest SNR galaxy in our sample for which a recessional velocity is obtainable. We varied the number of MC runs in an interval $N_{MC} \in [100,1000]$ in steps of 100 for this galaxy and observed stable values after 400 runs. The $1\sigma$ confidence interval of these MC simulations give the errors for each extracted property. We use the value of the initial best fit and calculate the errors in relation to the MC $1\sigma$ interval. In case the best fit value lies outside of this interval (see Figure \ref{figure:err_est_ex}), we use the mean and $1\sigma$ bounds of the MC simulation. In Appendix \ref{appendix} we show the residuals of the best fit value (velocity, age and metallicity) minus the mean/median of the respective MC realizations and indicate whether the best fit values lie within or outside the 1$\sigma$ from the MC simulation. While overall the best fit values are consistent with the mean and median of the MC realizations, there are some cases where the two values differ, in particular for the age.

\begin{figure}[!htb]
\centering
\includegraphics[width=\linewidth]{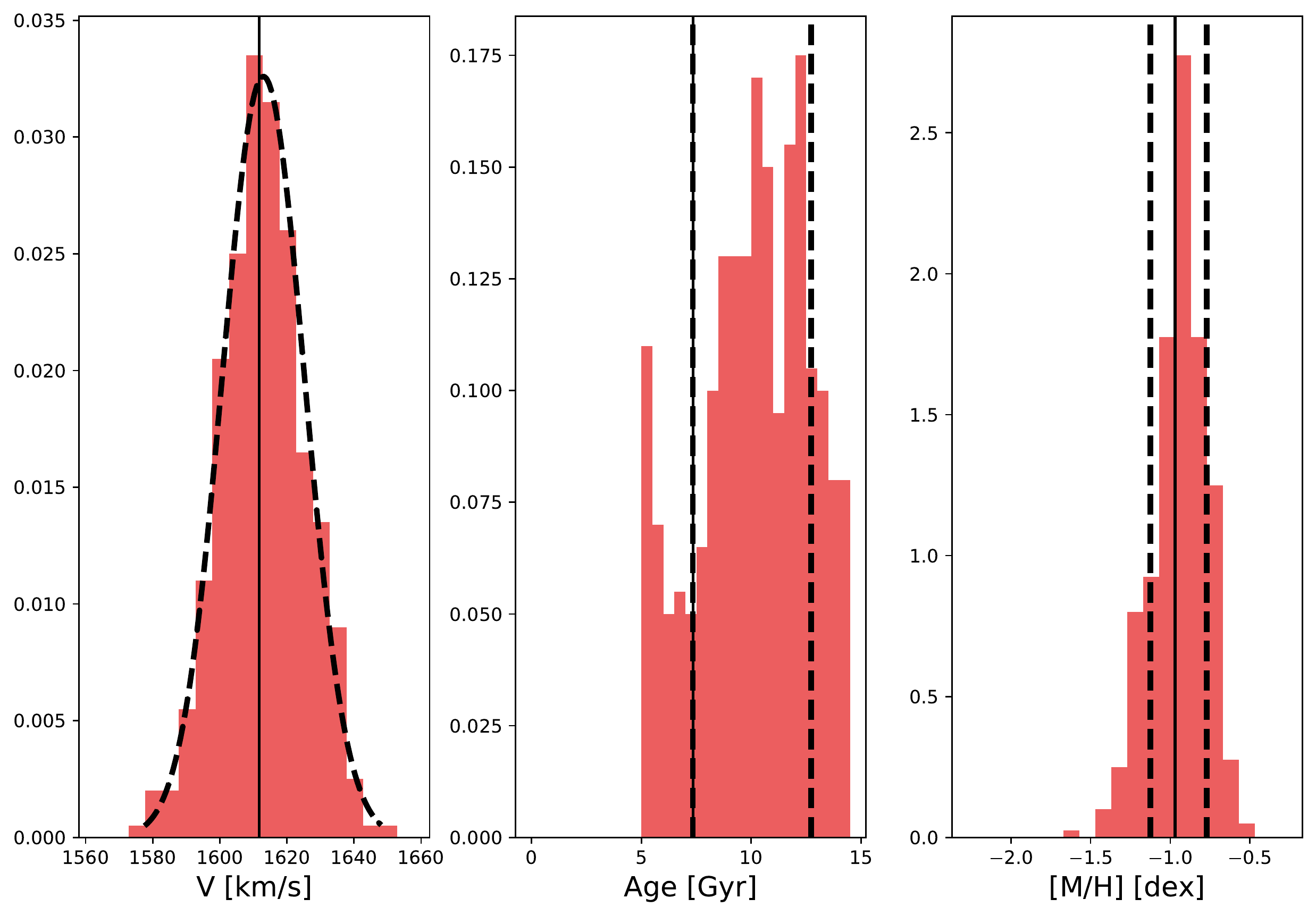}
\caption{Results from the MC error estimation with 400 iterations for the galaxy MATLAS-269. The histograms show the distributions of the fit values for the recessional velocity $V$, the age and the metallicity obtained from pPXF by randomly flipping the sign of the residual between the galaxy spectrum and the initial best-fit. The solid lines indicate the best-fit values of the original spectrum while the dashed lines show the standard deviation of the MC realizations. For the recessional velocity shown in the leftmost panel, we fit a Gaussian curve (dashed line) to the distribution of MC realizations.}
\label{figure:err_est_ex}
\end{figure}






\section{Results and discussion}
\label{sec:results}

In the following, we present the results of our analysis of the 56 non-cluster dwarf galaxies and compare them against dwarfs in the literature. First, we discuss the line-of-sight velocities obtained from the MUSE spectra and relate this new information to the assumptions made in the MATLAS studies \citep[e.g.,][]{habas2020newly,poulain2021structure,2021A&A...654A.105M,2021A&A...654A.161H,2022A&A...659A..14P}. Next, we illustrate the photometric properties of the subsample of MATLAS dwarfs observed with MUSE in relation to the MATLAS sample as a whole. We then examine the stellar population properties metallicity and age of this sample and compare with dwarf properties from other works. We summarize the derived properties of the dwarf sample studied in this work in Table \ref{tab:MUSE_dwarfs} in the Appendix \ref{appendix}.

\subsection{Line-of-sight velocity}

\begin{figure}[!htb]
\centering
\includegraphics[width=\linewidth]{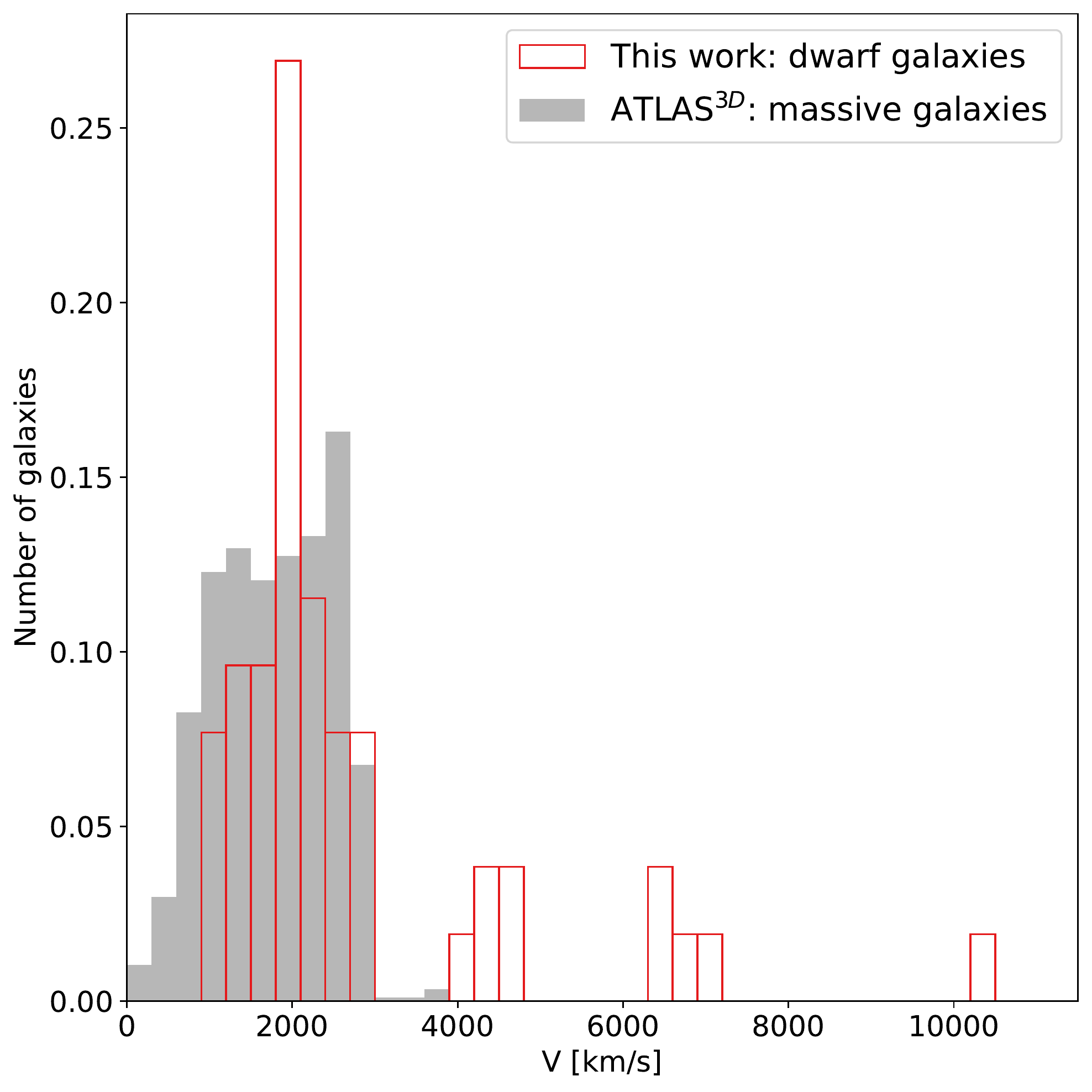}
\caption{Distribution of the measured dwarf line-of-sight velocities studied in this work. The first peak corresponds to the velocity space which is consistent with the hosts targeted in the ATLAS$^{3D}$ survey.}
\label{figure:radial_vel}
\end{figure}

We obtain recessional velocity estimates from all but four galaxies ($\sim$\,7\%) in our sample, for which the SNR is too low to identify or fit any spectral lines. The distribution of these velocities is shown in Figure \ref{figure:radial_vel}. The bulk of our sample ($\sim$\,75\%) shows velocities in an interval [1000, 3000]\,km/s. This is consistent with the distance probed by the MATLAS survey of $\sim$10\,-\,45\,Mpc with line-of-sight velocities $\in$ [-300, 3800]\,km/s. The dwarf galaxies in this velocity interval match well with the recessional velocities of nearby massive galaxies targeted in the ATLAS$^{3D}$ survey. The semi-automatic dwarf identification approach described in \citet{habas2020newly} thus shows a success rate of 75\% (79\% for the dEs) on this sample. We note that these numbers are lower limits for our dwarfs, since the four dwarfs for which we could not obtain a velocity estimation may still be satellites of nearby host galaxies and were classified as non-satellites for the sake of this calculation. We find that 10 dwarf galaxies identified in the MATLAS fields, and previously assumed satellites of the respective targeted ETG in the field, show velocities which are inconsistent with any host in the distance range probed by ATLAS$^{3D}$. These appear to lie further in the background. We note that 7 out of these 10 galaxies show strong emission lines, indicating ongoing star formation. We would expect such a correlation, since star forming objects appear brighter and are thus detectable in our fields, whereas distant quiescent ones may largely be too faint and thus elusive in this context. Another reason for this correlation is likely connected to the selection process of the MATLAS dwarfs during which a size cut was applied in order to remove background objects. Background dwarfs which remain in the sample after this cut are likely to be on the high end of the mass range (see Figure \ref{figure:mass_metal_clean}) and have a higher reservoir of gas and dust for star formation. Furthermore, star forming objects are challenging to classify and distance estimation based on visual intuition through surface brightness fluctuation is inapplicable for such objects.

\begin{figure}[!htb]
\centering
\includegraphics[width=\linewidth]{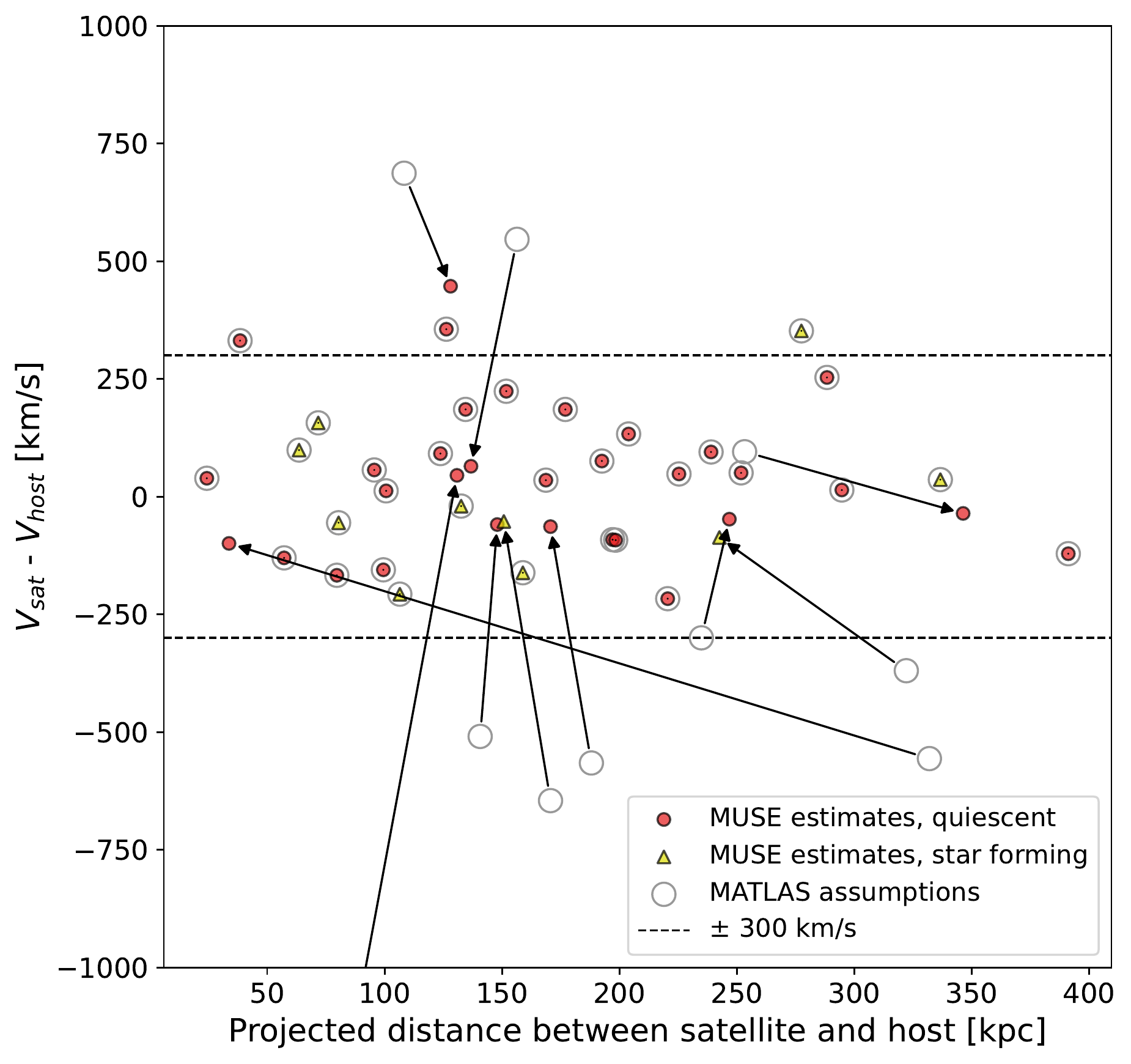}
\caption{Projected distance between satellite and assumed host galaxy in kpc vs recessional velocity difference between satellite and host ($\Delta V$ = $V_{sat}$-$V_{host}$) in km/s. We plot the projected distance and velocity difference to the host assumed in MATLAS (grey circles) compared with the updated host based on minimal $\Delta V$ through MUSE spectral fitting (red). These shifts are indicated by grey arrows. For red points circled in grey the assumed host stays the same with new velocity information on the dwarf. The dashed black lines show $\pm$\,300\,km/s.}
\label{figure:vel_proj_dist}
\end{figure}

We match the dwarf recessional velocities with the one of nearby ATLAS$^{3D}$ host galaxies and declare them satellites of the host with the smallest velocity difference $\Delta V$ = $V_{sat}$\,-\,$V_{host}$. We find a maximal $\Delta V_{max}$ = 460\,km/s and a maximal projected separation between host and satellite of $\Delta d_{max}$ = 391\,kpc. There is a clear gap in the velocity distribution shown in Figure \ref{figure:radial_vel}, between the dwarfs matching with hosts in the probed MATLAS distance range (first peak) and background dwarfs. We reassign the dwarfs from the assumed host in the MATLAS studies (targeted ETG in each field) to new hosts according to their best $\Delta V$ match and show the results of this procedure in Figure \ref{figure:vel_proj_dist}. We plot the projected distance between satellite and host in kpc on the x-axis and the difference in recessional velocities on the y-axis. The gray circles indicate the values following the MATLAS assumptions, which are shifted (gray arrows) towards the red points by assigning a better matching host galaxy with the new spectral information. For red points circled in grey the assumed MATLAS host does not change with new velocity information. We mark the $\pm$\,300\,km/s boundaries as dashed lines, which is a typical relative velocity cut for satellite populations. This is, however, not a strict cutoff as greater velocity dispersions are possible in group environments. If we instead adopt a cut of $\Delta V$ $\sim$ 500\,km/s as is done in \citet{habas2020newly}, all of our matched dwarfs lie within. 

\subsection{Photometric properties}
\label{phot_prop}

\begin{figure}[!htb]
\centering
\includegraphics[width=\linewidth]{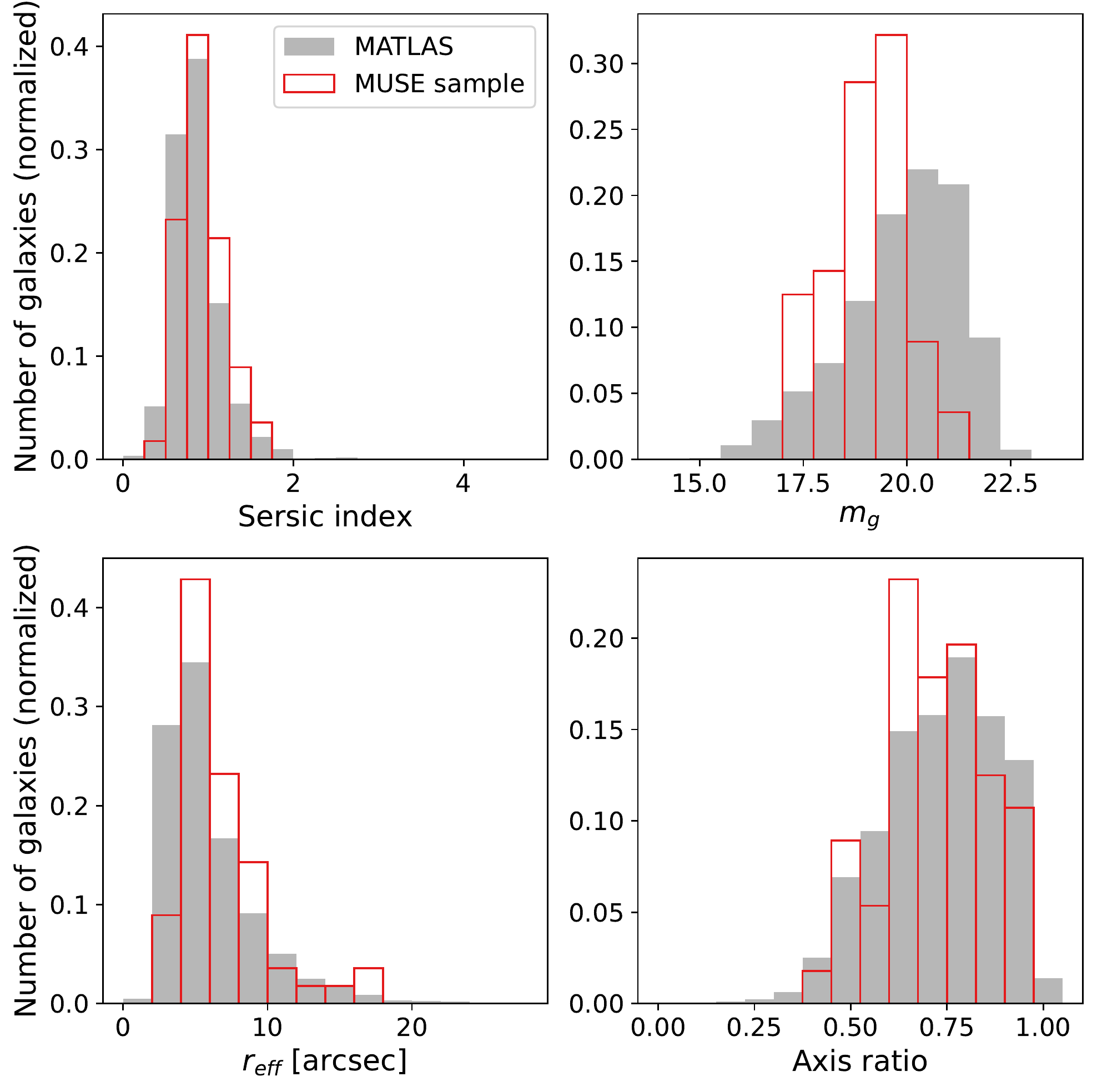}
\caption{Photometric properties of the MATLAS sub-sample targeted with MUSE (red) compared with the overall distribution of the MATLAS dwarf galaxies (grey) with robust \textsc{GALFIT} modelling. Top left: Sersic index, top right: apparent \emph{g}-band magnitude $m_{g}$, bottom left: effective radius $r_{eff}$ and bottom right: axis ratio.}
\label{figure:sample_properties}
\end{figure}

In order to illustrate whether our MATLAS subsample with MUSE observations is representative of the MATLAS sample overall, we compare their photometric properties from \textsc{galfit} \citep{peng2002detailed,peng2010detailed} modeling (see \citealp{habas2020newly} and \citealp{poulain2021structure}) in Figure \ref{figure:sample_properties}. We show the Sersic index, the \emph{g}-band apparent magnitude $m_{g}$, the effective radius $r_{eff}$ in arcsec and the axis ratio. The MATLAS sample consisting of 1589 dwarfs with successful \textsc{galfit} models and reliable $r_{eff}$ estimates is shown in gray, the subsample with MUSE observations is in red. Both samples are displayed normalized for improved visibility. The samples overlap well in general. We note a shift towards brighter magnitudes and larger effective radii for the MUSE sample. This is caused by our observational selection criteria, which are in place to ensure sufficient SNR for our main objectives in a single OB. We note that there are no Ultra Diffuse Galaxies (UDGs), i.e., dwarfs with excess effective radius in our MUSE sample.

\begin{figure}[!htb]
\centering
\includegraphics[width=\linewidth]{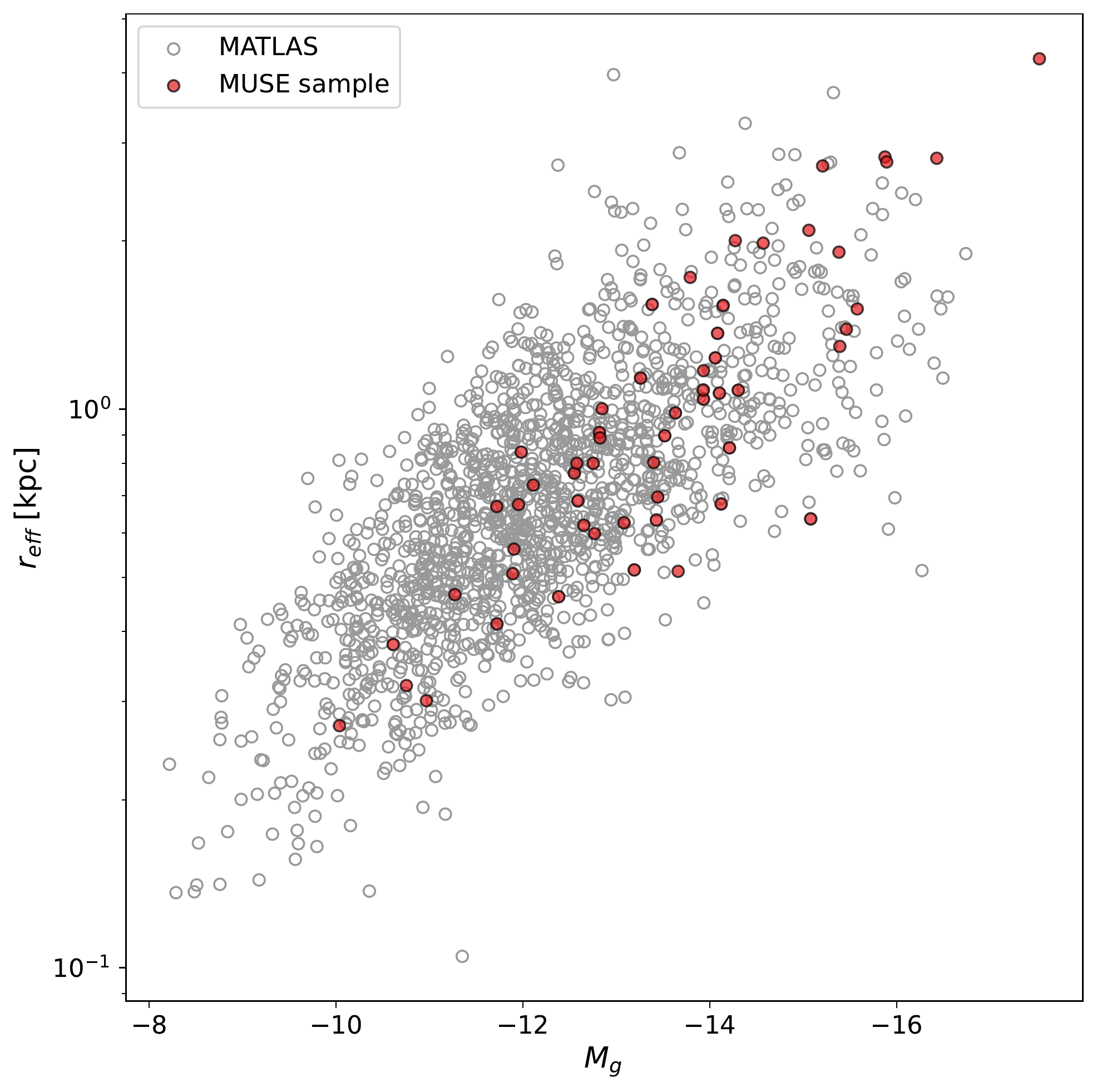}
\caption{Scaling relation absolute \emph{g}-band magnitude $M_{g}$ vs effective radius $r_{eff}$ in kpc. Comparison between the dwarf sample studied in this work (red) and the overall MATLAS sample (grey). To get these measurements for the sample with MUSE observations we use the distance of the associated host galaxy and if not available we use the dwarf velocity to estimate the distance via Hubble's law.}
\label{figure:scaling_relations}
\end{figure}

We show the scaling relation absolute magnitude $M_{g}$ vs effective radius $r_{eff}$ in Figure \ref{figure:scaling_relations} (see also \citealp{habas2020newly,poulain2021structure,2021A&A...654A.105M,2022A&A...659A..14P}). Since there are no distance or velocity estimates for the majority of the MATLAS dwarfs, $M_{g}$ was estimated by assuming the distance of central ETG of the field the dwarf appears in. A portion of the dwarf galaxies ($\approx$\,15\,\%) has a distance or velocity estimate from other surveys, in which case we use this estimate. For the MUSE sample we use the distance of the massive galaxy which best matches the MUSE estimate in terms of line-of-sight velocity. These host distances are taken from the ATLAS$^{3D}$ survey and are mostly based on redshift and surface brightness fluctuation (SBF) measurements \citep[see][]{2011MNRAS.413..813C}. For dwarfs which are not satellites of any host in the ATLAS$^{3D}$ catalog, we use the recessional velocity obtained via MUSE spectroscopy and transform it into the distance space via Hubble's law $D\,=\,V/H_{0}$. Here, $D$ denotes the distance to the dwarf, $v$ the line-of-sight velocity and $H_{0}$ the Hubble constant \citep[H$_0$ = (69.8\,$\pm$\,1.7)\,km/s/Mpc;][]{freedman2021measurements}. There is no notable difference between the subsample with MUSE observations (red) and the full MATLAS sample (gray) with \textsc{galfit} models. We therefore conclude that our MUSE sample is representative of the dwarf galaxies identified in the MATLAS survey. 

\subsection{Stellar populations}

\begin{figure*}[!htb]
\centering
\includegraphics[width=0.49\linewidth]{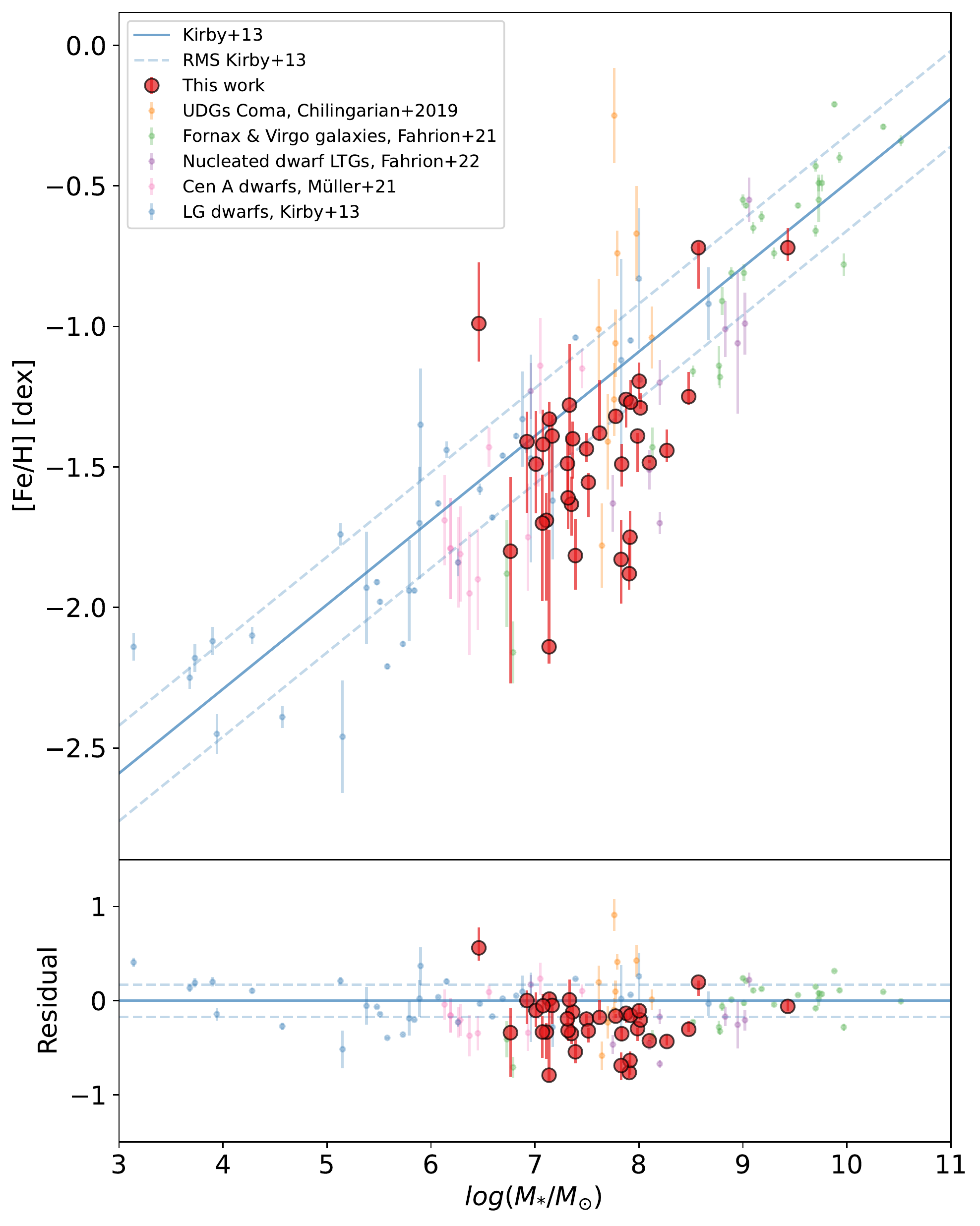}
\includegraphics[width=0.49\linewidth]{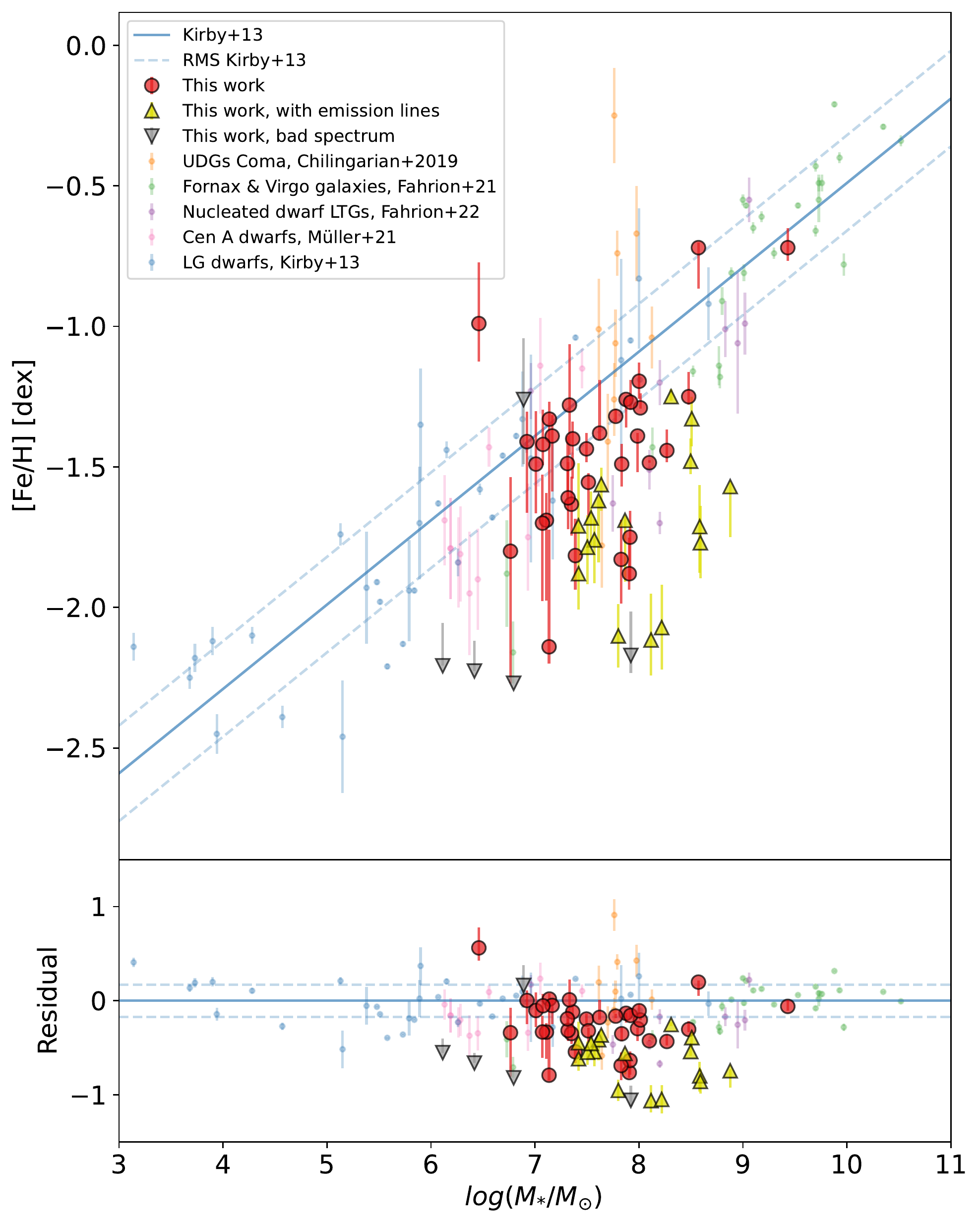}
\caption{Plotted is the universal stellar mass- metallicity relation from \citet{kirby2013universal} as the blue solid line with its rms as the two blue dashed lines. On the x-axis we plot the logarithm of the stellar mass in units of solar mass. Left: we plot the data from this work (red circles) but exclude star forming and low SNR galaxies from this sample. Right: full sample from this work. Star forming galaxies are shown as yellow upward pointing triangles, while low SNR galaxies are gray downward pointing triangles. We compare our results with galaxies from other works: LG dwarfs \citep[blue;][]{kirby2013universal}, Cen A dwarfs \citep[pink;][]{2021A&A...645A..92M}, galaxies in Fornax and Virgo \citep[green;][]{2021A&A...650A.137F}, nucleated dwarf LTGs \citep[purple;][]{2022A&A...667A.101F} and UDGs in the Coma cluster \citep[orange;][]{2019ApJ...884...79C}. Bottom: residual plots, i.e. metallicity dwarf - metallicity fit. The blue dashed lines indicate the rms of the fit from \citet{kirby2013universal}. The LG dwarf metallicities from \citet{kirby2013universal} are iron metallicities ([Fe/H]) while all other data points show total metallicities ([M/H]).}
\label{figure:mass_metal_clean}
\end{figure*}

Dwarf galaxies in the Local Universe follow the universal stellar mass-metallicity relation \citep{kirby2013universal}. In order to  relate our data to this observation, we estimate the stellar mass $M_{*}$ of our dwarf galaxies by first transforming the apparent \emph{g}-band magnitude $m_{g}$ from \citet{poulain2021structure} to the \emph{V}-band by using the transformation equation from \citet{lupton2005}:

\begin{equation}
    V = g - 0.5784*(g - r) - 0.0038
\end{equation}

and use the \emph{g}-\emph{r} colors from \citet{poulain2021structure}. For the galaxies without \emph{g}-\emph{r} values from \textsc{GALFIT}, we use the \emph{g}-\emph{r} estimates from Source Extractor on the MATLAS images by using an aperture of 3r$_{eff}$. We then transform the apparent \emph{V}-band magnitude to the absolute magnitude $M_{V}$. For this we estimate the distance to the dwarf galaxy as described in Section \ref{phot_prop}. Finally, we convert the absolute magnitude $M_{V}$ to $L_{V}$ and use the stellar mass-to-light estimate from pPXF to estimate the stellar mass $M_{*}$. 

In Figure \ref{figure:mass_metal_clean} we present the results from our sample (red) and compare it with the relation shown in \citet{kirby2013universal} and data points from other works. It should be noted that we show all LG dwarfs listed in Table 4 of \citet{kirby2013universal}. The presented fit, however, excludes the M31 dwarfs, since a different technique (coadded spectra) was used to estimate the metallicities and their uncertainties are larger when compared to the MW dwarfs. We note that our dwarf sample shows a wide range of different SNRs (see Figure \ref{figure:snr_apertures}). In order to gain robust estimates on the stellar populations high SNRs are needed \citep[see Figure A.1. in][]{2019A&A...628A..92F}. We are not able to reach high values for all galaxies but distinguish our results accordingly (see Figure \ref{figure:mass_metal_clean}, left vs. right).

On the left-hand plot in Figure \ref{figure:mass_metal_clean} we show only quiescent dwarf galaxies and exclude cases with low SNR ($\lesssim$ 8) spectra. On the right-hand plot, we show the full sample, including star forming galaxies (yellow) and galaxies with low SNR spectra (gray). As discussed in Section \ref{sec:fitting}, we mask all emission lines in star forming galaxies to determine the stellar population properties age and metallicity. This leaves only the calcium triplet (CaT) for the estimation of these properties. Since the SNR in the remaining absorption spectrum is not high enough to make robust estimates, we treat these cases separately and note reduced reliability. For all but one of the galaxies with low SNRs (gray) the metallicity estimation is on the lower end of the possible values from the considered SSPs. For these cases we cannot derive meaningful results.

We see a systematic shift to lower metallicites in our sample compared with the LG dwarf galaxies. This shift is consistent with observations in other works: \citet{2021A&A...650A.137F} (Fornax \& Virgo galaxies) and \citet{2022A&A...667A.101F} (nucleated dwarf LTGs) for this stellar mass range. As mentioned before (Section \ref{sec:intro}), however, this shift may be attributed to different strategies of measuring the metallicities between \citet{kirby2013universal}, this work and the other works cited in this study. \citet{kirby2013universal} perform spectroscopy on individual RGB stars in order to obtain the metallicity, whereas we and other studies used for comparison \citep{2021A&A...650A.137F,2022A&A...667A.101F,2021A&A...645A..92M,2019ApJ...884...79C} measure the metallicity with full spectrum fitting. We are thus sensitive to the entire stellar population properties, whereas \citet{kirby2013universal} present the mean metallicity over the RGB population in each galaxy. Another factor to consider is that the metallicities from \citet{kirby2013universal} are based on the iron absorption lines (Fe\,I), while we use lines across the entire spectral range probed by MUSE (incl. H$_{\beta}$, H$_{\alpha}$, the Mg Triplet, Fe I, Ca Triplet). The E-MILES 'base' models we are using assume that the integrated metallicity [M/H] is equal to the iron metallicity [Fe/H]. This assumption, however, only holds true at high metallicities ($\gtrsim$ -1 dex). Low metallicity stars -- abundant in our dwarf galaxies -- are alpha-enhanced which boosts [M/H] compared to [Fe/H]. Considering this assumption, the [Fe/H] as measured in \citet{kirby2013universal} should be lower than the [M/H] in our dwarf sample, which would make the discrepancy even stronger. We find that the SNR in our sample is too low to estimate [Fe/H] directly via line index measurements \citep{2010MNRAS.404.1639V} or to determine [Mg/Fe] as an additional fit parameter, which would help bridge the difference in the two measurement methods for the metallicity.

\citet{2020ApJ...896...13B} use MUSE to compare age and metallicity measurements from individual stars and full spectrum fitting for the nuclear star cluster (NSC) M54. Interestingly, they find that the two methods show excellent agreement and note only a 3\% difference in age and 0.2\,dex in metallicity. The values for full spectrum fitting suggest older and more metal-poor stellar populations when compared to the integrated spectra from individual stars. The study finds a metal-poor and a metal-rich component in the NSC. The two methods are consistent on the low-metallicity component, while full spectrum fitting returns a higher metallicity estimate for the high-metallicity component. Similar to our study, the authors discuss alpha-abundances and find that, alpha-enhancement in metal-poor stars cannot explain the difference in these measurements since the exact opposite behavior would be expected in this context.

Below the mass-metallicity plots in Figure \ref{figure:mass_metal_clean}, we show the residual, i.e., the difference between the \citet{kirby2013universal} relation and the data points. The gray dotted lines show the rms from \citet{kirby2013universal}. The systematic shift towards lower metallicities with the exception of three galaxies is even more apparent in this plot. A two-sample Kolmogorov–Smirnov (KS) test comparing the residual values for the LG dwarfs from \citet{kirby2013universal} and the residuals from the clean sample (quiescent, medium to high SNR) in this work, reveals a p-value of $p$ = 0.002. This indicates that the two samples show significant differences. While there is a significant scatter in other works (most notably \citealp{2019ApJ...884...79C}), it is interesting that all but three galaxies in our sample show on average lower metallicity values than the MW dwarfs.

\begin{figure}[!htb]
\centering
\includegraphics[width=\linewidth]{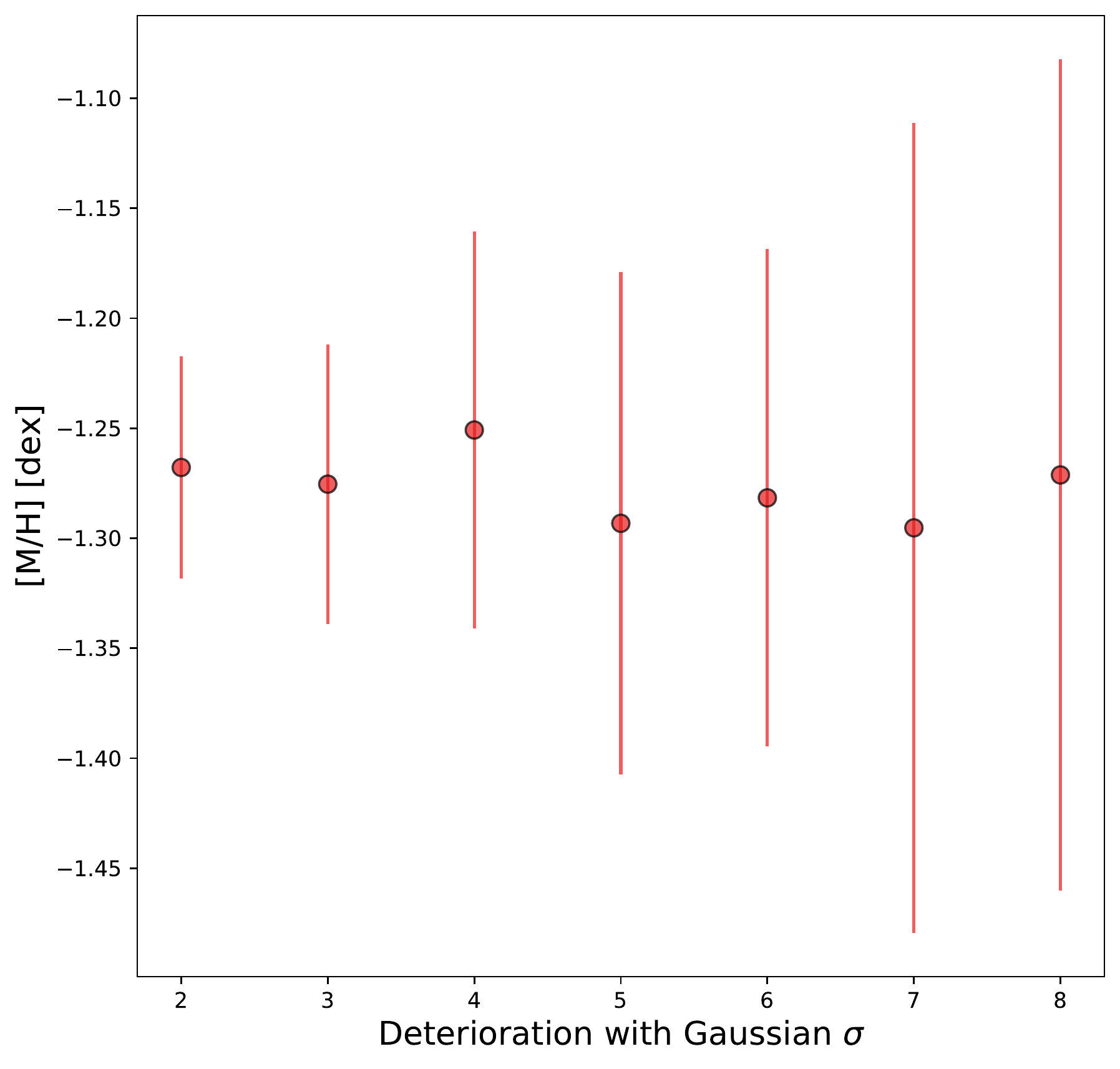}
\caption{Mean metallicity as a function of different degrees of spectrum deterioration for the highest SNR dwarf in the sample (MATLAS-553; SNR $\sim$ 62. The error bars on the y-axis show the standard deviation of 100 MC realizations for every $\sigma$. The best fit value for the metallicity is -1.29\,dex.}
\label{figure:metal_vs_sigma}
\end{figure}

In order to test whether the insufficient SNR in some of our dwarf spectra could be a contributing factor in the discrepancy between \citet{kirby2013universal} and this study, we deteriorate the quality of our highest SNR dwarf spectrum (MATLAS-553). To do this we first determine the best fit via pPXF and calculate the residuals between the best fit and the input galaxy spectrum at each wavelength. We then multiply the residuals with values from a Gaussian distribution with mean 0 and standard deviation $\sigma$ and add the products back to the best fit at every wavelength. We then run pPXF on this newly generated spectrum and note the returned metallicity. In order to increase the statistical significance of this test, we create 100 realizations for a range of $\sigma$ $\in$ [2,8] in steps of one. A higher value for $\sigma$ leads to a higher degree of SNR deterioration compared to the original spectrum. In Figure \ref{figure:metal_vs_sigma} we present the results of this test, which shows the mean metallicity from 100 realizations as a function of $\sigma$. The error bars show the standard deviation of the MC metallicity distributions. As we would expect, the errors increase as the SNR becomes lower. If the low SNR in our sample would push the metallicity estimates towards lower values, we would expect a decreasing trend in this plot. Since we note no obvious behavior in that regard, we can rule out the SNR in our sample as a leading factor in the observed offset.

Considering the results from \citet{2020ApJ...896...13B}, the offset in the quiescent medium-to-high SNR dwarfs is fully mitigated by shifting the entire sample by 0.2\,dex towards higher metallicities. A two-sample KS test yields a p-value of $p_{shift}$ = 0.8928 after this shift, suggesting the two samples are consistent with following the same distribution. If we add the star forming galaxies to this test (low SNR galaxies excluded), however, the offset is still statistically significant with a KS p-value of $p_{shift\,\,all}$ = 0.0432.

\begin{figure}[!htb]
\centering
\includegraphics[width=\linewidth]{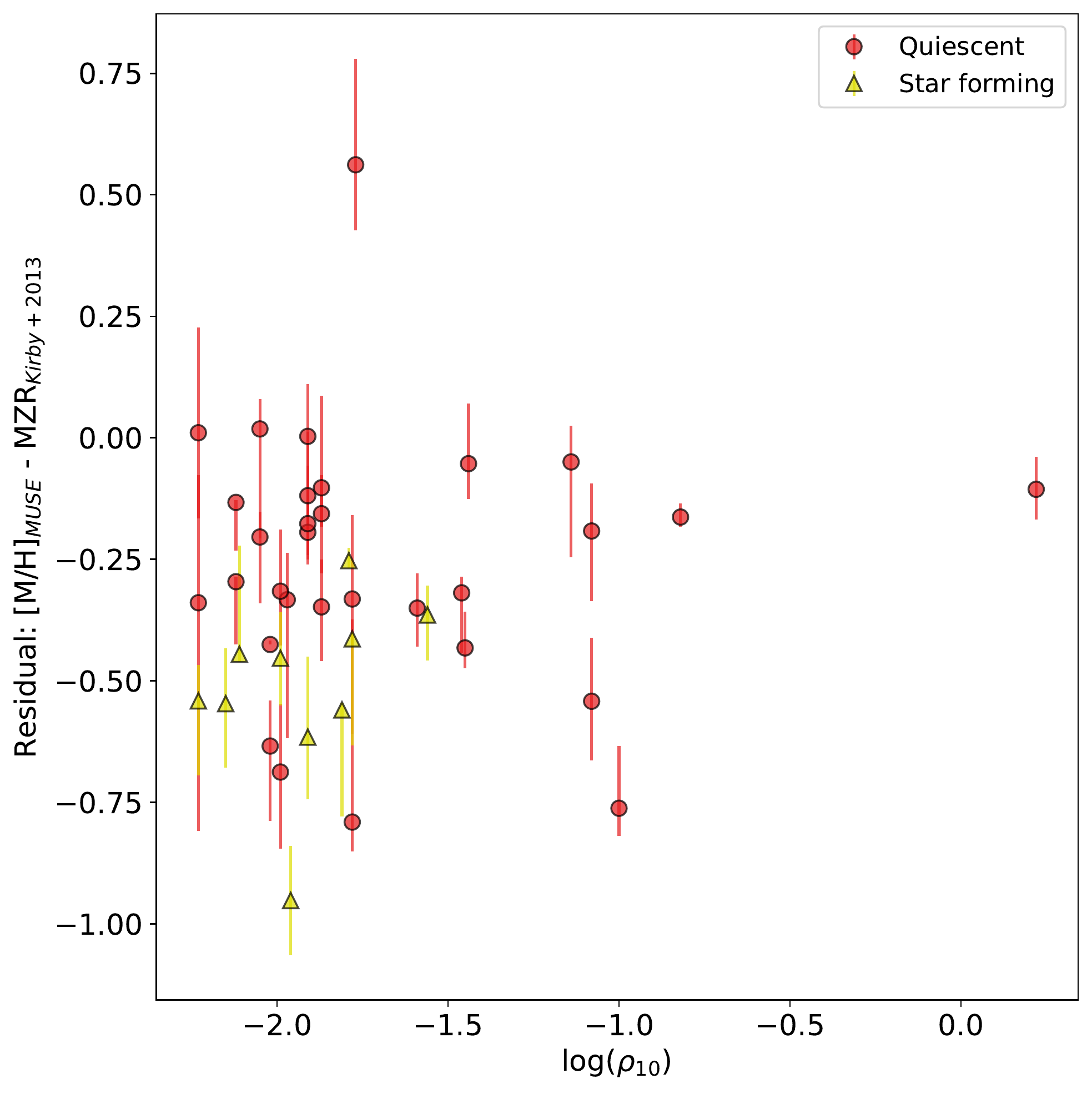}
\caption{Residuals between the metallicities for our dwarfs and the MZR from \citet{kirby2013universal} as a function of the local density parameter $\rho_{10}$.}
\label{figure:rho10_vs_residual}
\end{figure}

Another factor which could contribute to the offset in metallicities is the different environments in which the dwarf galaxies reside. We test this hypothesis by investigating the local density as described in \citet{2011MNRAS.416.1680C} as a function of the residuals between the relation presented in \citet{kirby2013universal} and the values from this study. The local density parameter

\begin{equation}
    \rho_{10} = N_{gal}/(\frac{4}{3} \pi r_{10}^3)
\end{equation}

is defined as the 10 nearest massive galaxies $N_{gal}$ of the assumed host galaxy divided by the sphere of radius $r_{10}$ which encloses these 10 neighbors. In \citet{habas2020newly} a correlation between this parameter and the morphology of the MATLAS dwarfs is found. We use the dwarf galaxies in our sample for which we could find a matching host in the ATLAS$^{3D}$ survey volume and compare its $\rho_{10}$ measure with the metallicity offset from the MZR. We show the results of this test in Figure \ref{figure:rho10_vs_residual}. Quiescent dwarfs are marked as red circles and star forming ones as yellow triangles. We note no apparent trend and can therefore not attribute the offset to the different density environments.

Finally, we test if the nucleus has any influence on the metallicity estimate and compare our results if we mask the nucleus or extract the spectra from the entire galaxy. We find very small differences between extracted metallicities with shifts in no particular direction. On average we find a difference of 0.002 dex towards lower metallicities with masked nuclei. For the dENs in our sample, the nucleus therefore does not contribute to the observed systematic offset.

\begin{figure}[!htb]
\centering
\includegraphics[width=\linewidth]{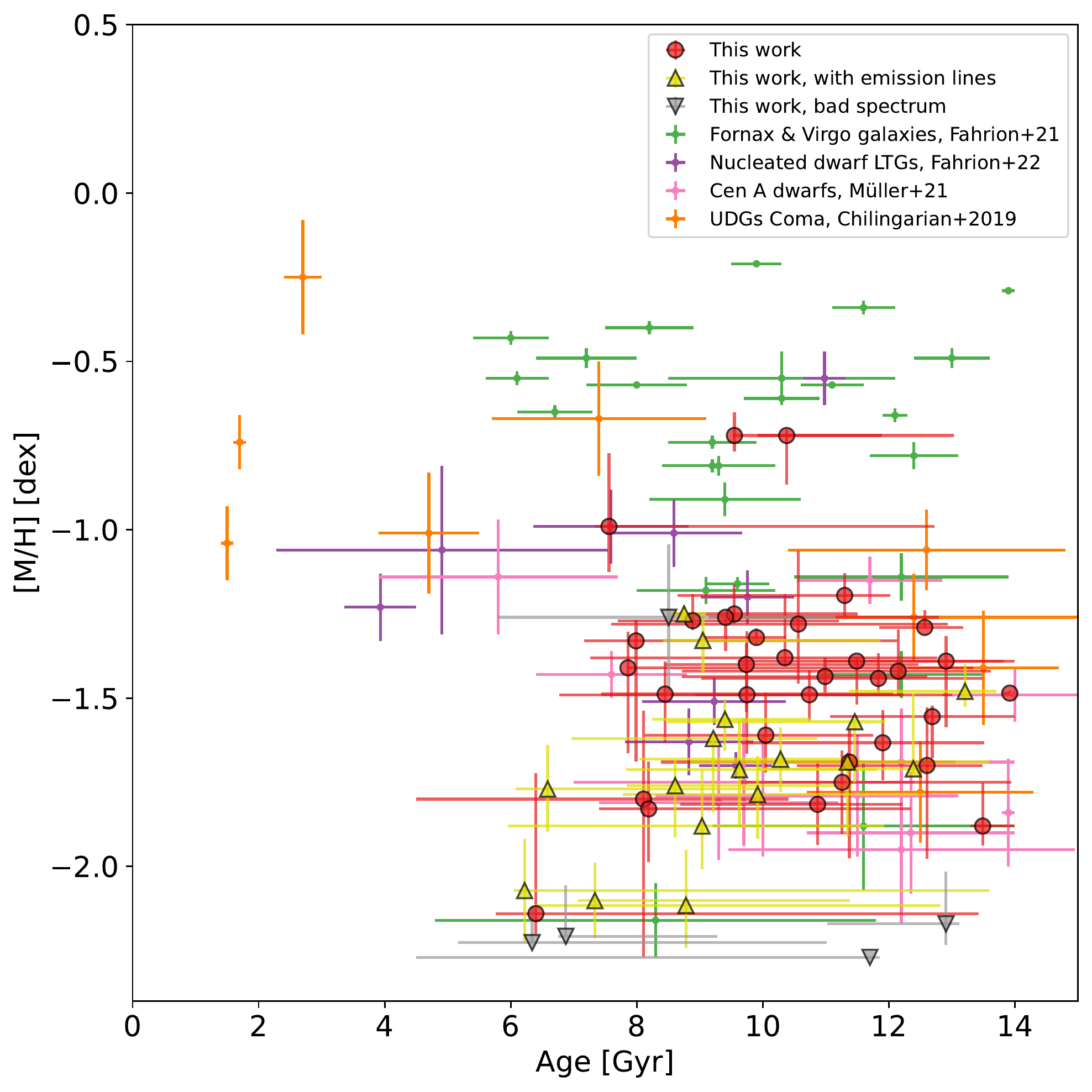}
\caption{Age vs. metallicity. The dwarf galaxies presented in this work are shown as red circles, yellow upwards pointing triangles (star forming) and gray downwards pointing triangles (low quality spectra). We compare with the results from other studies: galaxies in the Fornax and Virgo clusters \citep[green;][]{2021A&A...650A.137F}, nucleated dwarf LTGs \citep[purple;][]{2022A&A...667A.101F}, Cen A dwarfs \citep[pink;][]{2021A&A...645A..92M} and UDGs in the Coma cluster \citep[orange;][]{2019ApJ...884...79C}.}
\label{figure:age_metal}
\end{figure}

In Figure \ref{figure:age_metal} we show our results of the estimated age in Gyr versus the metallicity of our dwarf sample and compare with results from other works. It is important to note the large error bars for the age estimates, showing the difficulty of constraining this property with the quality type of spectra at hand. Overall we report an average old (6 to 14\,Gyr) and metal poor (-1.9 to -1.3 dex) stellar population for most of these dwarf galaxies with a few outliers on the upper and lower end for the metallicity. We once again compare our sample with other works: galaxies in the Fornax and Virgo clusters are shown in green \citep{2021A&A...650A.137F}, nucleated dwarf LTGs in purple \citep{2022A&A...667A.101F}, the dwarf galaxies around Centaurus A in pink \citep{2021A&A...645A..92M} and UDGs in the Coma cluster in orange \citep{2019ApJ...884...79C}. Even though the measurement errors are rather large, we can see that our sample is concentrated in the lower right corner of Figure \ref{figure:age_metal}, comparable with the Cen A dwarfs \citep{2021A&A...645A..92M}. Cluster dwarfs and nucleated dwarf LTGs appear to span a larger range of ages in comparison.



\section{Conclusions}
\label{sec:conc}

In this work, we analyze 56 dwarf galaxies based on MUSE spectroscopic observations. These galaxies are part of a large sample of dwarfs which have been previously identified in the MATLAS low-to-moderate density fields beyond the Local Volume. Through comparison with the overall photometric properties of the MATLAS dwarf sample, we find that our subsample observed with MUSE is representative of the dwarfs identified in MATLAS. Through full spectrum fitting with pPXF, we retrieve their line-of-sight velocity and stellar population properties age and metallicity. Our main results are the following:

\begin{enumerate}
  \item The bulk of the 56 dwarfs, namely 42 (75\,\%) show line-of-sight velocities which match the velocities of massive ATLAS$^{3D}$ galaxies in the field, i.e., $\sim$ 1000\,-\,3000\,km/s. This number increases to 79\,\% for dEs. Almost a third ($\sim$ 30\,\%) of the dwarf galaxies in the ATLAS$^{3D}$ velocity range show star forming activity. Based on the minimal velocity difference $\Delta V$ between massive and dwarf galaxy, we determine the satellite membership of the dwarf galaxies in our sample. We update the previously assumed association in Figure \ref{figure:vel_proj_dist} and conclude that our assumption (dwarfs are associated with the central, targeted ETG in each field) is correct in 57\,\% of the cases for the presented sample. 
  \item We find that $\sim$ 18\,\% (10) of the dwarfs in this sample are located further in the background, outside of the ATLAS$^{3D}$ survey volume. Of these, 70\,\% show emission lines, which is not unexpected in light of the semi-automatic identification approach for the MATLAS dwarfs \citep[see][]{habas2020newly}. There are no spectral lines apparent for 7\,\% (4) of the dwarfs in this sample, thus we cannot extract any velocity information.
  \item We demonstrate the viability the MUSE instrument for the study of low surface brightness dwarf galaxies in filler conditions and with a single observational block per galaxy. We determine the radius which optimizes the SNR of the extracted spectrum for each galaxy and relate it to the surface brightness in Figure \ref{figure:reff_SB}. The distribution of surface brightnesses at the optimized radius illustrates that there are significant gains in probing to a depth of $\sim$\,27\,mag arcsec$^{-2}$ but diminishing returns in terms of SNR thereafter.
  \item We find that the dwarfs presented in this work deviate from the universal stellar mass-metallicity relation shown in \citet{kirby2013universal}. The bulk of our sample is systematically offset towards lower metallicities, a property which, in this stellar mass range, can also be seen in dwarfs analyzed in other works. While the bulk of the sample with stellar masses in the range 10$^{6.5}$\,-\,10$^{8.5}$\,$M_{*}/M_{\odot}$ is more metal-poor than the LG dwarfs, only the two dwarfs with the highest stellar mass agree well with the universal mass-metallicity relation. We note that the star forming dwarfs in particular show a greater offset when compared to the quiescent ones. This could be due to the fact that only the CaT is left for this estimation, since all other lines show emissions and are therefore masked. The CaT may be insufficient to obtain a robust metallicity measurement.
  \item The overall shift towards lower metallicities at a given stellar mass cannot be attributed to insufficient SNRs in our dwarf spectra nor to the different density environments the dwarf galaxies reside in.
  \item We compare the age vs. metallicity of the dwarfs presented in this work with the results from other comparable studies in Figure \ref{figure:age_metal}. Although the error bars on the age are quite large, we can say that our sample is old (6\,-14\,Gyr) and mostly metal-poor (-1.3\,-\,1.9\,dex), which is consistent with the Centaurus A satellites \citep{2021A&A...645A..92M}. Dwarfs located in clusters are more metal-rich and span a larger range of ages \citep{2019ApJ...884...79C,2021A&A...650A.137F}.
\end{enumerate}

Our results suggest that there may exist a systematic deviation from the LG stellar mass-metallicity relation. This deviation may, however, be due to the difference in methodologies for deriving metallicities. \citet{kirby2013universal} base their results on individual RGB stars, while we and other works use full spectrum fitting. \citet{2020ApJ...896...13B} find a measurement shift of 0.2\,dex towards lower metallicities for full spectrum fitting when compared to the resolved stellar population analysis. By applying this shift towards higher metallicities, the offset is fully mitigated for the quiescent dwarfs but remains when also considering star forming dwarfs. This measurement difference is, however, based on the analysis of a single object and therefore does not hold statistical significance. In order to be able to paint a clearer picture on this matter, more methodology comparisons such as \citet{2020ApJ...896...13B} are needed. In addition, more high quality spectra of dwarfs, in particular, in the mass domain between the dwarf and massive galaxy regime will help clarify if the offset only occurs for dwarfs below a certain stellar mass threshold, since our highest mass galaxies as well as high mass galaxies from other works are consistent with the MZR. From our results it is plausible that full spectrum fitting of dwarf galaxies may lead to a steeper slope for the MZR or even a non-linear relation.

\begin{acknowledgements}
      We thank the referee for the constructive report, which helped to clarify and improve the manuscript. N.H. and O.M. are grateful to the Swiss National Science Foundation for financial support under the grant number PZ00P2\_202104. M.P. is supported by the Academy of Finland grant n:o 347089. S.L. acknowledges the support from the Sejong Science Fellowship Program by the National Research Foundation of Korea (NRF) grand funded by the Korea goverment (MSIT) (No. NRF-2021R1C1C2006790). N.H. and O.M. thank Guisseppina Battaglia for pointing out the work by Boecker et al. (2020). We thank the International Space Science Institute (ISSI) in Bern (Switzerland) for hosting our international team for a workshop on "Space Observations of Dwarf Galaxies from Deep Large Scale Surveys: The MATLAS Experience".
\end{acknowledgements}

%
%


\bibliographystyle{aa}
\bibliography{aanda}

\begin{thebibliography}{164}
\expandafter\ifx\csname natexlab\endcsname\relax\def\natexlab#1{#1}\fi

\bibitem[{Andrews \& Martini(2013)}]{andrews2013mass}
Andrews, B.~H. \& Martini, P. 2013, The Astrophysical Journal, 765, 140

\bibitem[{Bacon {et~al.}(2010)Bacon, Accardo, Adjali, Anwand, Bauer, Biswas,
  Blaizot, Boudon, Brau-Nogue, Brinchmann, {et~al.}}]{bacon2010muse}
Bacon, R., Accardo, M., Adjali, L., {et~al.} 2010, in Ground-based and Airborne
  Instrumentation for Astronomy III, Vol. 7735, SPIE, 131--139

\bibitem[{Bacon {et~al.}(2012)Bacon, Accardo, Adjali, Anwand, Bauer, Blaizot,
  Boudon, Brinchmann, Brotons, Caillier, {et~al.}}]{bacon2012news}
Bacon, R., Accardo, M., Adjali, L., {et~al.} 2012, The Messenger, 147, 4

\bibitem[{{Bell} {et~al.}(2011){Bell}, {Slater}, \&
  {Martin}}]{2011ApJ...742L..15B}
{Bell}, E.~F., {Slater}, C.~T., \& {Martin}, N.~F. 2011, \apjl, 742, L15

\bibitem[{Bennet {et~al.}(2019)Bennet, Sand, Crnojevi{\'c}, Spekkens,
  Karunakaran, Zaritsky, \& Mutlu-Pakdil}]{bennet2019m101}
Bennet, P., Sand, D., Crnojevi{\'c}, D., {et~al.} 2019, The Astrophysical
  Journal, 885, 153

\bibitem[{Bennet {et~al.}(2017)Bennet, Sand, Crnojevi{\'c}, Spekkens, Zaritsky,
  \& Karunakaran}]{bennet2017discovery}
Bennet, P., Sand, D., Crnojevi{\'c}, D., {et~al.} 2017, The Astrophysical
  Journal, 850, 109

\bibitem[{{Bennet} {et~al.}(2020){Bennet}, {Sand}, {Crnojevi{\'c}}, {Spekkens},
  {Karunakaran}, {Zaritsky}, \& {Mutlu-Pakdil}}]{2020ApJ...893L...9B}
{Bennet}, P., {Sand}, D.~J., {Crnojevi{\'c}}, D., {et~al.} 2020, \apjl, 893, L9

\bibitem[{{Bennet} {et~al.}(2017){Bennet}, {Sand}, {Crnojevi{\'c}}, {Spekkens},
  {Zaritsky}, \& {Karunakaran}}]{2017ApJ...850..109B}
{Bennet}, P., {Sand}, D.~J., {Crnojevi{\'c}}, D., {et~al.} 2017, \apj, 850, 109

\bibitem[{Bertin \& Arnouts(1996)}]{bertin1996sextractor}
Bertin, E. \& Arnouts, S. 1996, Astronomy and Astrophysics Supplement Series,
  117, 393

\bibitem[{{B{\'\i}lek} {et~al.}(2020){B{\'\i}lek}, {Duc}, {Cuillandre}, {Gwyn},
  {Cappellari}, {Bekaert}, {Bonfini}, {Bitsakis}, {Paudel}, {Krajnovi{\'c}},
  {Durrell}, \& {Marleau}}]{2020MNRAS.498.2138B}
{B{\'\i}lek}, M., {Duc}, P.-A., {Cuillandre}, J.-C., {et~al.} 2020, \mnras,
  498, 2138

\bibitem[{Binggeli \& Jerjen(1997)}]{binggeli1997photometric}
Binggeli, B. \& Jerjen, H. 1997, in Galaxy Scaling Relations: Origins,
  Evolution and Applications: Proceedings of the ESO Workshop Held at Garching,
  Germany, 18--20 November 1996, Springer, 103--112

\bibitem[{{Binggeli} {et~al.}(1990){Binggeli}, {Tarenghi}, \&
  {Sandage}}]{1990A&A...228...42B}
{Binggeli}, B., {Tarenghi}, M., \& {Sandage}, A. 1990, \aap, 228, 42

\bibitem[{{Boecker} {et~al.}(2020){Boecker}, {Alfaro-Cuello}, {Neumayer},
  {Mart{\'\i}n-Navarro}, \& {Leaman}}]{2020ApJ...896...13B}
{Boecker}, A., {Alfaro-Cuello}, M., {Neumayer}, N., {Mart{\'\i}n-Navarro}, I.,
  \& {Leaman}, R. 2020, \apj, 896, 13

\bibitem[{Brook {et~al.}(2014)Brook, Stinson, Gibson, Shen, Maccio, Obreja,
  Wadsley, \& Quinn}]{brook2014magicc}
Brook, C., Stinson, G., Gibson, B.~K., {et~al.} 2014, Monthly Notices of the
  Royal Astronomical Society, 443, 3809

\bibitem[{Brooks {et~al.}(2007)Brooks, Governato, Booth, Willman, Gardner,
  Wadsley, Stinson, \& Quinn}]{brooks2007origin}
Brooks, A.~M., Governato, F., Booth, C., {et~al.} 2007, The Astrophysical
  Journal, 655, L17

\bibitem[{Bullock \& Boylan-Kolchin(2017)}]{bullock2017small}
Bullock, J.~S. \& Boylan-Kolchin, M. 2017, Annual Review of Astronomy and
  Astrophysics, 55

\bibitem[{Buonanno {et~al.}(1985)Buonanno, Corsi, Fusi~Pecci, Hardy, \&
  Zinn}]{buonanno1985color}
Buonanno, R., Corsi, C., Fusi~Pecci, F., Hardy, E., \& Zinn, R. 1985, Astronomy
  and Astrophysics (ISSN 0004-6361), vol. 152, no. 1, Nov. 1985, p. 65-84.
  NSERC-supported research., 152, 65

\bibitem[{{Cappellari}(2017)}]{2017MNRAS.466..798C}
{Cappellari}, M. 2017, \mnras, 466, 798

\bibitem[{Cappellari {et~al.}(2006)Cappellari, Bacon, Bureau, Damen, Davies,
  De~Zeeuw, Emsellem, Falc{\'o}n-Barroso, Krajnovic, Kuntschner,
  {et~al.}}]{cappellari2006sauron}
Cappellari, M., Bacon, R., Bureau, M., {et~al.} 2006, Monthly Notices of the
  Royal Astronomical Society, 366, 1126

\bibitem[{{Cappellari} \& {Emsellem}(2004)}]{2004PASP..116..138C}
{Cappellari}, M. \& {Emsellem}, E. 2004, \pasp, 116, 138

\bibitem[{Cappellari {et~al.}(2011)Cappellari, Emsellem, Krajnovi{\'c},
  McDermid, Scott, Verdoes~Kleijn, Young, Alatalo, Bacon, Blitz,
  {et~al.}}]{cappellari2011atlas3d}
Cappellari, M., Emsellem, E., Krajnovi{\'c}, D., {et~al.} 2011, Monthly Notices
  of the Royal Astronomical Society, 413, 813

\bibitem[{{Cappellari} {et~al.}(2011{\natexlab{a}}){Cappellari}, {Emsellem},
  {Krajnovi{\'c}}, {McDermid}, {Scott}, {Verdoes Kleijn}, {Young}, {Alatalo},
  {Bacon}, {Blitz}, {Bois}, {Bournaud}, {Bureau}, {Davies}, {Davis}, {de
  Zeeuw}, {Duc}, {Khochfar}, {Kuntschner}, {Lablanche}, {Morganti}, {Naab},
  {Oosterloo}, {Sarzi}, {Serra}, \& {Weijmans}}]{2011MNRAS.413..813C}
{Cappellari}, M., {Emsellem}, E., {Krajnovi{\'c}}, D., {et~al.}
  2011{\natexlab{a}}, \mnras, 413, 813

\bibitem[{{Cappellari} {et~al.}(2011{\natexlab{b}}){Cappellari}, {Emsellem},
  {Krajnovi{\'c}}, {McDermid}, {Serra}, {Alatalo}, {Blitz}, {Bois}, {Bournaud},
  {Bureau}, {Davies}, {Davis}, {de Zeeuw}, {Khochfar}, {Kuntschner},
  {Lablanche}, {Morganti}, {Naab}, {Oosterloo}, {Sarzi}, {Scott}, {Weijmans},
  \& {Young}}]{2011MNRAS.416.1680C}
{Cappellari}, M., {Emsellem}, E., {Krajnovi{\'c}}, D., {et~al.}
  2011{\natexlab{b}}, \mnras, 416, 1680

\bibitem[{{Carlsten} {et~al.}(2019){Carlsten}, {Beaton}, {Greco}, \&
  {Greene}}]{2019ApJ...878L..16C}
{Carlsten}, S.~G., {Beaton}, R.~L., {Greco}, J.~P., \& {Greene}, J.~E. 2019,
  \apjl, 878, L16

\bibitem[{{Carlsten} {et~al.}(2022){Carlsten}, {Greene}, {Beaton}, {Danieli},
  \& {Greco}}]{2022ApJ...933...47C}
{Carlsten}, S.~G., {Greene}, J.~E., {Beaton}, R.~L., {Danieli}, S., \& {Greco},
  J.~P. 2022, \apj, 933, 47

\bibitem[{{Chiboucas} {et~al.}(2013){Chiboucas}, {Jacobs}, {Tully}, \&
  {Karachentsev}}]{2013AJ....146..126C}
{Chiboucas}, K., {Jacobs}, B.~A., {Tully}, R.~B., \& {Karachentsev}, I.~D.
  2013, \aj, 146, 126

\bibitem[{{Chilingarian} {et~al.}(2019){Chilingarian}, {Afanasiev}, {Grishin},
  {Fabricant}, \& {Moran}}]{2019ApJ...884...79C}
{Chilingarian}, I.~V., {Afanasiev}, A.~V., {Grishin}, K.~A., {Fabricant}, D.,
  \& {Moran}, S. 2019, \apj, 884, 79

\bibitem[{Chiosi {et~al.}(2020)Chiosi, D’Onofrio, Merlin, Piovan, \&
  Marziani}]{chiosi2020parallelism}
Chiosi, C., D’Onofrio, M., Merlin, E., Piovan, L., \& Marziani, P. 2020,
  Astronomy \& Astrophysics, 643, A136

\bibitem[{{Cohen} {et~al.}(2018){Cohen}, {van Dokkum}, {Danieli}, {Romanowsky},
  {Abraham}, {Merritt}, {Zhang}, {Mowla}, {Kruijssen}, {Conroy}, \&
  {Wasserman}}]{2018ApJ...868...96C}
{Cohen}, Y., {van Dokkum}, P., {Danieli}, S., {et~al.} 2018, \apj, 868, 96

\bibitem[{Courteau {et~al.}(2007)Courteau, Dutton, van~den Bosch, MacArthur,
  Dekel, McIntosh, \& Dale}]{courteau2007scaling}
Courteau, S., Dutton, A.~A., van~den Bosch, F.~C., {et~al.} 2007, Scaling
  Relations of Spiral Galaxies., 671 (1): 203--225

\bibitem[{Cowie {et~al.}(1996)Cowie, Songaila, Hu, \& Cohen}]{cowie1996new}
Cowie, L.~L., Songaila, A., Hu, E.~M., \& Cohen, J. 1996, arXiv preprint
  astro-ph/9606079

\bibitem[{{Crnojevi{\'c}} {et~al.}(2019){Crnojevi{\'c}}, {Sand}, {Bennet},
  {Pasetto}, {Spekkens}, {Caldwell}, {Guhathakurta}, {McLeod}, {Seth}, {Simon},
  {Strader}, \& {Toloba}}]{2019ApJ...872...80C}
{Crnojevi{\'c}}, D., {Sand}, D.~J., {Bennet}, P., {et~al.} 2019, \apj, 872, 80

\bibitem[{{Danieli} {et~al.}(2017){Danieli}, {van Dokkum}, {Merritt},
  {Abraham}, {Zhang}, {Karachentsev}, \& {Makarova}}]{2017ApJ...837..136D}
{Danieli}, S., {van Dokkum}, P., {Merritt}, A., {et~al.} 2017, \apj, 837, 136

\bibitem[{{Danieli} {et~al.}(2022){Danieli}, {van Dokkum}, {Trujillo-Gomez},
  {Kruijssen}, {Romanowsky}, {Carlsten}, {Shen}, {Li}, {Abraham}, {Brodie},
  {Conroy}, {Gannon}, \& {Greco}}]{2022ApJ...927L..28D}
{Danieli}, S., {van Dokkum}, P., {Trujillo-Gomez}, S., {et~al.} 2022, \apjl,
  927, L28

\bibitem[{{Davis} {et~al.}(2021){Davis}, {Nierenberg}, {Peter}, {Garling},
  {Greco}, {Kochanek}, {Utomo}, {Casey}, {Pogge}, {Roberts}, {Sand}, \&
  {Sardone}}]{2021MNRAS.500.3854D}
{Davis}, A.~B., {Nierenberg}, A.~M., {Peter}, A. H.~G., {et~al.} 2021, \mnras,
  500, 3854

\bibitem[{{Djorgovski} \& {Davis}(1987)}]{1987ApJ...313...59D}
{Djorgovski}, S. \& {Davis}, M. 1987, \apj, 313, 59

\bibitem[{{Dressler} {et~al.}(1987){Dressler}, {Lynden-Bell}, {Burstein},
  {Davies}, {Faber}, {Terlevich}, \& {Wegner}}]{1987ApJ...313...42D}
{Dressler}, A., {Lynden-Bell}, D., {Burstein}, D., {et~al.} 1987, \apj, 313, 42

\bibitem[{{Drlica-Wagner} {et~al.}(2020){Drlica-Wagner}, {Bechtol}, {Mau},
  {McNanna}, {Nadler}, {Pace}, {Li}, {Pieres}, {Rozo}, {Simon}, {Walker},
  {Wechsler}, {Abbott}, {Allam}, {Annis}, {Bertin}, {Brooks}, {Burke},
  {Rosell}, {Carrasco Kind}, {Carretero}, {Costanzi}, {da Costa}, {De Vicente},
  {Desai}, {Diehl}, {Doel}, {Eifler}, {Everett}, {Flaugher}, {Frieman},
  {Garc{\'\i}a-Bellido}, {Gaztanaga}, {Gruen}, {Gruendl}, {Gschwend},
  {Gutierrez}, {Honscheid}, {James}, {Krause}, {Kuehn}, {Kuropatkin}, {Lahav},
  {Maia}, {Marshall}, {Melchior}, {Menanteau}, {Miquel}, {Palmese}, {Plazas},
  {Sanchez}, {Scarpine}, {Schubnell}, {Serrano}, {Sevilla-Noarbe}, {Smith},
  {Suchyta}, {Tarle}, \& {DES Collaboration}}]{2020ApJ...893...47D}
{Drlica-Wagner}, A., {Bechtol}, K., {Mau}, S., {et~al.} 2020, \apj, 893, 47

\bibitem[{{Drlica-Wagner} {et~al.}(2021){Drlica-Wagner}, {Carlin}, {Nidever},
  {Ferguson}, {Kuropatkin}, {Adam{\'o}w}, {Cerny}, {Choi}, {Esteves},
  {Mart{\'\i}nez-V{\'a}zquez}, {Mau}, {Miller}, {Mutlu-Pakdil}, {Neilsen},
  {Olsen}, {Pace}, {Riley}, {Sakowska}, {Sand}, {Santana-Silva}, {Tollerud},
  {Tucker}, {Vivas}, {Zaborowski}, {Zenteno}, {Abbott}, {Allam}, {Bechtol},
  {Bell}, {Bell}, {Bilaji}, {Bom}, {Carballo-Bello}, {Crnojevi{\'c}}, {Cioni},
  {Diaz-Ocampo}, {de Boer}, {Erkal}, {Gruendl}, {Hernandez-Lang}, {Hughes},
  {James}, {Johnson}, {Li}, {Mao}, {Mart{\'\i}nez-Delgado}, {Massana},
  {McNanna}, {Morgan}, {Nadler}, {No{\"e}l}, {Palmese}, {Peter}, {Rykoff},
  {S{\'a}nchez}, {Shipp}, {Simon}, {Smercina}, {Soares-Santos}, {Stringfellow},
  {Tavangar}, {van der Marel}, {Walker}, {Wechsler}, {Wu}, {Yanny},
  {Fitzpatrick}, {Huang}, {Jacques}, {Nikutta}, {Scott}, \& {Astro Data
  Lab}}]{2021ApJS..256....2D}
{Drlica-Wagner}, A., {Carlin}, J.~L., {Nidever}, D.~L., {et~al.} 2021, \apjs,
  256, 2

\bibitem[{Duc {et~al.}(2015)Duc, Cuillandre, Karabal, Cappellari, Alatalo,
  Blitz, Bournaud, Bureau, Crocker, Davies, {et~al.}}]{duc2015atlas3d}
Duc, P.-A., Cuillandre, J.-C., Karabal, E., {et~al.} 2015, Monthly Notices of
  the Royal Astronomical Society, 446, 120

\bibitem[{D’Onofrio {et~al.}(2021)D’Onofrio, Marziani, \&
  Chiosi}]{d2021past}
D’Onofrio, M., Marziani, P., \& Chiosi, C. 2021, Frontiers in Astronomy and
  Space Sciences, 157

\bibitem[{Eigenthaler {et~al.}(2018)Eigenthaler, Puzia, Taylor,
  Ordenes-Briceno, Munoz, Ribbeck, Alamo-Martinez, Zhang, {\'A}ngel,
  Capaccioli, {et~al.}}]{eigenthaler2018next}
Eigenthaler, P., Puzia, T.~H., Taylor, M.~A., {et~al.} 2018, The Astrophysical
  Journal, 855, 142

\bibitem[{{Emsellem} {et~al.}(2019){Emsellem}, {van der Burg}, {Fensch},
  {Je{\v{r}}{\'a}bkov{\'a}}, {Zanella}, {Agnello}, {Hilker}, {M{\"u}ller},
  {Rejkuba}, {Duc}, {Durrell}, {Habas}, {Lelli}, {Lim}, {Marleau}, {Peng}, \&
  {S{\'a}nchez-Janssen}}]{2019A&A...625A..76E}
{Emsellem}, E., {van der Burg}, R. F.~J., {Fensch}, J., {et~al.} 2019, \aap,
  625, A76

\bibitem[{Erb {et~al.}(2006)Erb, Steidel, Shapley, Pettini, Reddy, \&
  Adelberger}]{erb2006stellar}
Erb, D.~K., Steidel, C.~C., Shapley, A.~E., {et~al.} 2006, The Astrophysical
  Journal, 646, 107

\bibitem[{Faber \& Jackson(1976)}]{faber1976velocity}
Faber, S. \& Jackson, R.~E. 1976, The Astrophysical Journal, 204, 668

\bibitem[{{Fahrion} {et~al.}(2022){Fahrion}, {Bulichi}, {Hilker}, {Leaman},
  {Lyubenova}, {M{\"u}ller}, {Neumayer}, {Pinna}, {Rejkuba}, \& {van de
  Ven}}]{2022A&A...667A.101F}
{Fahrion}, K., {Bulichi}, T.-E., {Hilker}, M., {et~al.} 2022, \aap, 667, A101

\bibitem[{{Fahrion} {et~al.}(2019{\natexlab{a}}){Fahrion}, {Georgiev},
  {Hilker}, {Lyubenova}, {van de Ven}, {Alfaro-Cuello}, {Corsini}, {Sarzi},
  {McDermid}, \& {de Zeeuw}}]{2019A&A...625A..50F}
{Fahrion}, K., {Georgiev}, I., {Hilker}, M., {et~al.} 2019{\natexlab{a}}, \aap,
  625, A50

\bibitem[{{Fahrion} {et~al.}(2021){Fahrion}, {Lyubenova}, {van de Ven},
  {Hilker}, {Leaman}, {Falc{\'o}n-Barroso}, {Bittner}, {Coccato}, {Corsini},
  {Gadotti}, {Iodice}, {McDermid}, {Mart{\'\i}n-Navarro}, {Pinna}, {Poci},
  {Sarzi}, {de Zeeuw}, \& {Zhu}}]{2021A&A...650A.137F}
{Fahrion}, K., {Lyubenova}, M., {van de Ven}, G., {et~al.} 2021, \aap, 650,
  A137

\bibitem[{{Fahrion} {et~al.}(2019{\natexlab{b}}){Fahrion}, {Lyubenova}, {van de
  Ven}, {Leaman}, {Hilker}, {Mart{\'\i}n-Navarro}, {Zhu}, {Alfaro-Cuello},
  {Coccato}, {Corsini}, {Falc{\'o}n-Barroso}, {Iodice}, {McDermid}, {Sarzi}, \&
  {de Zeeuw}}]{2019A&A...628A..92F}
{Fahrion}, K., {Lyubenova}, M., {van de Ven}, G., {et~al.} 2019{\natexlab{b}},
  \aap, 628, A92

\bibitem[{Fahrion {et~al.}(2020)Fahrion, M{\"u}ller, Rejkuba, Hilker,
  Lyubenova, van~de Ven, Georgiev, Lelli, Pawlowski, \&
  Jerjen}]{fahrion2020metal}
Fahrion, K., M{\"u}ller, O., Rejkuba, M., {et~al.} 2020, Astronomy \&
  Astrophysics, 634, A53

\bibitem[{{Fensch} {et~al.}(2019){Fensch}, {van der Burg},
  {Je{\v{r}}{\'a}bkov{\'a}}, {Emsellem}, {Zanella}, {Agnello}, {Hilker},
  {M{\"u}ller}, {Rejkuba}, {Duc}, {Durrell}, {Habas}, {Lim}, {Marleau}, {Peng},
  \& {S{\'a}nchez Janssen}}]{2019A&A...625A..77F}
{Fensch}, J., {van der Burg}, R. F.~J., {Je{\v{r}}{\'a}bkov{\'a}}, T., {et~al.}
  2019, \aap, 625, A77

\bibitem[{Ferguson {et~al.}(1990)Ferguson, Sandage, Hollenbach, \&
  AJ}]{ferguson1990nasa}
Ferguson, H., Sandage, A., Hollenbach, D., \& AJ, T.~H. 1990

\bibitem[{{Ferguson} \& {Binggeli}(1994)}]{1994A&ARv...6...67F}
{Ferguson}, H.~C. \& {Binggeli}, B. 1994, \aapr, 6, 67

\bibitem[{Ferguson \& Sandage(1989)}]{ferguson1989spatial}
Ferguson, H.~C. \& Sandage, A. 1989, The Astrophysical Journal, 346, L53

\bibitem[{Ferrarese {et~al.}(2012)Ferrarese, Cote, Cuillandre, Gwyn, Peng,
  MacArthur, Duc, Boselli, Mei, Erben, {et~al.}}]{ferrarese2012next}
Ferrarese, L., Cote, P., Cuillandre, J.-C., {et~al.} 2012, The Astrophysical
  Journal Supplement Series, 200, 4

\bibitem[{Fitzpatrick \& Graves(2015)}]{fitzpatrick2015early}
Fitzpatrick, P.~J. \& Graves, G.~J. 2015, Monthly Notices of the Royal
  Astronomical Society, 447, 1383

\bibitem[{{Font} {et~al.}(2011){Font}, {Benson}, {Bower}, {Frenk}, {Cooper},
  {De Lucia}, {Helly}, {Helmi}, {Li}, {McCarthy}, {Navarro}, {Springel},
  {Starkenburg}, {Wang}, \& {White}}]{2011MNRAS.417.1260F}
{Font}, A.~S., {Benson}, A.~J., {Bower}, R.~G., {et~al.} 2011, \mnras, 417,
  1260

\bibitem[{Freedman(2021)}]{freedman2021measurements}
Freedman, W.~L. 2021, The Astrophysical Journal, 919, 16

\bibitem[{{Frenk} \& {White}(2012)}]{2012AnP...524..507F}
{Frenk}, C.~S. \& {White}, S.~D.~M. 2012, Annalen der Physik, 524, 507

\bibitem[{{Gallazzi} {et~al.}(2006){Gallazzi}, {Charlot}, {Brinchmann}, \&
  {White}}]{2006MNRAS.370.1106G}
{Gallazzi}, A., {Charlot}, S., {Brinchmann}, J., \& {White}, S. D.~M. 2006,
  \mnras, 370, 1106

\bibitem[{{Gallazzi} {et~al.}(2005){Gallazzi}, {Charlot}, {Brinchmann},
  {White}, \& {Tremonti}}]{2005MNRAS.362...41G}
{Gallazzi}, A., {Charlot}, S., {Brinchmann}, J., {White}, S. D.~M., \&
  {Tremonti}, C.~A. 2005, \mnras, 362, 41

\bibitem[{Garnett(2002)}]{garnett2002luminosity}
Garnett, D.~R. 2002, The Astrophysical Journal, 581, 1019

\bibitem[{{Geha} {et~al.}(2017){Geha}, {Wechsler}, {Mao}, {Tollerud}, {Weiner},
  {Bernstein}, {Hoyle}, {Marchi}, {Marshall}, \&
  {Mu{\~n}oz}}]{2017ApJ...847....4G}
{Geha}, M., {Wechsler}, R.~H., {Mao}, Y.-Y., {et~al.} 2017, \apj, 847, 4

\bibitem[{{Gonz{\'a}lez Delgado} {et~al.}(2014){Gonz{\'a}lez Delgado}, {Cid
  Fernandes}, {Garc{\'\i}a-Benito}, {P{\'e}rez}, {de Amorim},
  {Cortijo-Ferrero}, {Lacerda}, {L{\'o}pez Fern{\'a}ndez}, {S{\'a}nchez}, {Vale
  Asari}, {Alves}, {Bland-Hawthorn}, {Galbany}, {Gallazzi}, {Husemann},
  {Bekeraite}, {Jungwiert}, {L{\'o}pez-S{\'a}nchez}, {de Lorenzo-C{\'a}ceres},
  {Marino}, {Mast}, {Moll{\'a}}, {del Olmo}, {S{\'a}nchez-Bl{\'a}zquez}, {van
  de Ven}, {V{\'\i}lchez}, {Walcher}, {Wisotzki}, {Ziegler}, \& {CALIFA
  Collaboration}}]{2014ApJ...791L..16G}
{Gonz{\'a}lez Delgado}, R.~M., {Cid Fernandes}, R., {Garc{\'\i}a-Benito}, R.,
  {et~al.} 2014, \apjl, 791, L16

\bibitem[{{Greco} {et~al.}(2018){Greco}, {Greene}, {Strauss}, {Macarthur},
  {Flowers}, {Goulding}, {Huang}, {Kim}, {Komiyama}, {Leauthaud}, {Leisman},
  {Lupton}, {Sif{\'o}n}, \& {Wang}}]{2018ApJ...857..104G}
{Greco}, J.~P., {Greene}, J.~E., {Strauss}, M.~A., {et~al.} 2018, \apj, 857,
  104

\bibitem[{{Gu{\'e}rou} {et~al.}(2017){Gu{\'e}rou}, {Krajnovi{\'c}}, {Epinat},
  {Contini}, {Emsellem}, {Bouch{\'e}}, {Bacon}, {Michel-Dansac}, {Richard},
  {Weilbacher}, {Schaye}, {Marino}, {den Brok}, \&
  {Erroz-Ferrer}}]{2017A&A...608A...5G}
{Gu{\'e}rou}, A., {Krajnovi{\'c}}, D., {Epinat}, B., {et~al.} 2017, \aap, 608,
  A5

\bibitem[{Habas {et~al.}(2020)Habas, Marleau, Duc, Durrell, Paudel, Poulain,
  S{\'a}nchez-Janssen, Sreejith, Ramasawmy, Stemock, {et~al.}}]{habas2020newly}
Habas, R., Marleau, F.~R., Duc, P.-A., {et~al.} 2020, Monthly Notices of the
  Royal Astronomical Society, 491, 1901

\bibitem[{{Heesters} {et~al.}(2021){Heesters}, {Habas}, {Marleau},
  {M{\"u}ller}, {Duc}, {Poulain}, {Durrell}, {S{\'a}nchez-Janssen}, \&
  {Paudel}}]{2021A&A...654A.161H}
{Heesters}, N., {Habas}, R., {Marleau}, F.~R., {et~al.} 2021, \aap, 654, A161

\bibitem[{{Hou} {et~al.}(2014){Hou}, {Yu}, \& {Lu}}]{2014ApJ...791....8H}
{Hou}, J., {Yu}, Q., \& {Lu}, Y. 2014, \apj, 791, 8

\bibitem[{{Ibata} {et~al.}(2007){Ibata}, {Martin}, {Irwin}, {Chapman},
  {Ferguson}, {Lewis}, \& {McConnachie}}]{2007ApJ...671.1591I}
{Ibata}, R., {Martin}, N.~F., {Irwin}, M., {et~al.} 2007, \apj, 671, 1591

\bibitem[{Ibata {et~al.}(2013)Ibata, Lewis, Conn, Irwin, McConnachie, Chapman,
  Collins, Fardal, Ferguson, Ibata, {et~al.}}]{ibata2013vast}
Ibata, R.~A., Lewis, G.~F., Conn, A.~R., {et~al.} 2013, Nature, 493, 62

\bibitem[{{Ibata} {et~al.}(2014){Ibata}, {Lewis}, {McConnachie}, {Martin},
  {Irwin}, {Ferguson}, {Babul}, {Bernard}, {Chapman}, {Collins}, {Fardal},
  {Mackey}, {Navarro}, {Pe{\~n}arrubia}, {Rich}, {Tanvir}, \&
  {Widrow}}]{2014ApJ...780..128I}
{Ibata}, R.~A., {Lewis}, G.~F., {McConnachie}, A.~W., {et~al.} 2014, \apj, 780,
  128

\bibitem[{{Irwin} {et~al.}(2009){Irwin}, {Hoffman}, {Spekkens}, {Haynes},
  {Giovanelli}, {Linder}, {Catinella}, {Momjian}, {Koribalski}, {Davies},
  {Brinks}, {de Blok}, {Putman}, \& {van Driel}}]{2009ApJ...692.1447I}
{Irwin}, J.~A., {Hoffman}, G.~L., {Spekkens}, K., {et~al.} 2009, \apj, 692,
  1447

\bibitem[{Kewley \& Ellison(2008)}]{kewley2008metallicity}
Kewley, L.~J. \& Ellison, S.~L. 2008, The Astrophysical Journal, 681, 1183

\bibitem[{{Kim} {et~al.}(2011){Kim}, {Kim}, {Hwang}, {Lee}, {Chun}, \&
  {Ann}}]{2011MNRAS.412.1881K}
{Kim}, E., {Kim}, M., {Hwang}, N., {et~al.} 2011, \mnras, 412, 1881

\bibitem[{{Kim} {et~al.}(2018){Kim}, {Peter}, \&
  {Hargis}}]{2018PhRvL.121u1302K}
{Kim}, S.~Y., {Peter}, A. H.~G., \& {Hargis}, J.~R. 2018, \prl, 121, 211302

\bibitem[{Kirby {et~al.}(2013)Kirby, Cohen, Guhathakurta, Cheng, Bullock, \&
  Gallazzi}]{kirby2013universal}
Kirby, E.~N., Cohen, J.~G., Guhathakurta, P., {et~al.} 2013, The Astrophysical
  Journal, 779, 102

\bibitem[{{Koposov} {et~al.}(2008){Koposov}, {Belokurov}, {Evans}, {Hewett},
  {Irwin}, {Gilmore}, {Zucker}, {Rix}, {Fellhauer}, {Bell}, \&
  {Glushkova}}]{2008ApJ...686..279K}
{Koposov}, S., {Belokurov}, V., {Evans}, N.~W., {et~al.} 2008, \apj, 686, 279

\bibitem[{K{\"o}ppen {et~al.}(2007)K{\"o}ppen, Weidner, \&
  Kroupa}]{koppen2007possible}
K{\"o}ppen, J., Weidner, C., \& Kroupa, P. 2007, Monthly Notices of the Royal
  Astronomical Society, 375, 673

\bibitem[{{Kormendy}(1977)}]{1977ApJ...218..333K}
{Kormendy}, J. 1977, \apj, 218, 333

\bibitem[{{Kroupa}(2001)}]{2001MNRAS.322..231K}
{Kroupa}, P. 2001, \mnras, 322, 231

\bibitem[{Kunkel \& Demers(1976)}]{kunkel1976magellanic}
Kunkel, W.~E. \& Demers, S. 1976, in The Galaxy and the Local Group, Vol. 182,
  241

\bibitem[{Kuntschner {et~al.}(2001)Kuntschner, Lucey, Smith, Hudson, \&
  Davies}]{kuntschner2001dependence}
Kuntschner, H., Lucey, J.~R., Smith, R.~J., Hudson, M.~J., \& Davies, R.~L.
  2001, Monthly Notices of the Royal Astronomical Society, 323, 615

\bibitem[{La~Barbera {et~al.}(2008)La~Barbera, Busarello, Merluzzi, De~La~Rosa,
  Coppola, \& Haines}]{la2008sdss}
La~Barbera, F., Busarello, G., Merluzzi, P., {et~al.} 2008, The Astrophysical
  Journal, 689, 913

\bibitem[{{La Marca} {et~al.}(2022){La Marca}, {Peletier}, {Iodice},
  {Paolillo}, {Choque Challapa}, {Venhola}, {Forbes}, {Cantiello}, {Hilker},
  {Rejkuba}, {Arnaboldi}, {Spavone}, {D'Ago}, {Raj}, {Ragusa}, {Mirabile},
  {Rampazzo}, {Spiniello}, {Mieske}, \& {Schipani}}]{2022A&A...659A..92L}
{La Marca}, A., {Peletier}, R., {Iodice}, E., {et~al.} 2022, \aap, 659, A92

\bibitem[{{Lee} {et~al.}(2006){Lee}, {Skillman}, {Cannon}, {Jackson}, {Gehrz},
  {Polomski}, \& {Woodward}}]{2006ApJ...647..970L}
{Lee}, H., {Skillman}, E.~D., {Cannon}, J.~M., {et~al.} 2006, \apj, 647, 970

\bibitem[{Lequeux {et~al.}(1979)Lequeux, Peimbert, Rayo, Serrano, \&
  Torres-Peimbert}]{lequeux1979chemical}
Lequeux, J., Peimbert, M., Rayo, J., Serrano, A., \& Torres-Peimbert, S. 1979,
  Astronomy and Astrophysics, vol. 80, no. 2, Dec. 1979, p. 155-166., 80, 155

\bibitem[{{Li} {et~al.}(2010){Li}, {De Lucia}, \&
  {Helmi}}]{2010MNRAS.401.2036L}
{Li}, Y.-S., {De Lucia}, G., \& {Helmi}, A. 2010, \mnras, 401, 2036

\bibitem[{Lian {et~al.}(2018)Lian, Thomas, \& Maraston}]{lian2018modelling}
Lian, J., Thomas, D., \& Maraston, C. 2018, Monthly Notices of the Royal
  Astronomical Society, 481, 4000

\bibitem[{{Lu} {et~al.}(2017){Lu}, {Benson}, {Wetzel}, {Mao}, {Tonnesen},
  {Peter}, {Boylan-Kolchin}, \& {Wechsler}}]{2017ApJ...846...66L}
{Lu}, Y., {Benson}, A., {Wetzel}, A., {et~al.} 2017, \apj, 846, 66

\bibitem[{{Lu} {et~al.}(2014){Lu}, {Wechsler}, {Somerville}, {Croton},
  {Porter}, {Primack}, {Behroozi}, {Ferguson}, {Koo}, {Guo}, {Safarzadeh},
  {Finlator}, {Castellano}, {White}, {Sommariva}, \&
  {Moody}}]{2014ApJ...795..123L}
{Lu}, Y., {Wechsler}, R.~H., {Somerville}, R.~S., {et~al.} 2014, \apj, 795, 123

\bibitem[{Lupton(2005)}]{lupton2005}
Lupton, R. 2005, Transformations between SDSS magnitudes and other systems

\bibitem[{Lynden-Bell(1976)}]{lynden1976dwarf}
Lynden-Bell, D. 1976, Monthly Notices of the Royal Astronomical Society, 174,
  695

\bibitem[{Ma {et~al.}(2016)Ma, Hopkins, Faucher-Gigu{\`e}re, Zolman, Muratov,
  Kere{\v{s}}, \& Quataert}]{ma2016origin}
Ma, X., Hopkins, P.~F., Faucher-Gigu{\`e}re, C.-A., {et~al.} 2016, Monthly
  Notices of the Royal Astronomical Society, 456, 2140

\bibitem[{{Magorrian} {et~al.}(1998){Magorrian}, {Tremaine}, {Richstone},
  {Bender}, {Bower}, {Dressler}, {Faber}, {Gebhardt}, {Green}, {Grillmair},
  {Kormendy}, \& {Lauer}}]{1998AJ....115.2285M}
{Magorrian}, J., {Tremaine}, S., {Richstone}, D., {et~al.} 1998, \aj, 115, 2285

\bibitem[{{Maiolino} \& {Mannucci}(2019)}]{2019A&ARv..27....3M}
{Maiolino}, R. \& {Mannucci}, F. 2019, \aapr, 27, 3

\bibitem[{{Mannucci} {et~al.}(2010){Mannucci}, {Cresci}, {Maiolino}, {Marconi},
  \& {Gnerucci}}]{2010MNRAS.408.2115M}
{Mannucci}, F., {Cresci}, G., {Maiolino}, R., {Marconi}, A., \& {Gnerucci}, A.
  2010, \mnras, 408, 2115

\bibitem[{{Mao} {et~al.}(2021){Mao}, {Geha}, {Wechsler}, {Weiner}, {Tollerud},
  {Nadler}, \& {Kallivayalil}}]{2021ApJ...907...85M}
{Mao}, Y.-Y., {Geha}, M., {Wechsler}, R.~H., {et~al.} 2021, \apj, 907, 85

\bibitem[{{Marleau} {et~al.}(2021){Marleau}, {Habas}, {Poulain}, {Duc},
  {M{\"u}ller}, {Lim}, {Durrell}, {S{\'a}nchez-Janssen}, {Paudel}, {Ahad},
  {Chougule}, {B{\'\i}lek}, \& {Fensch}}]{2021A&A...654A.105M}
{Marleau}, F.~R., {Habas}, R., {Poulain}, M., {et~al.} 2021, \aap, 654, A105

\bibitem[{{Martin} {et~al.}(2006){Martin}, {Ibata}, {Irwin}, {Chapman},
  {Lewis}, {Ferguson}, {Tanvir}, \& {McConnachie}}]{2006MNRAS.371.1983M}
{Martin}, N.~F., {Ibata}, R.~A., {Irwin}, M.~J., {et~al.} 2006, \mnras, 371,
  1983

\bibitem[{{Martin} {et~al.}(2016){Martin}, {Ibata}, {Lewis}, {McConnachie},
  {Babul}, {Bate}, {Bernard}, {Chapman}, {Collins}, {Conn}, {Crnojevi{\'c}},
  {Fardal}, {Ferguson}, {Irwin}, {Mackey}, {McMonigal}, {Navarro}, \&
  {Rich}}]{2016ApJ...833..167M}
{Martin}, N.~F., {Ibata}, R.~A., {Lewis}, G.~F., {et~al.} 2016, \apj, 833, 167

\bibitem[{{Mateo}(1998)}]{1998ARA&A..36..435M}
{Mateo}, M.~L. 1998, \araa, 36, 435

\bibitem[{{McClure} \& {van den Bergh}(1968)}]{1968AJ.....73.1008M}
{McClure}, R.~D. \& {van den Bergh}, S. 1968, \aj, 73, 1008

\bibitem[{{McConnachie}(2012)}]{2012AJ....144....4M}
{McConnachie}, A.~W. 2012, AJ, 144, 4

\bibitem[{{McConnachie} {et~al.}(2018){McConnachie}, {Ibata}, {Martin},
  {Ferguson}, {Collins}, {Gwyn}, {Irwin}, {Lewis}, {Mackey}, {Davidge},
  {Arias}, {Conn}, {C{\^o}t{\'e}}, {Crnojevic}, {Huxor}, {Penarrubia},
  {Spengler}, {Tanvir}, {Valls-Gabaud}, {Babul}, {Barmby}, {Bate}, {Bernard},
  {Chapman}, {Dotter}, {Harris}, {McMonigal}, {Navarro}, {Puzia}, {Rich},
  {Thomas}, \& {Widrow}}]{2018ApJ...868...55M}
{McConnachie}, A.~W., {Ibata}, R., {Martin}, N., {et~al.} 2018, \apj, 868, 55

\bibitem[{{McConnachie} {et~al.}(2009){McConnachie}, {Irwin}, {Ibata},
  {Dubinski}, {Widrow}, {Martin}, {C{\^o}t{\'e}}, {Dotter}, {Navarro},
  {Ferguson}, {Puzia}, {Lewis}, {Babul}, {Barmby}, {Bienaym{\'e}}, {Chapman},
  {Cockcroft}, {Collins}, {Fardal}, {Harris}, {Huxor}, {Mackey},
  {Pe{\~n}arrubia}, {Rich}, {Richer}, {Siebert}, {Tanvir}, {Valls-Gabaud}, \&
  {Venn}}]{2009Natur.461...66M}
{McConnachie}, A.~W., {Irwin}, M.~J., {Ibata}, R.~A., {et~al.} 2009, \nat, 461,
  66

\bibitem[{Moll{\'a} {et~al.}(2015)Moll{\'a}, Cavichia, Gavil{\'a}n, \&
  Gibson}]{molla2015galactic}
Moll{\'a}, M., Cavichia, O., Gavil{\'a}n, M., \& Gibson, B.~K. 2015, Monthly
  Notices of the Royal Astronomical Society, 451, 3693

\bibitem[{Mouhcine {et~al.}(2011)Mouhcine, Kriwattanawong, \&
  James}]{mouhcine2011galaxy}
Mouhcine, M., Kriwattanawong, W., \& James, P. 2011, Monthly Notices of the
  Royal Astronomical Society, 412, 1295

\bibitem[{{Mould} {et~al.}(1983){Mould}, {Kristian}, \& {Da
  Costa}}]{1983ApJ...270..471M}
{Mould}, J.~R., {Kristian}, J., \& {Da Costa}, G.~S. 1983, \apj, 270, 471

\bibitem[{{M{\"u}ller} {et~al.}(2021){M{\"u}ller}, {Durrell}, {Marleau}, {Duc},
  {Lim}, {Posti}, {Agnello}, {S{\'a}nchez-Janssen}, {Poulain}, {Habas},
  {Emsellem}, {Paudel}, {van der Burg}, \& {Fensch}}]{2021ApJ...923....9M}
{M{\"u}ller}, O., {Durrell}, P.~R., {Marleau}, F.~R., {et~al.} 2021, \apj, 923,
  9

\bibitem[{M{\"u}ller {et~al.}(2021)M{\"u}ller, Fahrion, Rejkuba, Hilker, Lelli,
  Lutz, Pawlowski, Coccato, Anand, \& Jerjen}]{muller2021properties}
M{\"u}ller, O., Fahrion, K., Rejkuba, M., {et~al.} 2021, Astronomy \&
  Astrophysics, 645, A92

\bibitem[{{M{\"u}ller} {et~al.}(2021){M{\"u}ller}, {Fahrion}, {Rejkuba},
  {Hilker}, {Lelli}, {Lutz}, {Pawlowski}, {Coccato}, {Anand}, \&
  {Jerjen}}]{2021A&A...645A..92M}
{M{\"u}ller}, O., {Fahrion}, K., {Rejkuba}, M., {et~al.} 2021, \aap, 645, A92

\bibitem[{{M{\"u}ller} \& {Jerjen}(2020)}]{2020A&A...644A..91M}
{M{\"u}ller}, O. \& {Jerjen}, H. 2020, \aap, 644, A91

\bibitem[{M{\"u}ller {et~al.}(2016)M{\"u}ller, Jerjen, Pawlowski, \&
  Binggeli}]{muller2016testing}
M{\"u}ller, O., Jerjen, H., Pawlowski, M.~S., \& Binggeli, B. 2016, Astronomy
  \& Astrophysics, 595, A119

\bibitem[{{M{\"u}ller} {et~al.}(2020){M{\"u}ller}, {Marleau}, {Duc}, {Habas},
  {Fensch}, {Emsellem}, {Poulain}, {Lim}, {Agnello}, {Durrell}, {Paudel},
  {S{\'a}nchez-Janssen}, \& {van der Burg}}]{2020A&A...640A.106M}
{M{\"u}ller}, O., {Marleau}, F.~R., {Duc}, P.-A., {et~al.} 2020, \aap, 640,
  A106

\bibitem[{M{\"u}ller {et~al.}(2020)M{\"u}ller, Marleau, Duc, Habas, Fensch,
  Emsellem, Poulain, Lim, Agnello, Durrell, {et~al.}}]{muller2020spectroscopic}
M{\"u}ller, O., Marleau, F.~R., Duc, P.-A., {et~al.} 2020, Astronomy \&
  Astrophysics, 640, A106

\bibitem[{M{\"u}ller {et~al.}(2018)M{\"u}ller, Pawlowski, Jerjen, \&
  Lelli}]{muller2018whirling}
M{\"u}ller, O., Pawlowski, M.~S., Jerjen, H., \& Lelli, F. 2018, Science, 359,
  534

\bibitem[{M{\"u}ller {et~al.}(2021)M{\"u}ller, Pawlowski, Lelli, Fahrion,
  Rejkuba, Hilker, Kanehisa, Libeskind, \& Jerjen}]{muller2021coherent}
M{\"u}ller, O., Pawlowski, M.~S., Lelli, F., {et~al.} 2021, Astronomy \&
  Astrophysics, 645, L5

\bibitem[{M{\"u}ller {et~al.}(2019)M{\"u}ller, Rejkuba, Pawlowski, Ibata,
  Lelli, Hilker, \& Jerjen}]{muller2019dwarf}
M{\"u}ller, O., Rejkuba, M., Pawlowski, M.~S., {et~al.} 2019, Astronomy \&
  Astrophysics, 629, A18

\bibitem[{{Munshi} {et~al.}(2021){Munshi}, {Brooks}, {Applebaum},
  {Christensen}, {Quinn}, \& {Sligh}}]{2021ApJ...923...35M}
{Munshi}, F., {Brooks}, A.~M., {Applebaum}, E., {et~al.} 2021, \apj, 923, 35

\bibitem[{{Mutlu-Pakdil} {et~al.}(2022){Mutlu-Pakdil}, {Sand}, {Crnojevi{\'c}},
  {Jones}, {Caldwell}, {Guhathakurta}, {Seth}, {Simon}, {Spekkens}, {Strader},
  \& {Toloba}}]{2022ApJ...926...77M}
{Mutlu-Pakdil}, B., {Sand}, D.~J., {Crnojevi{\'c}}, D., {et~al.} 2022, \apj,
  926, 77

\bibitem[{{Nadler} {et~al.}(2019){Nadler}, {Mao}, {Green}, \&
  {Wechsler}}]{2019ApJ...873...34N}
{Nadler}, E.~O., {Mao}, Y.-Y., {Green}, G.~M., \& {Wechsler}, R.~H. 2019, \apj,
  873, 34

\bibitem[{{Panter} {et~al.}(2008){Panter}, {Jimenez}, {Heavens}, \&
  {Charlot}}]{2008MNRAS.391.1117P}
{Panter}, B., {Jimenez}, R., {Heavens}, A.~F., \& {Charlot}, S. 2008, \mnras,
  391, 1117

\bibitem[{{Park} {et~al.}(2017){Park}, {Moon}, {Zaritsky}, {Pak}, {Lee}, {Kim},
  {Kim}, \& {Cha}}]{2017ApJ...848...19P}
{Park}, H.~S., {Moon}, D.-S., {Zaritsky}, D., {et~al.} 2017, \apj, 848, 19

\bibitem[{Pasquali {et~al.}(2010)Pasquali, Gallazzi, Fontanot, Van Den~Bosch,
  De~Lucia, Mo, \& Yang}]{pasquali2010ages}
Pasquali, A., Gallazzi, A., Fontanot, F., {et~al.} 2010, Monthly Notices of the
  Royal Astronomical Society, 407, 937

\bibitem[{Pawlowski {et~al.}(2012)Pawlowski, Pflamm-Altenburg, \&
  Kroupa}]{pawlowski2012vpos}
Pawlowski, M., Pflamm-Altenburg, J., \& Kroupa, P. 2012, Monthly Notices of the
  Royal Astronomical Society, 423, 1109

\bibitem[{Peng {et~al.}(2002)Peng, Ho, Impey, \& Rix}]{peng2002detailed}
Peng, C.~Y., Ho, L.~C., Impey, C.~D., \& Rix, H.-W. 2002, The Astronomical
  Journal, 124, 266

\bibitem[{Peng {et~al.}(2010)Peng, Ho, Impey, \& Rix}]{peng2010detailed}
Peng, C.~Y., Ho, L.~C., Impey, C.~D., \& Rix, H.-W. 2010, The Astronomical
  Journal, 139, 2097

\bibitem[{Peng {et~al.}(2015)Peng, Maiolino, \&
  Cochrane}]{peng2015strangulation}
Peng, Y., Maiolino, R., \& Cochrane, R. 2015, Nature, 521, 192

\bibitem[{Poulain {et~al.}(2021)Poulain, Marleau, Habas, Duc,
  S{\'a}nchez-Janssen, Durrell, Paudel, Ahad, Chougule, M{\"u}ller,
  {et~al.}}]{poulain2021structure}
Poulain, M., Marleau, F.~R., Habas, R., {et~al.} 2021, Monthly Notices of the
  Royal Astronomical Society

\bibitem[{{Poulain} {et~al.}(2022){Poulain}, {Marleau}, {Habas}, {Duc},
  {S{\'a}nchez-Janssen}, {Durrell}, {Paudel}, {M{\"u}ller}, {Lim},
  {B{\'\i}lek}, \& {Fensch}}]{2022A&A...659A..14P}
{Poulain}, M., {Marleau}, F.~R., {Habas}, R., {et~al.} 2022, \aap, 659, A14

\bibitem[{{Prole} {et~al.}(2021){Prole}, {van der Burg}, {Hilker}, \&
  {Spitler}}]{2021MNRAS.500.2049P}
{Prole}, D.~J., {van der Burg}, R.~F.~J., {Hilker}, M., \& {Spitler}, L.~R.
  2021, \mnras, 500, 2049

\bibitem[{S{\'a}nchez-Bl{\'a}zquez {et~al.}(2006)S{\'a}nchez-Bl{\'a}zquez,
  Gorgas, Cardiel, \& Gonz{\'a}lez}]{sanchez2006stellar}
S{\'a}nchez-Bl{\'a}zquez, P., Gorgas, J., Cardiel, N., \& Gonz{\'a}lez, J.
  2006, Astronomy \& Astrophysics, 457, 809

\bibitem[{{Sandage}(1972)}]{1972ApJ...176...21S}
{Sandage}, A. 1972, \apj, 176, 21

\bibitem[{{Sawala} {et~al.}(2022){Sawala}, {Cautun}, {Frenk}, {Helly},
  {Jasche}, {Jenkins}, {Johansson}, {Lavaux}, {McAlpine}, \&
  {Schaller}}]{2022NatAs.tmp..273S}
{Sawala}, T., {Cautun}, M., {Frenk}, C., {et~al.} 2022, Nature Astronomy
  [\eprint[arXiv]{2205.02860}]

\bibitem[{Sawala {et~al.}(2012)Sawala, Scannapieco, \& White}]{sawala2012local}
Sawala, T., Scannapieco, C., \& White, S. 2012, Monthly Notices of the Royal
  Astronomical Society, 420, 1714

\bibitem[{Sheth {et~al.}(2006)Sheth, Jimenez, Panter, \&
  Heavens}]{sheth2006environment}
Sheth, R.~K., Jimenez, R., Panter, B., \& Heavens, A.~F. 2006, The
  Astrophysical Journal, 650, L25

\bibitem[{{Simon}(2019)}]{2019ARA&A..57..375S}
{Simon}, J.~D. 2019, \araa, 57, 375

\bibitem[{{Soto} {et~al.}(2016){Soto}, {Lilly}, {Bacon}, {Richard}, \&
  {Conseil}}]{2016MNRAS.458.3210S}
{Soto}, K.~T., {Lilly}, S.~J., {Bacon}, R., {Richard}, J., \& {Conseil}, S.
  2016, \mnras, 458, 3210

\bibitem[{Steyrleithner {et~al.}(2020)Steyrleithner, Hensler, \&
  Boselli}]{steyrleithner2020effect}
Steyrleithner, P., Hensler, G., \& Boselli, A. 2020, Monthly Notices of the
  Royal Astronomical Society, 494, 1114

\bibitem[{{Stierwalt} {et~al.}(2009){Stierwalt}, {Haynes}, {Giovanelli},
  {Kent}, {Martin}, {Saintonge}, {Karachentsev}, \&
  {Karachentseva}}]{2009AJ....138..338S}
{Stierwalt}, S., {Haynes}, M.~P., {Giovanelli}, R., {et~al.} 2009, \aj, 138,
  338

\bibitem[{{Su} {et~al.}(2021){Su}, {Salo}, {Janz}, {Laurikainen}, {Venhola},
  {Peletier}, {Iodice}, {Hilker}, {Cantiello}, {Napolitano}, {Spavone}, {Raj},
  {van de Ven}, {Mieske}, {Paolillo}, {Capaccioli}, {Valentijn}, \&
  {Watkins}}]{2021A&A...647A.100S}
{Su}, A.~H., {Salo}, H., {Janz}, J., {et~al.} 2021, \aap, 647, A100

\bibitem[{Sybilska {et~al.}(2017)Sybilska, Lisker, Kuntschner, Vazdekis, van~de
  Ven, Peletier, Falc{\'o}n-Barroso, Vijayaraghavan, \&
  Janz}]{sybilska2017helena}
Sybilska, A., Lisker, T., Kuntschner, H., {et~al.} 2017, Monthly Notices of the
  Royal Astronomical Society, 470, 815

\bibitem[{Tammann(1994)}]{tammann1994dwarf}
Tammann, G. 1994, in European Southern Observatory Conference and Workshop
  Proceedings, Vol.~49, 3

\bibitem[{{Tanoglidis} {et~al.}(2021){Tanoglidis}, {Drlica-Wagner}, {Wei},
  {Li}, {S{\'a}nchez}, {Zhang}, {Peter}, {Feldmeier-Krause}, {Prat}, {Casey},
  {Palmese}, {S{\'a}nchez}, {DeRose}, {Conselice}, {Gagnon}, {Abbott},
  {Aguena}, {Allam}, {Avila}, {Bechtol}, {Bertin}, {Bhargava}, {Brooks},
  {Burke}, {Rosell}, {Kind}, {Carretero}, {Chang}, {Costanzi}, {da Costa}, {De
  Vicente}, {Desai}, {Diehl}, {Doel}, {Eifler}, {Everett}, {Evrard},
  {Flaugher}, {Frieman}, {Garc{\'\i}a-Bellido}, {Gerdes}, {Gruendl},
  {Gschwend}, {Gutierrez}, {Hartley}, {Hollowood}, {Huterer}, {James},
  {Krause}, {Kuehn}, {Kuropatkin}, {Maia}, {March}, {Marshall}, {Menanteau},
  {Miquel}, {Ogando}, {Paz-Chinch{\'o}n}, {Romer}, {Roodman}, {Sanchez},
  {Scarpine}, {Serrano}, {Sevilla-Noarbe}, {Smith}, {Suchyta}, {Tarle},
  {Thomas}, {Tucker}, {Walker}, \& {DES Collaboration}}]{2021ApJS..252...18T}
{Tanoglidis}, D., {Drlica-Wagner}, A., {Wei}, K., {et~al.} 2021, \apjs, 252, 18

\bibitem[{Tassis {et~al.}(2008)Tassis, Kravtsov, \& Gnedin}]{tassis2008scaling}
Tassis, K., Kravtsov, A.~V., \& Gnedin, N.~Y. 2008, The Astrophysical Journal,
  672, 888

\bibitem[{Teeninga {et~al.}(2015)Teeninga, Moschini, Trager, \&
  Wilkinson}]{teeninga2015improved}
Teeninga, P., Moschini, U., Trager, S.~C., \& Wilkinson, M.~H. 2015, in
  International Symposium on Mathematical Morphology and Its Applications to
  Signal and Image Processing, Springer, 157--168

\bibitem[{Thomas {et~al.}(2005)Thomas, Maraston, \& Bender}]{thomas2005epochs}
Thomas, D., Maraston, C., \& Bender, R. 2005, in Multiwavelength Mapping of
  Galaxy Formation and Evolution: Proceedings of the ESO Workshop Held at
  Venice, Italy, 13-16 October 2003, Springer, 296--301

\bibitem[{Thomas {et~al.}(2010)Thomas, Maraston, Schawinski, Sarzi, \&
  Silk}]{thomas2010environment}
Thomas, D., Maraston, C., Schawinski, K., Sarzi, M., \& Silk, J. 2010, Monthly
  Notices of the Royal Astronomical Society, 404, 1775

\bibitem[{Tolstoy {et~al.}(2009)Tolstoy, Hill, \& Tosi}]{tolstoy2009star}
Tolstoy, E., Hill, V., \& Tosi, M. 2009, Annual Review of Astronomy and
  Astrophysics, 47, 371

\bibitem[{Trager {et~al.}(2000)Trager, Faber, Worthey, \&
  Gonz{\'a}lez}]{trager2000stellar}
Trager, S., Faber, S., Worthey, G., \& Gonz{\'a}lez, J.~J. 2000, The
  Astronomical Journal, 120, 165

\bibitem[{Tremonti {et~al.}(2004)Tremonti, Heckman, Kauffmann, Brinchmann,
  Charlot, White, Seibert, Peng, Schlegel, Uomoto,
  {et~al.}}]{tremonti2004origin}
Tremonti, C.~A., Heckman, T.~M., Kauffmann, G., {et~al.} 2004, The
  Astrophysical Journal, 613, 898

\bibitem[{Trussler {et~al.}(2020)Trussler, Maiolino, Maraston, Peng, Thomas,
  Goddard, \& Lian}]{trussler2020both}
Trussler, J., Maiolino, R., Maraston, C., {et~al.} 2020, Monthly Notices of the
  Royal Astronomical Society, 491, 5406

\bibitem[{Tully \& Fisher(1977)}]{tully1977new}
Tully, R.~B. \& Fisher, J.~R. 1977, Astronomy and Astrophysics, vol. 54, no. 3,
  Feb. 1977, p. 661-673., 54, 661

\bibitem[{Vaduvescu {et~al.}(2007)Vaduvescu, McCall, \&
  Richer}]{vaduvescu2007chemical}
Vaduvescu, O., McCall, M.~L., \& Richer, M.~G. 2007, The Astronomical Journal,
  134, 604

\bibitem[{Vazdekis {et~al.}(2016)Vazdekis, Koleva, Ricciardelli, R{\"o}ck, \&
  Falc{\'o}n-Barroso}]{vazdekis2016uv}
Vazdekis, A., Koleva, M., Ricciardelli, E., R{\"o}ck, B., \&
  Falc{\'o}n-Barroso, J. 2016, Monthly Notices of the Royal Astronomical
  Society, 463, 3409

\bibitem[{{Vazdekis} {et~al.}(2010){Vazdekis}, {S{\'a}nchez-Bl{\'a}zquez},
  {Falc{\'o}n-Barroso}, {Cenarro}, {Beasley}, {Cardiel}, {Gorgas}, \&
  {Peletier}}]{2010MNRAS.404.1639V}
{Vazdekis}, A., {S{\'a}nchez-Bl{\'a}zquez}, P., {Falc{\'o}n-Barroso}, J.,
  {et~al.} 2010, \mnras, 404, 1639

\bibitem[{Venhola {et~al.}(2019)Venhola, Peletier, Laurikainen, Salo, Iodice,
  Mieske, Hilker, Wittmann, Paolillo, Cantiello, {et~al.}}]{venhola2019fornax}
Venhola, A., Peletier, R., Laurikainen, E., {et~al.} 2019, Astronomy \&
  astrophysics, 625, A143

\bibitem[{Vincenzo {et~al.}(2016)Vincenzo, Matteucci, Belfiore, \&
  Maiolino}]{vincenzo2016modern}
Vincenzo, F., Matteucci, F., Belfiore, F., \& Maiolino, R. 2016, Monthly
  Notices of the Royal Astronomical Society, 455, 4183

\bibitem[{Xia \& Yu(2019{\natexlab{a}})}]{xia2019stellar}
Xia, M. \& Yu, Q. 2019{\natexlab{a}}, The Astrophysical Journal, 874, 105

\bibitem[{Xia \& Yu(2019{\natexlab{b}})}]{xia2019origin}
Xia, M. \& Yu, Q. 2019{\natexlab{b}}, The Astrophysical Journal, 880, 5

\bibitem[{Zahid {et~al.}(2012)Zahid, Bresolin, Kewley, Coil, \&
  Dav{\'e}}]{zahid2012metallicities}
Zahid, H., Bresolin, F., Kewley, L., Coil, A.~L., \& Dav{\'e}, R. 2012, The
  Astrophysical Journal, 750, 120

\bibitem[{{Zaritsky} {et~al.}(2019){Zaritsky}, {Donnerstein}, {Dey},
  {Kadowaki}, {Zhang}, {Karunakaran}, {Mart{\'\i}nez-Delgado}, {Rahman}, \&
  {Spekkens}}]{2019ApJS..240....1Z}
{Zaritsky}, D., {Donnerstein}, R., {Dey}, A., {et~al.} 2019, \apjs, 240, 1

\bibitem[{Zhang {et~al.}(2018)Zhang, Puzia, Peng, Liu, C{\^o}t{\'e}, Ferrarese,
  Duc, Eigenthaler, Lim, Lan{\c{c}}on, {et~al.}}]{zhang2018stellar}
Zhang, H.-X., Puzia, T.~H., Peng, E.~W., {et~al.} 2018, The Astrophysical
  Journal, 858, 37

\end{thebibliography}

\begin{appendix}

\section{Supplementary materials}
\label{appendix}

\subsection{Dwarf galaxy cutouts}

\begin{figure*}[hbt!]
\begin{center}
\includegraphics[width=\linewidth]{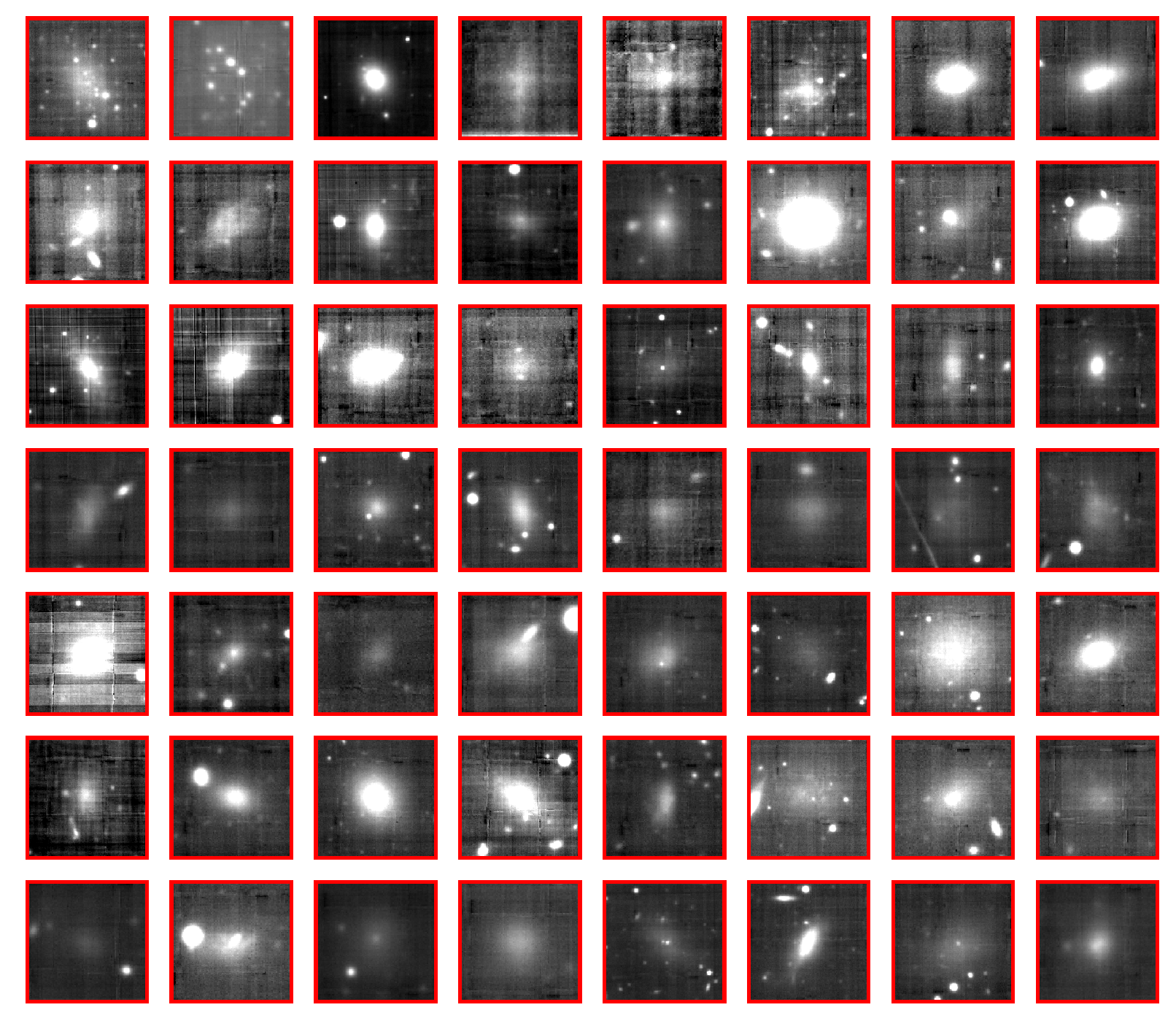}
\caption{Cutouts of all dwarf galaxies produced by collapsing the MUSE data cubes along the wavelength axis, resulting in 2D images. The size of the cutouts are chosen to be the diameter of the dwarf's visual appearance in the MUSE image plus a margin of 2\,arcsec.}
\label{figure:cutouts}
\end{center}
\end{figure*}

In Figure \ref{figure:cutouts} we present cutouts of the 56 dwarf galaxies studied in this work. We collapse the MUSE datacubes along the wavelength axis by taking the median of each spaxel to produce the images.

\subsection{Error estimation}
\label{appendix:MC}

\begin{figure*}[!htb]
\begin{center}
\includegraphics[width=0.49\linewidth]{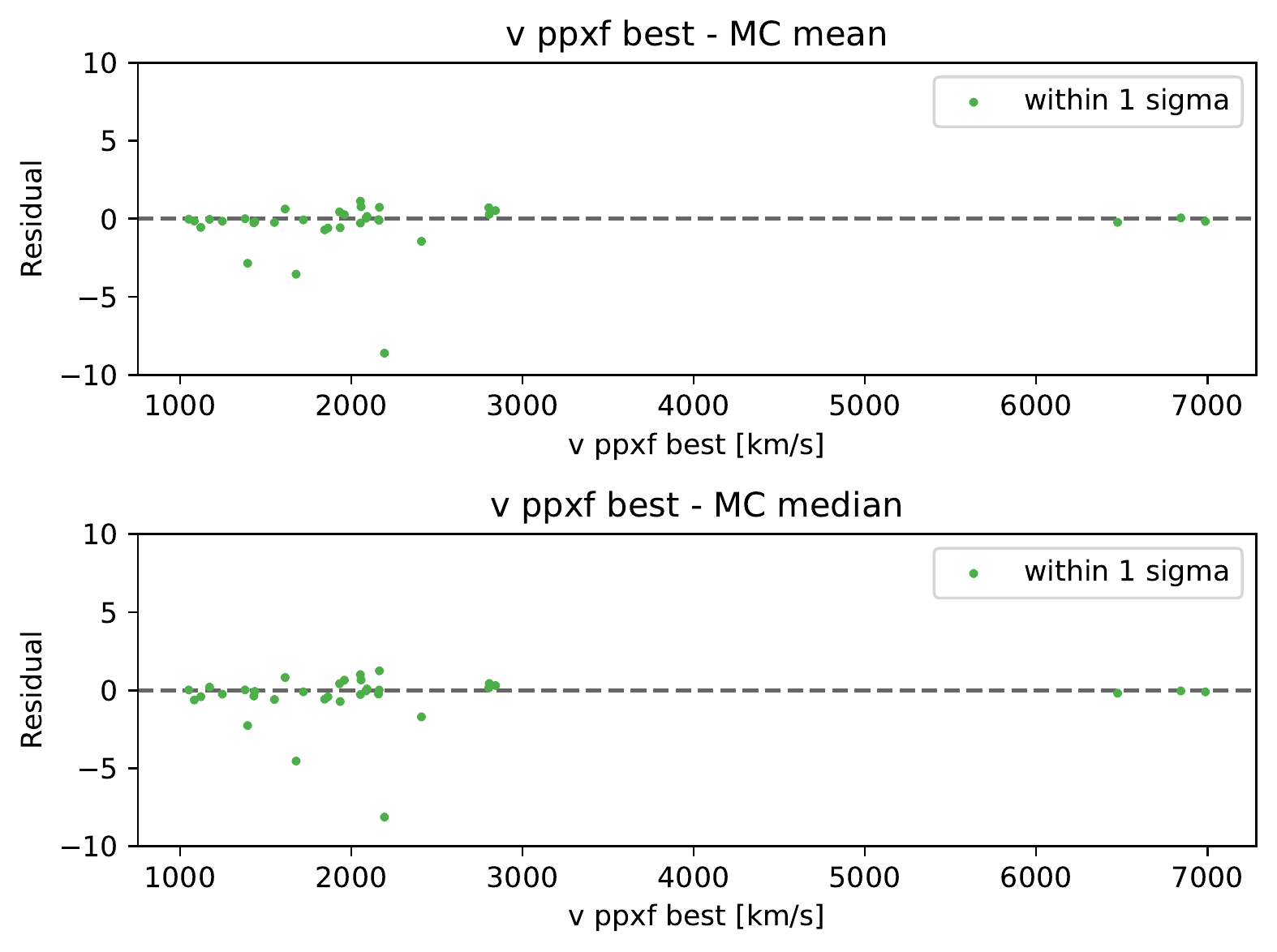}
\includegraphics[width=0.49\linewidth]{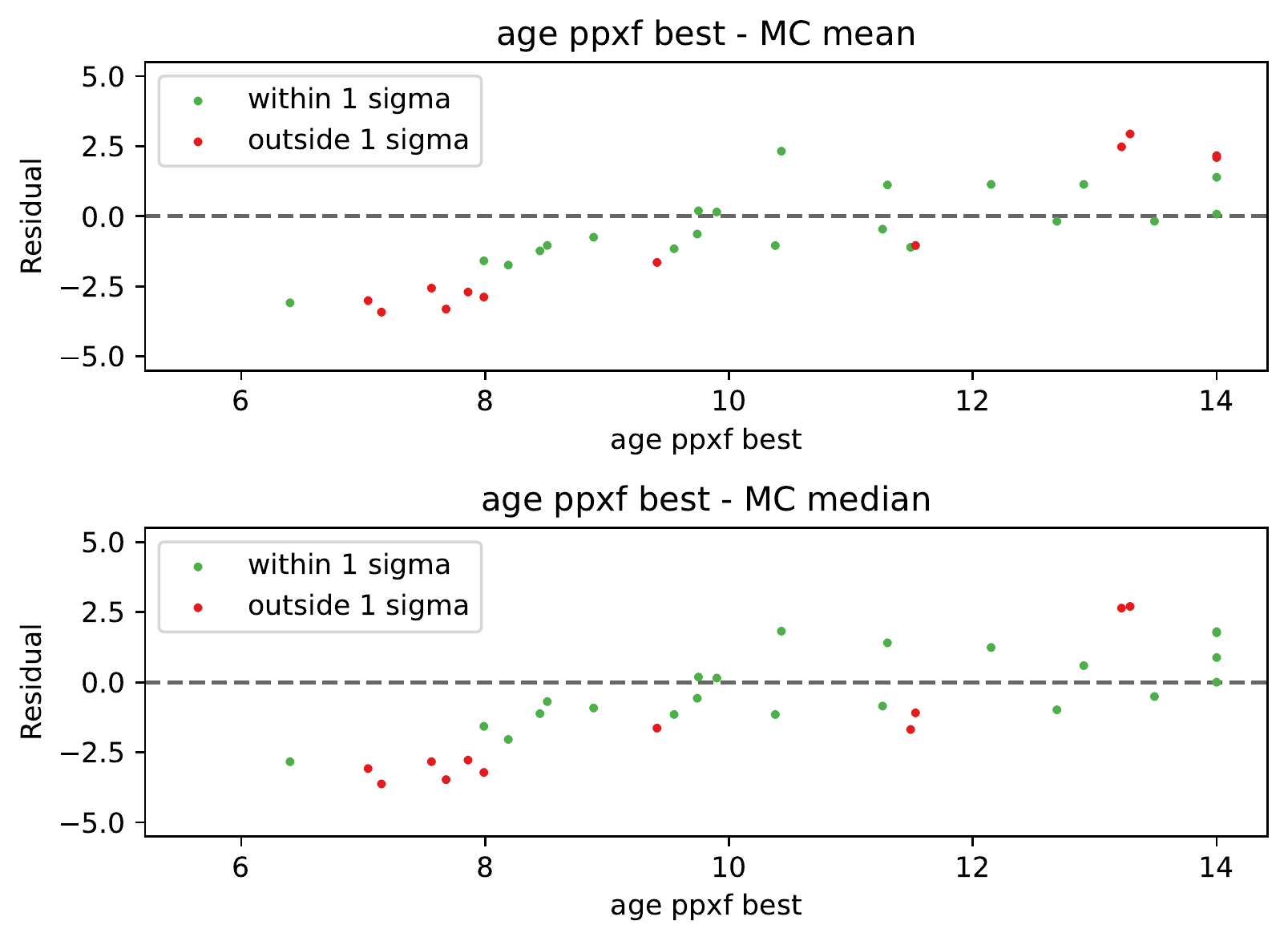}
\includegraphics[width=0.49\linewidth]{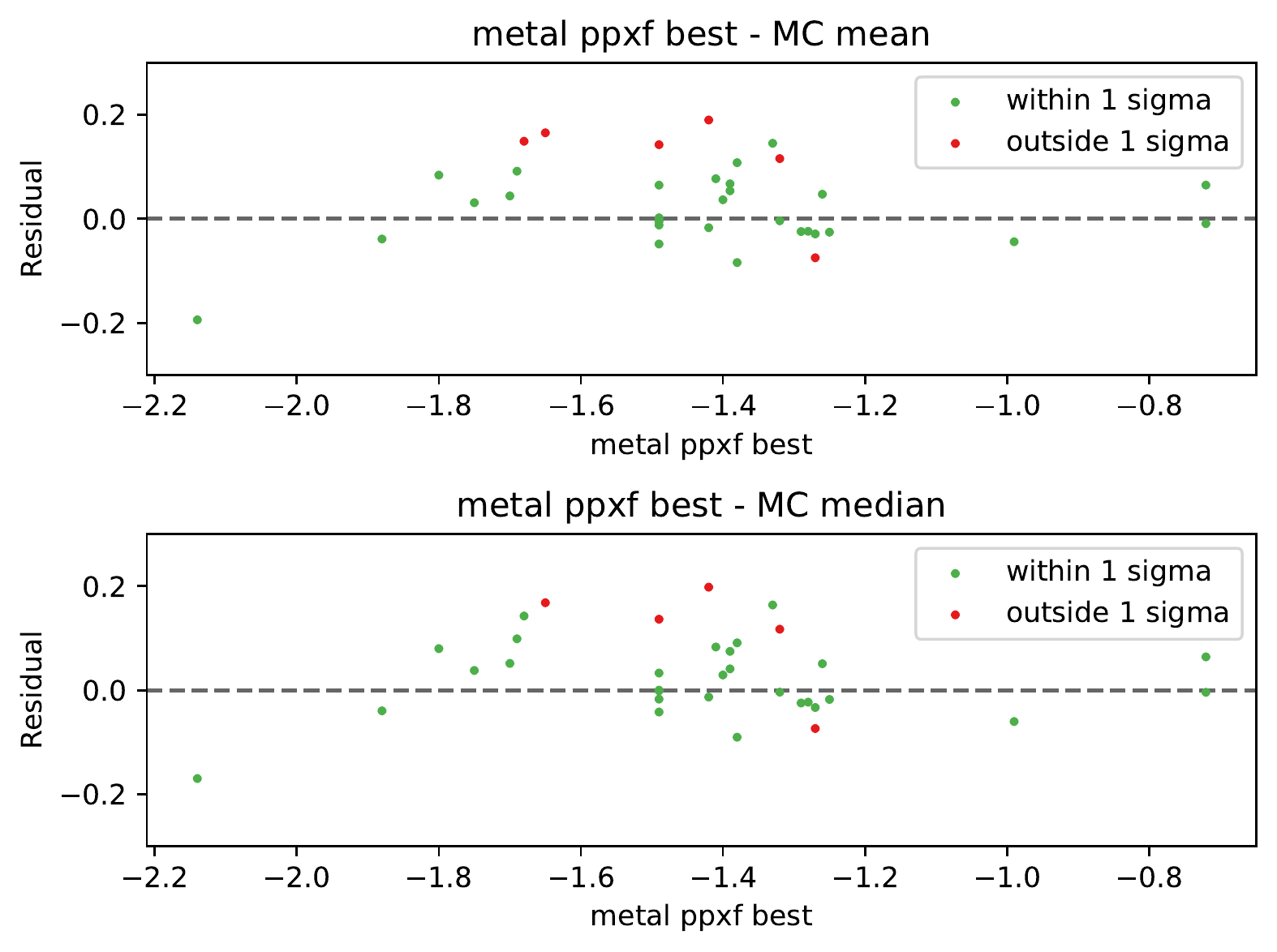}
\caption{Residual plots of the pPXF best fit values subtracted by the mean and median of the MC realizations for the parameters: recessional velocity ($v$; top left), age (top right), metallicity (bottom center). Green dots indicate that the best fit value resides inside the 1$\sigma$ bounds of the MC realizations, red means the value is outside of these bounds.}
\label{figure:error_est}
\end{center}
\end{figure*}

In this section we show the results of the error estimation via MC simulations. The signs of the residuals between galaxy spectrum and best fit are flipped randomly and refitted with pPXF. This is done in 400 realizations per galaxy which results in distributions for the extracted properties velocity, age and metallicity. In some cases the best fit value lies outside of the 1$\sigma$ confidence interval from the MC realizations (see Figure \ref{figure:err_est_ex}). In Figure \ref{figure:error_est} we show the residual of the best fit minus the mean/median of the MC realizations and indicate by color whether or not the best fit value lies within (green) or outside (red) the 1$\sigma$ bounds of the MC simulations.

\subsection{Data table}
\label{appendix:data}

We summarize the properties of the 56 dwarfs studied in this work in Table \ref{tab:MUSE_dwarfs}. We obtained the values from \citet{habas2020newly}, \citet{poulain2021structure} and this work.

\renewcommand{\arraystretch}{1.5}
\setlength{\tabcolsep}{2.5pt}

\clearpage
\onecolumn
\begin{ThreePartTable}
\begin{TableNotes}
\footnotesize
\item Note. Column (1): MATLAS ID of dwarf galaxies. Dwarfs identified in the MATLAS survey are numbered from 1 to 2210, e.g., MATLAS-2019. Column (2): morphology of the dwarf galaxy. Column (3): right ascension in degrees. Column (4): declination in degrees. Column (5): assumed host galaxy based on minimal difference in line-of-sight velocities between satellite and massive host. Column (6) distance of the assumed host if present. If no host could be assigned, the distance results from the dwarf recessional velocity via Hubble's law. Column (7): Absolute \emph{g}-band magnitude. We use the apparent \emph{g}-band magnitude from \citet{poulain2021structure} and the distance from column (6). Column (8): Effective radius in arcseconds via \textsc{galfit} modelling from \citet{poulain2021structure}. Column (9): emission line flag. Values 1 for emission lines present, 0 for no emission lines. Column (10): background flag. Values 1 if the dwarf galaxy resides outside of the ATLAS$^{3D}$ target volume and 0 if it lies inside. This is based on the dwarf line-of-sight velocity compared with the velocities of the massive ATLAS$^{3D}$ galaxies (see Figure \ref{figure:radial_vel}). Column (11): dwarf line-of-sight velocity in km/s. Column (12): dwarf age in Gyr. Column (13): dwarf metallicity in dex. Column (14): stellar mass-to-light ratio in the \emph{V}-band. Column (15): signal-to-noise ratio of the continuum spectrum. Column (16): quality of the spectrum. Values 1 for clear spectral lines, 0 if the noise dominates the lines and the measured values are uncertain.
\end{TableNotes}

\begin{longtable}{llrrlrrrrlllllrr}
\caption{Properties of the dwarf galaxies studied in this work.}\label{tab:MUSE_dwarfs}\\
\toprule
ID & Morph & RA & Dec & Host & D & M$_{g}$ & R$_{eff}$ & Em & Bkg & V & Age & [M/H] & ML & SNR & Q \\
 & & [deg] & [deg] &  & [Mpc] & [mag] & [''] &  &  & [km/s] & [Gyr] & [dex] &  &  & \\
(1) & (2) & (3) & (4) & (5) & (6) & (7) & (8) & (9) & (10) & (11) & (12) & (13) & (14) & (15) & (16) \\
\midrule
\endfirsthead
\caption{\textit{(Continued)} Properties of the dwarf galaxies studied in this work.}\\
\toprule
ID & Morph & RA & Dec & Host & D & M$_{g}$ & R$_{eff}$ & Em & Bkg & V & Age & [M/H] & ML & SNR & Q \\
 &  & [deg] & [deg] &  & [Mpc] & [mag] & [''] &  &  & [km/s] & [Gyr] & [dex] &  &  &  \\
(1) & (2) & (3) & (4) & (5) & (6) & (7) & (8) & (9) & (10) & (11) & (12) & (13) & (14) & (15) & (16) \\
\midrule
\endhead
\bottomrule
\multicolumn{3}{r}{\textit{Continued on next page}} \\
\endfoot
\bottomrule
\insertTableNotes
\endlastfoot

1297 &    dE & 184.7614 &  5.0981 &   NGC4255 &      31.0 &   17.9 & 13.1 &         0 &          0 &   1931.3$^{+3.9}_{-5.0}$ & 11.8$^{+1.7}_{-2.8}$ & -1.44$^{+0.07}_{-0.04}$ & 2.2$^{+0.0}_{-0.6}$ &        24.0 &           1 \\
2019 &    dE & 226.3340 &  1.8127 &   NGC5846 &      21.0 &   17.8 & 17.2 &         0 &          0 &   2158.8$^{+5.2}_{-5.1}$ & 13.5$^{+0.5}_{-0.2}$ & -1.88$^{+0.13}_{-0.06}$ & 2.1$^{+0.1}_{-0.0}$ &        24.3 &           1 \\
  10 &   dEN &  18.7949 & -1.4718 &   NGC0448 &      30.0 &   17.3 &  4.4 &         1 &          0 &   1852.5$^{+1.4}_{-1.9}$ &  8.8$^{+0.4}_{-0.2}$ & -1.25$^{+0.03}_{-0.01}$ & 1.6$^{+0.1}_{-0.0}$ &        61.9 &           1 \\
  35 &    dE &  20.4945 &  3.9712 &   NGC0474 &      31.0 &   19.9 &  5.1 &         0 &          0 &  2193.8$^{+27.5}_{-9.7}$ &  6.4$^{+7.0}_{-0.6}$ & -2.14$^{+0.42}_{-0.06}$ & 1.2$^{+0.8}_{-0.1}$ &         9.3 &           1 \\
 205 &    dI &  42.3268 & -1.5714 &           &      37.0 &   18.9 &  6.5 &         0 &            & 2603.7$^{+12.6}_{-17.0}$ & 12.9$^{+0.2}_{-1.9}$ & -2.17$^{+0.16}_{-0.06}$ & 2.1$^{+0.0}_{-0.3}$ &        13.4 &           0 \\
 223 &    dI &  43.1790 & -1.3266 &           &      60.0 &   18.0 &  9.8 &         1 &          1 &  4167.3$^{+9.8}_{-11.6}$ &  9.6$^{+2.2}_{-1.8}$ & -1.71$^{+0.15}_{-0.16}$ & 1.7$^{+0.2}_{-0.3}$ &        19.0 &           1 \\
 300 &    dE &  49.9520 & -2.3088 &           &      93.0 &   19.4 &  3.1 &         0 &          1 &   6476.1$^{+4.7}_{-4.5}$ &  9.5$^{+2.0}_{-0.9}$ & -1.25$^{+0.09}_{-0.03}$ & 1.6$^{+0.3}_{-0.0}$ &        21.1 &           1 \\
 303 &    dE &  50.0618 & -1.7127 &   NGC1289 &      38.0 &   19.0 &  5.6 &         0 &          0 &   2806.1$^{+3.3}_{-3.6}$ &  9.4$^{+3.4}_{-0.3}$ &   -1.26$^{+0.0}_{-0.1}$ & 1.8$^{+0.4}_{-0.0}$ &        25.0 &           1 \\
 290 &    dE &  49.5108 & -1.6533 &   NGC1289 &      38.0 &   18.8 &  8.2 &         0 &          0 &   2842.5$^{+8.1}_{-9.4}$ & 11.5$^{+2.5}_{-0.5}$ & -1.39$^{+0.01}_{-0.13}$ & 1.9$^{+0.3}_{-0.1}$ &        15.9 &           1 \\
 313 &    dI &  71.6313 & -5.2406 &           &      67.0 &   19.1 &  6.4 &         1 &          1 &  4709.9$^{+6.5}_{-21.2}$ &  8.8$^{+4.0}_{-2.4}$ & -2.12$^{+0.16}_{-0.13}$ & 1.2$^{+0.7}_{-0.1}$ &        13.2 &           1 \\
 445 &    dE & 134.2161 & -3.2605 &   NGC2699 &      26.0 &   18.9 &  4.1 &         0 &          0 &   2053.2$^{+3.4}_{-2.7}$ & 11.0$^{+1.6}_{-2.3}$ & -1.44$^{+0.06}_{-0.05}$ & 1.5$^{+0.6}_{-0.2}$ &        32.4 &           1 \\
 443 &    dE & 134.1938 & -3.2526 &   NGC2699 &      26.0 &   19.0 &  4.9 &         0 &          0 &   1959.5$^{+5.3}_{-5.9}$ & 10.4$^{+2.4}_{-3.1}$ & -1.38$^{+0.19}_{-0.02}$ & 2.2$^{+0.0}_{-0.7}$ &        13.8 &           1 \\
 444 &   dEN & 134.2092 & -2.9154 &   NGC2699 &      26.0 &   19.4 &  4.9 &         0 &          0 &   2091.8$^{+4.7}_{-5.1}$ &  9.7$^{+2.7}_{-1.3}$ &  -1.4$^{+0.06}_{-0.14}$ & 1.7$^{+0.3}_{-0.1}$ &        21.3 &           1 \\
 553 &   dEN & 145.4222 & -3.7319 &   NGC2974 &      21.0 &   17.5 &  6.7 &         0 &          0 &   1719.9$^{+1.7}_{-1.6}$ & 12.6$^{+0.6}_{-0.7}$ &  -1.29$^{+0.05}_{-0.0}$ & 2.0$^{+0.3}_{-0.1}$ &        63.0 &           1 \\
 574 &    dE & 146.1150 & -3.3048 &   NGC2974 &      21.0 &   19.5 &  7.2 &         0 &          0 &   1935.1$^{+9.1}_{-7.8}$ &  8.0$^{+4.1}_{-0.8}$ & -1.33$^{+0.06}_{-0.36}$ & 1.5$^{+0.5}_{-0.1}$ &        12.8 &           1 \\
1476 &   dEN & 191.5301 & -3.2690 &   NGC4691 &      16.0 &   17.4 &  6.6 &         0 &          0 &   1050.7$^{+1.4}_{-1.5}$ &  9.9$^{+0.5}_{-0.9}$ & -1.32$^{+0.03}_{-0.02}$ & 1.8$^{+0.1}_{-0.1}$ &        55.0 &           1 \\
1486 &   dEN & 191.7061 & -2.5005 &           &      64.0 &   17.6 &  9.1 &         1 &          1 &   4474.8$^{+0.5}_{-8.8}$ & 11.5$^{+0.5}_{-3.0}$ & -1.57$^{+0.01}_{-0.18}$ & 1.9$^{+0.0}_{-0.3}$ &        34.1 &           1 \\
 324 &    dE &  72.6729 & -3.8295 &           &      65.0 &   18.7 &  4.1 &         1 &          1 &   4563.4$^{+1.7}_{-9.4}$ & 13.2$^{+0.5}_{-1.8}$ & -1.48$^{+0.08}_{-0.05}$ & 2.1$^{+0.1}_{-0.3}$ &        24.4 &           1 \\
 323 &    dE &  72.6134 & -3.6137 &           &      64.0 &   18.5 &  4.9 &         1 &          1 &   4492.5$^{+5.3}_{-5.2}$ &  9.0$^{+2.8}_{-0.6}$ &  -1.33$^{+0.08}_{-0.1}$ & 1.7$^{+0.3}_{-0.1}$ &        22.9 &           1 \\
 321 &    dE &  72.4610 & -3.5938 &           &      14.0 &   19.9 &  4.8 &         0 &            &  963.0$^{+16.2}_{-24.3}$ &  6.3$^{+4.7}_{-1.2}$ &  -2.23$^{+0.11}_{-0.0}$ & 1.2$^{+0.5}_{-0.2}$ &         8.4 &           0 \\
1408 &   dEN & 190.2965 & -5.0977 &   NGC4546 &      14.0 &   18.8 &  8.5 &         0 &          0 &   1083.1$^{+7.0}_{-7.5}$ & 12.9$^{+0.9}_{-3.3}$ &  -1.39$^{+0.08}_{-0.2}$ & 2.1$^{+0.2}_{-0.4}$ &        14.5 &           1 \\
  15 &    dE &  18.9577 & -1.3915 &           &      14.0 &   19.8 &  4.3 &         0 &            & 1015.0$^{+81.4}_{-67.2}$ & 11.7$^{+0.2}_{-7.2}$ &   -2.27$^{+0.0}_{-0.0}$ & 2.1$^{+0.0}_{-0.9}$ &        13.3 &           0 \\
  29 &    dE &  20.1466 &  3.1456 &   NGC0474 &      31.0 &   19.2 &  7.6 &         1 &          0 & 2153.4$^{+10.2}_{-10.6}$ &  9.2$^{+1.6}_{-2.3}$ & -1.62$^{+0.01}_{-0.22}$ & 1.9$^{+0.2}_{-0.7}$ &        13.2 &           1 \\
 222 &    dE &  43.1665 & -1.1599 &           &      98.0 &   19.6 &  4.0 &         0 &          1 &   6845.4$^{+3.2}_{-3.5}$ & 10.4$^{+2.6}_{-0.5}$ & -0.72$^{+0.02}_{-0.15}$ & 2.3$^{+0.4}_{-0.1}$ &        23.5 &           1 \\
 273 &    dE &  48.9098 & -2.9508 &   NGC1266 &      30.0 &   19.6 &  6.3 &         1 &          0 &  2522.1$^{+5.6}_{-16.1}$ & 12.4$^{+0.3}_{-2.5}$ & -1.71$^{+0.22}_{-0.02}$ & 2.0$^{+0.0}_{-0.3}$ &        15.8 &           1 \\
 269 &    dE &  48.6046 & -2.9236 &   NGC1253 &      16.0 &   20.4 &  4.9 &         0 &          0 & 1613.6$^{+11.9}_{-12.7}$ &  7.6$^{+5.2}_{-0.2}$ & -0.99$^{+0.22}_{-0.13}$ & 1.6$^{+0.9}_{-0.0}$ &         8.7 &           1 \\
 448 &    dE & 134.2757 & -3.3425 &   NGC2699 &      26.0 &   20.4 &  3.2 &         0 &          0 &  2053.2$^{+7.7}_{-10.5}$ &  7.9$^{+5.5}_{-0.0}$ & -1.41$^{+0.11}_{-0.25}$ & 1.5$^{+0.7}_{-0.0}$ &        13.9 &           1 \\
1232 &    dE & 184.1771 &  6.6895 &   NGC4215 &      32.0 &   18.9 &  6.4 &         0 &          0 &   2086.6$^{+3.9}_{-4.1}$ & 10.7$^{+2.3}_{-1.8}$ & -1.49$^{+0.07}_{-0.08}$ & 2.1$^{+0.1}_{-0.5}$ &        26.0 &           1 \\
 320 &    dE &  72.4170 & -4.2329 &           &      14.0 &   20.6 &  4.1 &         0 &            &  960.2$^{+15.1}_{-12.6}$ &  6.9$^{+2.4}_{-0.1}$ &  -2.21$^{+0.15}_{-0.0}$ & 1.2$^{+0.2}_{-0.0}$ &         6.7 &           0 \\
 318 &    dE &  72.1070 & -3.8733 & PGC016060 &      38.0 &   20.3 &  3.7 &         0 &          0 &  2803.0$^{+9.2}_{-11.6}$ & 11.4$^{+1.7}_{-3.0}$ &  -1.69$^{+0.1}_{-0.28}$ & 1.1$^{+1.0}_{-0.4}$ &        11.6 &           1 \\
 585 &    dE & 146.4547 & -0.5469 &    IC0560 &      27.0 &   18.8 & 11.7 &         1 &          0 &  1832.9$^{+12.8}_{-9.2}$ &  9.0$^{+2.9}_{-3.1}$ & -1.88$^{+0.17}_{-0.13}$ & 1.1$^{+0.7}_{-0.0}$ &        14.2 &           1 \\
 578 &    dE & 146.2100 &  0.1444 &           &      92.0 &   19.6 &  6.1 &         1 &          1 & 6463.2$^{+22.1}_{-33.7}$ &  6.2$^{+7.4}_{-0.2}$ & -2.07$^{+0.15}_{-0.15}$ & 1.2$^{+0.8}_{-0.0}$ &        13.8 &           1 \\
 218 &   dEN &  42.9988 & -1.1746 &           &     100.0 &   17.5 &  8.8 &         0 &          1 &   6988.7$^{+2.3}_{-1.8}$ &  9.6$^{+2.3}_{-0.0}$ & -0.72$^{+0.07}_{-0.05}$ & 2.3$^{+0.4}_{-0.0}$ &        25.7 &           1 \\
 428 &   dEN & 133.6689 & -3.1090 &   NGC2695 &      32.0 &   19.7 &  5.2 &         0 &          0 & 2164.3$^{+10.2}_{-12.0}$ & 10.6$^{+2.4}_{-3.0}$ & -1.28$^{+0.22}_{-0.18}$ & 1.5$^{+0.8}_{-0.0}$ &        11.2 &           1 \\
 429 &    dE & 133.7407 & -2.9390 &   NGC2695 &      32.0 &   21.2 &  3.0 &         0 &          0 & 1677.5$^{+44.1}_{-36.4}$ &  8.1$^{+2.3}_{-3.6}$ &  -1.8$^{+0.26}_{-0.47}$ & 1.7$^{+0.1}_{-0.7}$ &         4.9 &           1 \\
 420 &    dE & 133.3598 & -2.5094 &   NGC2695 &      32.0 &   19.0 &  4.6 &         1 &          0 &  1869.1$^{+5.3}_{-12.7}$ &  8.6$^{+2.8}_{-0.8}$ & -1.76$^{+0.07}_{-0.15}$ & 1.5$^{+0.3}_{-0.1}$ &        18.1 &           1 \\
 544 &    dE & 145.1226 &  5.2314 &   NGC2962 &      34.0 &   19.1 &  5.4 &         1 &          0 &  2123.7$^{+13.4}_{-1.5}$ &  9.9$^{+2.7}_{-2.1}$ & -1.79$^{+0.11}_{-0.13}$ & 1.1$^{+0.8}_{-0.3}$ &        17.3 &           1 \\
 645 &    dE & 153.3468 &  3.3098 &   NGC3156 &      22.0 &   19.7 &  6.4 &         0 &          0 &  1394.6$^{+13.4}_{-7.4}$ &  9.8$^{+2.5}_{-3.0}$ & -1.49$^{+0.19}_{-0.18}$ & 1.7$^{+0.3}_{-0.4}$ &        10.8 &           1 \\
 652 &   dEN & 153.6326 &  3.3670 &   NGC3156 &      22.0 &   17.6 & 10.1 &         0 &          0 &   1246.6$^{+2.3}_{-2.1}$ &  8.9$^{+2.3}_{-1.2}$ & -1.27$^{+0.08}_{-0.03}$ & 1.6$^{+0.3}_{-0.2}$ &        37.1 &           1 \\
 642 &    dE & 153.1371 &  3.7079 &   NGC3156 &      22.0 &   19.3 &  4.4 &         0 &          0 &   1121.2$^{+4.9}_{-3.8}$ & 11.9$^{+1.6}_{-2.2}$ &  -1.63$^{+0.1}_{-0.11}$ & 2.2$^{+0.1}_{-0.6}$ &        22.2 &           1 \\
 928 &    dE & 169.8325 &  2.7918 &   NGC3640 &      26.0 &   18.2 &  8.5 &         0 &          0 &   1551.1$^{+6.9}_{-6.4}$ & 11.3$^{+2.7}_{-1.6}$ & -1.75$^{+0.09}_{-0.15}$ & 1.8$^{+0.3}_{-0.2}$ &        16.6 &           1 \\
 974 &    dE & 170.4586 &  2.9454 &   NGC3641 &      26.0 &   19.3 &  4.8 &         0 &          0 &   1844.5$^{+5.1}_{-3.6}$ & 10.0$^{+1.7}_{-1.9}$ & -1.61$^{+0.13}_{-0.11}$ & 1.4$^{+0.5}_{-0.1}$ &        22.3 &           1 \\
 992 &    dE & 170.7233 &  3.2479 &   NGC3640 &      26.0 &   17.9 &  6.7 &         0 &          0 &   1431.0$^{+2.7}_{-2.1}$ & 13.9$^{+0.0}_{-0.1}$ &  -1.49$^{+0.01}_{-0.0}$ & 2.2$^{+0.0}_{-0.0}$ &        44.0 &           1 \\
1400 &    dE & 190.2123 &  7.9309 &   NGC4623 &      17.0 &   17.1 & 16.2 &         0 &          0 &   2162.6$^{+2.1}_{-1.7}$ & 11.3$^{+0.7}_{-2.7}$ &  -1.2$^{+0.07}_{-0.06}$ & 2.0$^{+0.1}_{-0.3}$ &        31.8 &           1 \\
 829 &    dE & 164.2337 &  9.4990 &           &     148.0 &   20.0 &  3.9 &         1 &          1 & 10366.3$^{+22.0}_{-2.1}$ &  6.6$^{+3.4}_{-0.5}$ & -1.77$^{+0.13}_{-0.13}$ & 1.7$^{+0.2}_{-0.6}$ &        11.8 &           1 \\
2103 &    dE & 239.0837 &  6.1882 &   NGC6017 &      29.0 &   18.0 & 14.2 &         1 &          0 &   1735.4$^{+6.0}_{-8.4}$ & 11.3$^{+2.2}_{-2.5}$ & -1.69$^{+0.01}_{-0.22}$ & 1.5$^{+0.6}_{-0.1}$ &        20.1 &           1 \\
1793 &   dEN & 213.4112 & -3.3354 &   NGC5507 &      28.0 &   19.4 &  6.4 &         0 &          0 &   1863.3$^{+7.0}_{-6.1}$ & 12.7$^{+1.3}_{-1.6}$ & -1.55$^{+0.03}_{-0.12}$ & 2.0$^{+0.2}_{-0.2}$ &        16.7 &           1 \\
  20 &   dEN &  19.5843 &  3.4333 &   NGC0474 &      31.0 &   20.7 &  4.5 &         0 &          0 & 2409.8$^{+15.8}_{-13.4}$ & 12.6$^{+0.9}_{-2.1}$ &  -1.7$^{+0.17}_{-0.28}$ & 2.2$^{+0.0}_{-0.4}$ &         9.1 &           1 \\
1497 &    dE & 191.8999 & -1.6508 &   NGC4690 &      40.0 &   21.0 &  4.3 &         0 &          0 & 2634.9$^{+15.8}_{-16.5}$ &  8.5$^{+2.6}_{-2.7}$ & -1.26$^{+0.22}_{-0.23}$ & 1.2$^{+0.8}_{-0.1}$ &         6.0 &           0 \\
1781 &   dEN & 212.7999 & -5.1036 &   NGC5493 &      39.0 &   18.6 &  5.8 &         1 &          0 &  2763.9$^{+0.1}_{-22.9}$ &  7.3$^{+4.1}_{-0.3}$ &  -2.1$^{+0.11}_{-0.11}$ & 1.2$^{+0.3}_{-0.0}$ &        17.0 &           1 \\
2086 &   dEN & 226.8657 &  1.3651 &   NGC5845 &      25.0 &   19.4 &  6.6 &         0 &          0 &   1379.6$^{+5.8}_{-5.8}$ &  8.4$^{+3.6}_{-1.0}$ &  -1.49$^{+0.1}_{-0.14}$ & 1.6$^{+0.4}_{-0.2}$ &        17.4 &           1 \\
2088 &    dE & 226.8973 &  1.9944 &   NGC5839 &      22.0 &   19.8 &  4.8 &         0 &          0 &   1172.1$^{+6.0}_{-6.4}$ & 12.2$^{+1.5}_{-3.4}$ & -1.42$^{+0.12}_{-0.07}$ & 2.0$^{+0.2}_{-0.3}$ &        17.6 &           1 \\
2091 &   dEN & 227.0116 &  2.2354 &   NGC5845 &      25.0 &   19.2 &  8.2 &         0 &          0 &   1436.3$^{+6.3}_{-5.7}$ & 10.9$^{+1.3}_{-2.2}$ & -1.82$^{+0.13}_{-0.12}$ & 1.5$^{+0.5}_{-0.1}$ &        18.9 &           1 \\
2098 &    dE & 238.7639 &  0.4664 &   NGC6010 &      31.0 &   19.0 &  4.3 &         1 &          0 &   1814.8$^{+5.0}_{-2.8}$ & 10.3$^{+1.3}_{-2.2}$ &  -1.68$^{+0.09}_{-0.1}$ & 1.3$^{+0.6}_{-0.3}$ &        31.1 &           1 \\
2094 &    dE & 238.3115 &  0.7097 &   NGC6010 &      31.0 &   18.4 &  8.3 &         0 &          0 &   2057.0$^{+4.7}_{-6.4}$ &  8.2$^{+4.2}_{-0.8}$ & -1.83$^{+0.14}_{-0.16}$ & 1.5$^{+0.4}_{-0.1}$ &        17.4 &           1 \\
1393 &   dEN & 189.9051 & -4.8152 &   NGC4602 &      34.0 &   19.3 &  4.8 &         1 &          0 &   2452.3$^{+8.5}_{-3.7}$ &  9.4$^{+1.3}_{-1.2}$ & -1.56$^{+0.06}_{-0.09}$ & 1.7$^{+0.1}_{-0.2}$ &        24.2 &           1 \\

\end{longtable}
\end{ThreePartTable}

\end{appendix}

\end{document}